\documentclass[%
 reprint,
nofootinbib,
 amsmath,amssymb,
 prd
]{revtex4-1}

\usepackage{graphicx}% Include figure files
\usepackage{dcolumn}% Align table columns on decimal point
\usepackage{bm}% bold math
\usepackage{xcolor}
\usepackage[normalem]{ulem}

\usepackage{color}

\usepackage{graphicx}% Include figure files
\usepackage{dcolumn}% Align table columns on decimal point
\usepackage{bm}% bold math
\usepackage{hyperref}% add hypertext capabilities
\usepackage{mathtools}
\usepackage{eufrak}
\usepackage{amssymb}
\usepackage{amsmath}
\usepackage{verbatim}
\usepackage{enumitem}
\usepackage{cprotect}
\usepackage{fancyvrb}
\usepackage{slashed}
\usepackage{lineno}
\usepackage{hyperref}
\hypersetup{
    colorlinks,
    citecolor=red,
    filecolor=red,
    linkcolor=red,
    urlcolor=red
}

\def\RE{{\rm Re}} % \def\RE{\mathop{\Re{\rm e}}\nolimits}
\def\IM{{\rm Im}}
\def\Tr{\mathop{\rm Tr}\nolimits}
\def\vecsigma{\boldsymbol{\sigma}}

\def\g{\Delta}

\def\un{\mathbf{1}} % \def\un{\mathbbm{1}}

\def\pv{{\bf{p}}}

\def\vv{{\bf{v}}}
\def\rv{{\bf{r}}}
\def\kv{{\bf{k}}}
\def\nv{{\bf{n}}}

\def\Lv{{\bf{L}}}
\def\Mv{{\bf M}}
\def\Nv{{\bf N}}

\def\Rv{{\bf R}}
\def\Sv{{\bf S}}
\def\Vv{{\bf V}}

\def\xv{{\bf x}}
\def\yv{{\bf y}}
\def\zv{{\bf z}}

\def\yu{\hat{\bf y}}
\def\zu{\hat{\bf z}}

\def\ni{\noindent}

\def\ie{{\sl i.e.}}
\def\eg{{\sl e.g.}}

\def\kt{     { \bf{k}_{\rm T}  }    }

\def\kpt{  \kv'_{\rm T} }
\def\ktkt{\kv^2_{\rm T}}
\def\kptkpt{{\kv'}^2_{\rm T}}

\def\pt{     { \bf{p}_{\rm T}  }    }
\def\ptpt{\pv^2_{\rm T}}

\def\T{_{\rm{T}}}
\def\L{_{\rm{L}}}

\def\bt{b_{\rm{T}}}
\def\bl{b_{\rm{L}}}
\def\LT{_{\rm{LT}}}
\def\npl{_{\rm npl}}

\def\mkp{{\bf m}} 
\def\nkp{{\bf n}} 
\def\lkp{{\bf l}} 
\def\XS{{\bf X}}
\def\YS{{\bf Y}}
\def\ZS{{\bf Z}}

\def\Mx{\rm{\textbf{m}}}
\def\Nx{\rm{\textbf{n}}}
\def\Lx{\rm{\textbf{l}}}

\def\q{\mathfrak{q}}

\def\ET{\epsilon}

\def\h{\mathfrak{h}}

\def\Mh{M}

\def\VMM{{\rm VM}}

\def\fL{f_{\rm L}}

\def\Sn{S_n}
\def\Sm{S_m}
\def\Sl{S_l}
\def\NG{N}

\def\be{\begin{equation}}
\def\ee{\end{equation}}
\def\inv{^{-1}}

\def\ptv{p_{\rm{T}}}
\def\gevc{\rm{GeV}/c}
\def\glgt{|G_{\rm L}/G_{\rm T}|}
\def\thetalt{\theta_{\rm LT}}
\def\Mhh{M_{12}}
\def\z{z_{12}}

\def\zone{z_{1}}
\def\ztwo{z_{2}}
\def\ptone{p_{1\rm T}}
\def\pttwo{p_{2\rm T}}

\def\zh{z}
\def\fvmps{f_{\rm VM/\rm PS}}
\def\case{$G\L\equiv G\T$ case.}
\begin{document}

%\preprint{APS/123-QED}

\title{Production of vector mesons in the String+${}^3P_0$ model of polarized quark fragmentation}

\author{A. Kerbizi$^{\, 1}$, X. Artru$^{\, 2}$ and A. Martin$^{\, 1}$}
\affiliation{\\ $^{1}$ {\small INFN Sezione di Trieste and Dipartimento di Fisica, Universit\`a degli Studi di Trieste,}\\
{\small Via Valerio 2, 34127 Trieste, Italy}\\
$^{2}$ \small Université de Lyon, Institut de Physique des deux Infinis (IP2I Lyon), \\ \small Université Lyon 1 and CNRS, France
}
 
\date{\today}% It is always \today, today,

\begin{abstract}
The production of vector mesons in the fragmentation process of polarized quarks is studied within the recursive String+${}^3P_0$ model, improving a previous version of the model in which the production of pseudoscalar mesons only was considered. Two types of couplings of the vector meson to quarks are introduced, their coupling constants being the additional free parameters of the model. The angular distribution of the decay products of the polarized vector meson is deduced from the spin density matrix of the meson and the spin information is propagated along the fragmentation chain taking into account the entanglement of spin states. The new model is implemented in a stand alone Monte Carlo program utilized to investigate in detail kinematic distributions and transverse spin asymmetries. The sensitivity of these observables to the new free parameters is discussed and the Monte Carlo results are compared with experimental data on transverse spin asymmetries.
\end{abstract}

\maketitle
%\tableofcontents

\section{Introduction}
The quark (and gluon) fragmentation process is one of the most intriguing and interesting phenomenon of Quantum Chromodynamics. It belongs to the soft (long-distance), non-perturbative domain and it is usually encoded in fragmentation functions (FFs). FFs are thought to be universal functions, \ie{}, common to all high energy collision processes producing jets of hadrons (for a review see Ref. \cite{Metz-Vossen-FFs}). The most studied FF is $D_{1q}^h(z,\ptv)$ which describes the fragmentation of an unpolarized quark $q$ in a not analyzed hadron $h$. The variable $z$ is the fraction of the quark energy carried by the hadron and $\ptv$ is the transverse momentum of the hadron with respect to the quark momentum. The $Q^2$ dependence of the fragmentation functions is not considered in this work.

Particularly interesting is the spin-dependent fragmentation function $H_{1q}^{h\perp}(z,\ptv)$ which describes the Collins effect in the fragmentation of a transversely polarized quark $q$ in a not analyzed hadron \cite{collins}. The effect is an azimuthal distribution of the form
\begin{eqnarray}\label{eq:distribution collins}
\frac{d^3N_{h}}{d\zh\,dp_{T}\,d\phi_h} &\propto&  1 + a^{q\uparrow \rightarrow h+X}\,|\textbf{S}_{\rm{T}}|\,\sin (\phi_h-\phi_S).
\end{eqnarray}
The angles $\phi_h$ and $\phi_{S}$ are respectively the azimuthal angles of the hadron transverse momentum and of the fragmenting quark transverse polarization $\textbf{S}_{\rm T}$ around the quark momentum. The combination $\phi_C=\phi_h-\phi_S$ is the Collins angle of the hadron and the amplitude $a^{q\uparrow\rightarrow h+X}$ of the $\sin\phi_C$ modulation for a fully polarized quark is the Collins analysing power. It is conventionally written as $a^{q\uparrow\rightarrow h+X}= -\ptv\,H_{1q}^{h\perp}/(\zh \,M\,D_{1q}^h)$, $M$ being the hadrons mass.

The Collins FF appears, coupled to the transversity parton distribution function (transversity PDF) $h_1^q$, in the so-called Collins asymmetry in semi-inclusive deep inelastic scattering (SIDIS) off transversely polarized nucleons. Neglecting the intrinsic quark transverse momentum, the asymmetry can be written as 
\begin{equation}\label{eq:Acoll}
    A_{Coll}=\frac{\sum_{q} e_q^2\,h_1^q\times a^{q\uparrow\rightarrow h+X}\,D_{1q}^{h}}{\sum_{q} e_q^2\,f_1^q\times D_{1q}^{h}},
\end{equation}
where $q=u,d,s,\bar{u},\bar{d}$ or $\bar{s}$, and $f_1^q$ is the unpolarized PDF.
The Collins asymmetry has been measured by HERMES \cite{hermes-ssa} on protons, by COMPASS on deuterons \cite{COMPASS-2005} and on protons \cite{COMPASS-collins-sivers}, and at Jefferson Lab on $\rm {}^3He$ \cite{jlab-ssa} and found different from zero for the proton target. The Collins effect can be directly accessed in $e^+e^-$ annihilation to hadrons, assuming that one knows the directions of the fragmenting quark and antiquark, by measuring the asymmetry \cite{belle-2019} 
\begin{equation}\label{eq:AUL ee general}
    a_{12}= \hat{a}_{\rm NN} \times \frac{\sum_{q}\,e_q^2\,\, a^{q\uparrow\rightarrow h_1+X}\,D_{1q}^{ h_1} \times a^{\bar{q}\uparrow\rightarrow h_2+X}\,D_{1\bar{q}}^{h_2}}{\sum_{q} \, e_q^2\,\,D_{1q}^{h_1}\times D_{1\bar{q}}^{h_2}},
\end{equation}    
where $h_1$ and $h_2$ are two back-to-back hadrons, and $\hat{a}_{\rm NN}$ is the elementary quark double transverse spin asymmetry \cite{Artru-Collins}. By combining in phenomenological analyses SIDIS and $e^+e^-$ data it has been possible to extract both the Collins FF and the transversity PDF \cite{M.B.B,Anselmino-belle-2015,Kang:2015msa}.

It is important to have a simulation model of quark fragmentation, implemented in a Monte Carlo (MC) program, reproducing the Collins effect as well as other effects like the \textit{dihadron asymmetry} \cite{Collins:1993kq,Jaffe-Jin-Tang,Bianconi-IFF} in the fragmentation of transversely polarized quarks and the \textit{jet handedness} \cite{Nachtmann,Efremov,Hayward:2021psm-Handedness} in the fragmentation of longitudinally polarized quarks. 
A promising model for the polarized quark fragmentation is the recursive String+${}^3P_0$ model \cite{DS09,DS11,DS13, kerbizi-2018, kerbizi-2019}. This model extends the Lund Model of string fragmentation \cite{Lund1983} with the inclusion of the quark spin degree of freedom. It respects confinement, it is \textit{left-right (LR) symmetric} \footnote{\textit{LR  symmetry} should better mean \textit{[Quark] Line Reversal symmetry}, namely the reversal of the quark fragmentation chain.} \cite{Lund1983} and is based on quantum amplitudes instead of probabilities. The basic assumption which explains the spin effects is that at each string breaking the quark-antiquark pairs are produced in the ${}^3P_0$ state \cite{Andersson:3P0}, namely with total spin $S=1$, relative orbital angular momentum $L=1$ and total angular momentum $J=0$.
Two slightly different versions of the String+${}^3P_0$ model have been proposed, M18 \cite{kerbizi-2018} and M19 \cite{kerbizi-2019}, the difference being the choice of an input function. Both of them are restricted to the production of pseudoscalar mesons (PS) and have been implemented in stand alone MC programs which gave similar results. In particular, they both provide a satisfactory description of the main properties of the measured Collins and dihadron asymmetries and produce also the jet-handedness effect. While M18 is more general than M19, the latter is more simple and more suitable for further developments. It has been interfaced to the hadronization part of the PYTHIA 8 event generator \cite{pythia8} to fully exploit its predictive power and to have a more complete description of the polarized SIDIS process \cite{Kerbizi:StringSpinner}.

For a more complete description of the fragmentation process, vector meson (VM) production must be considered. Hadrons coming from VM decays in fact give an important contribution to the sample of the observed hadrons.
The VM production was first included in the String+${}^3P_0$ model for the process $pp^{\uparrow}\rightarrow \rho X$ \cite{Czyzewski-vm} limited to the production of leading vector mesons which were treated as unpolarized. The main difficulty of including polarized VMs in the polarized quark fragmentation process is to take properly into account the spin correlations among the initial quark, the VM and the leftover quark in the recurring process $q^{\uparrow}\rightarrow h^{\uparrow}+q'^{\uparrow}$.

In this paper we present the new String+${}^3P_0$ model (M20), extending M19, in which the production of VMs in the polarized quark fragmentation chain is taken into account. The new model has been first presented in Ref. \cite{Kerbizi:PhD} and it is partly based on the work of Ref. \cite{DS09}.
It is assumed that vector mesons are coupled to quarks with coupling constants $G\L$ and $G\T$ for longitudinally and transversely polarized vector mesons respectively. Besides the ratio $\fvmps=|G\L|^2+2|G\T|^2$ between the abbundances of vector and pseudoscalar mesons, we have essentially two new free parameters for the spin effects, $\glgt$ governing the relative fraction of longitudinal and transverse vector mesons and $\thetalt=\arg{\glgt}$ governing the oblique polarizations, namely the interference between longitudinal and transverse polarizations.
The model is formulated at the amplitude level which automatically preserves positivity and allows to propagate the spin information along the fragmentation chain respecting quantum entanglement following the prescriptions of Refs. \cite{collins-corr,knowles-corr}.
At present M20 has been implemented in a stand alone MC program which allows to study in depth the model predictions.

The article is organized as follows. The theoretical aspects of the new model are described in Sec. \ref{section theory}. Section \ref{sec:MC implementation} describes the stand alone MC implementation of M20. The results of the simulations including the sensitivity to the free parameters are presented in Sec. \ref{sec:results}. New dihadron asymmetries arising from the possible oblique polarization of VMs are presented in Sec. \ref{sec:new 2h asymmetries}. The comparison with the existing SIDIS and $e^+e^-$ data are presented in Sec. \ref{sec:comparison}. The conclusions are given in Sec. \ref{sec:conclusions}.

\section{Vector meson production in the String+$^3P_0$ model}\label{section theory}
The fragmentation process $q_A\bar{q}_B\rightarrow h_1h_2\dots h_r\dots h_N$, where $q_A$ is a quark, $\bar{q}_B$ either an antiquark in $e^+e^-$ annihilation or the target remnant in SIDIS and $h_1h_2\dots h_r\dots h_N$ the primary produced hadrons, is phenomenologically described as the decay of a relativistic string, stretched between $q_A$ and $\bar{q}_B$ \cite{Artru-Mennessier,Lund1983}.
The string decay appears, in the infinite momentum frame, as a recursive series of elementary splittings $q\rightarrow h+q'$, $q$ is the recurring splitted quark, $h=q\bar{q}'$ the emitted hadron in the splitting and $q'$ the leftover quark. The label $r$ indicates the rank and the rank one hadron contains $q_A$. We denote by $k$, $p$ and $k'$ the four-momenta of $q$, $h$ and $q'$. We will use the null-plane components $p^{\pm}=p^0\pm p^z$ and $k^{\pm}=k^0\pm k^z$. 
The $\zu$ axis or \textit{string axis} points towards the direction of the initial quark $q_A$ in the string rest frame. The hadron momentum can then be expressed in terms of the longitudinal splitting variable $Z=p^+/k^+$ and the hadron transverse momentum $\pt=\kt-\kpt$ with respect to the string axis, $\kt$ and $\kpt$ being the transverse momenta of $q$ and $q'$ respectively. The mass shell condition writes $p^+p^-=\ET^2$ where $\ET^2=M^2+\ptpt$ is the transverse energy squared of the hadron and $M$ its mass.
The quark spin information is encoded in $2\times 2$ density matrices $\hat\rho(q) = (\un + \vecsigma \cdot \Sv_q)/2$~ where $\Sv_q$ is the quark polarization vector.

The general formalism of the String+$^3P_0$ model presented in Ref. \cite{kerbizi-2018} can include the production of mesons of arbitrary spin. The spin variable $s_h$ of the meson enters the quark-meson-quark vertex $\Gamma_{h,s_h}(\kpt,\kt)$, which is a 2$\times$2 matrix in quark spin space. $s_h$ refers to the helicity, the spin along 
a chosen transverse axis or, for spin 1, specifies a linearly polarized state. To make a full Monte Carlo 
simulation of quark fragmentation with PS and VM production, we must add two items to the prescriptions of M18 or M19, namely  
\begin{itemize}\itemsep0em 
\item[-] the generation of $\Mh$ from a continuous resonant mass spectrum,
\item[-] the simulation of the decay process. 
\end{itemize}
The second point deserves special attention, because the spin state of the $(hq')$ system is generally entangled. One cannot simulate separately the decay of $h$ and the fragmentation of the leftover quark $q'$. 

 \subsection{The $q\!\uparrow\,\to \VMM\!\uparrow + q'$ splitting function}\label{sec:elementary splitting for vm}

\subsubsection{General formula}

Let us start by including the emision of VMs in the formalism of M19 without treating the subsequent decay. To label the VM spin state, we replace $s_h$ by the 3-vector
$\Vv$ which is the space part of the covariant amplitude $A^\mu$ in the VM rest frame, as specified in subsection \ref{defineV}.   
$\Vv$ is real for linear polarization, complex for vector polarization and normalized by $\Vv\!\cdot\!\Vv^*=1$.
The probability density of emitting $h$ in the elementary splitting is given by the \textit{splitting function}, which, when summing over the spin states of $q'$, writes (cf. Eq. (36) of \cite{kerbizi-2018})
% to be used in Eq. (\ref{split-dis})   
%
\begin{eqnarray}  
\label{polar-split-mat}   
&&F_{q',h,q}(\Mh,\Vv,Z,\pt;\kt,\Sv_q) 
\equiv  \frac{dN(q\uparrow\to h\uparrow+q')}{d\Mh^2\,d^2\pt \, dZ/Z}
\nonumber \\ &&=
 \Tr \left[  T(\q',\h,\q) \ \rho(q)  \ T^\dagger(\q',\h,\q) \right] .
\end{eqnarray}

The gothic letter $\h \equiv \{h, p, s_h\}$ = \{hadron species, 4-momentum, spin state\} represents the meson state, whereas  $\q\equiv\{q,k\}$ = 
\{quark flavor, quark momentum\} represents the quark state, spin excluded. 
$T$ is the 2$\times$2 \textit{splitting matrix}, given by
\begin{eqnarray} \label{eq: T final vm}  
&& T(\q',\h,\q)  =  
C_{q',h,q} \, D_h(\Mh)\,  \check{g}(\ET^2)  \,  [(1-Z)/\ET^2]^{a/2} 
\nonumber\\ &&\times  \,  \exp[ -\bl \,\ET^2/(2Z)] 
\,\Delta_{q'}(\kpt) \,  \Gamma_{h,\Vv} \, \hat u_q^{-1/2}(\ktkt) \,.  
\end{eqnarray}
%
%$\Mh$ and $\ET\equiv(\Mh^2+\pt^2)^{1/2}$ are the hadron mass and transverse energy~; $Z\equiv p^+/k^+$.

The coefficient $C_{q',h,q}$ is proportional to the $(\bar q' q)$ wave function in flavor space; 
 $a$ and $\bl$ correspond to the parameters $a$ and $b$ of the Lund Model. % {\it string fragility} parameter.      
$\check g(\ET^2) $ is a model input function which, like in M19, we take 
\begin{eqnarray} \label{Na} % (\ref{Na}) 
\nonumber &&\check g^2(\ET^2) =1/ N_a(\ET^2), \\
&& N_a(\ET^2) = \int_0^1 \frac{dZ}{Z} \, \left(\frac{1-Z}{\ET^2}\right)^a \, \exp( -\bl \,\ET^2/Z)\,.
\end{eqnarray}
The 2$\times$2 matrix    
\be    \label{gamma(q)}    
\g_q(\kt)   =  (\mu + \sigma_z\, \vecsigma\cdot\kt) \, f\T(\ktkt)  
\ee
contains the spin and $\kt$ dependence of the quark propagator in the String+$^3P_0$ model, $\mu$ being a complex mass parameter and
$f\T(\ktkt)$ a fast decreasing function of $\ktkt$, mainly responsible for the transverse momentum cutoff. We take $f\T(\ktkt)=\exp{\left(-\bt\ktkt/2\right)}$, with $\bt$ a free parameter already present in the Lund Model.

$D_h(\Mh)$ has the denominator of the vector meson propagator. We take the Breit-Wigner form
\begin{equation}\label{eq:BW distribution}
    D_h(\Mh)=\frac{ \sqrt{N_{D}} }{\Mh^2-\overline{m}_h^2+i\,\overline{m}_h\gamma_{h}} \,;
\end{equation}
where $\overline{m}_h$ is the position of the resonance peak and $\gamma_{h}$ the resonance width, both set to the values in PDG \cite{PDG2019}. $N_{D}$ is the normalization constant of the mass distribution $|D_h(M)|^2$ of the resonance.

The 2$\times$2 matrix $\Gamma_{h,\Vv}$ sets the coupling of the vector meson to the quark line. Its most simple form is  \cite{DS09}
 \begin{eqnarray}\label{eq: Gamma vm}
 \Gamma_{h,\Vv} &=& G\T\,\boldsymbol{\sigma}\T\cdot \textbf{V}\T^* \, \sigma_z + G\L\,V_z^* \times \un \,.
 %  = \Gamma_{h,\alpha}V_\alpha^*
 \end{eqnarray}
% It replaces the $\Gamma_h=\sigma_z$ of pseudoscalar mesons. % emission.
 $G\L$ and $G\T$ are the coupling constants for longitudinal and transverse linear polarizations of the VM. 
 This decomposition is analogous to that in $G_{\rm M}$ and $G_{\rm E}$ of the nucleon form factor and that in $^3S_1$ and $^3D_1$ of the deuteron 
wave function. In a covariant quark-multiperipheral model the analogue couplings would be $\gamma^{\mu}$ and $\sigma^{\mu\nu}p_{\nu}$.
We allow $G\L/G\T$ to be complex, as a result of different quantum actions of the initial string for the L and T polarizations. In the following we will use as parameters
\be \label{fraction[L]}
\glgt 
\,, \quad \theta\LT = \arg(G\L/G\T) \,,
\ee
which are the new free parameters of M20. A relevant quantity is the fraction
\be \label{fraction[L]}
\fL = \frac{|G\L|^2}{2|G\T|^2+|G\L|^2}\,;
\ee
for $\Sv_q={\bf 0}$ it is $\fL=\hat\rho_{\rm ll}$ (see Eq. 
(\ref{eq:rho vm MNL decomposition})), hence $\fL$ is the fraction of the longitudinally polarized vector mesons.

$\hat u_q$ is a 2$\times$2 matrix given by Eq. (47) of \cite{kerbizi-2018}, which we decompose as 
\begin{eqnarray} \label{eq:uq}
%\begin{aligned} 
\hat u_q(\kt)  &=& \sum_{h} \hat u_{q,h}(\kt) \,,
\\
 \hat u_{q,h}(\kt) &=&
 |C_{q',h,q}|^2   \int d^2{\kpt} \, \check g^2(\ET^2)  
\, N_a(\ET^2)% f\T^2(\kptkpt) \, 
\nonumber \\
&\times & 
\sum_{s_h}  \Gamma^\dagger_{h,s_h}  \
\g^\dagger_{q'}(\kpt)\,\g_{q'}(\kpt) \ 
\Gamma_{h,s_h}  \,.
\label{uchapeau}  
%\end{aligned} 
\end{eqnarray}
%   with $\hat u_0 > \hat u_1$. 
% 
% $\hat u_{q,h}(\kt)$ 
$\hat u_{q,h}(\kt)$ is the contribution of hadron species $h$. For vector mesons, ${s_h}=\Vv$ and $\sum_\Vv$ is made over three orthonormal basic vectors \footnote{Due to the fluctuacting mass of the VM, one should insert $\int d\Mh^2 |D_h(M)|^2$ in Eq. (\ref{uchapeau}) before the integral over $\kpt$.}. For pseudoscalar mesons it is $s_h\equiv0$ and $\Gamma_h=\sigma_z$.  
%...........................
%For meson  and Eq. (\ref{eq: Gamma vm}) with $s_h\equiv\Vv$ for vector mesons, we have

With our choice $\check g^2(\ET^2) \, N_a(\ET^2) = 1$ (choice of M19) both $\hat u_{q}(\kt)$ and $\hat u_{q,h}(\kt)$ become proportional to the unit matrix and independent of $\kt$:
\begin{eqnarray}\label{eq: uq M20}
    \hat{u}_{q,h}  &=& 
    \textbf{1} \times    |C_{q',h,q}|^2 \left(|\mu|^2+\langle\ktkt\rangle_{f\T} \right) 
   \nonumber \\ &\times&
    \left\{ \begin{array}{cl}
    1 & \text{(PS case)} \\
       (2|G\T|^2+|G\L|^2) &  \text{(VM case)}  % \int d\Mh \, |D_h(\Mh)|^2
    \end{array} \right.
\end{eqnarray}
with the notation
\be
\langle\ktkt\rangle_{f\T} \equiv \int d^2\kt \,\ktkt \, f\T^2(\ktkt) \, \big/ \int d^2\kt \, f\T^2(\ktkt) \,,
\ee
where $f\T$ is the function appearing in Eq. (\ref{gamma(q)}).
So, from now $\hat u_q(\ktkt)$ and $\hat{u}_{q,h}(\ktkt)$ will be considered as constant numbers and we will omit ``$\textbf{1} \times $'' which appears in Eq. (\ref{eq: uq M20}). The relative probability of getting the hadron species $h$ in the splitting $q\to h+q'$ is then 
\be \label{Proba(h)}
P(q\to h+q') = \hat{u}_{q,h}/\hat u_q \,.
\ee
It is independent of $\kt$ and of the polarization of $q$, contrary to other choices of the function $\check g(\ET^2)$.

 \subsubsection{Frame for the polarization vector $\Vv$} \label{defineV}% [of the vector meson]}
%\subsection{[defineV] Boost compositions bringing the meson at rest}% erivation of $\Vv$ from $A^\mu$ and different bases of linear polarization} 
    % $A^\mu\exp-ip_h\cdot X$ i

\begin{figure} % [b!]  % =Fig.1 %%%%%%%%%%   FIGURE   %%%%%%%%%%%%%
\begin{minipage}{.5\textwidth}
 \includegraphics[width=0.9\textwidth]{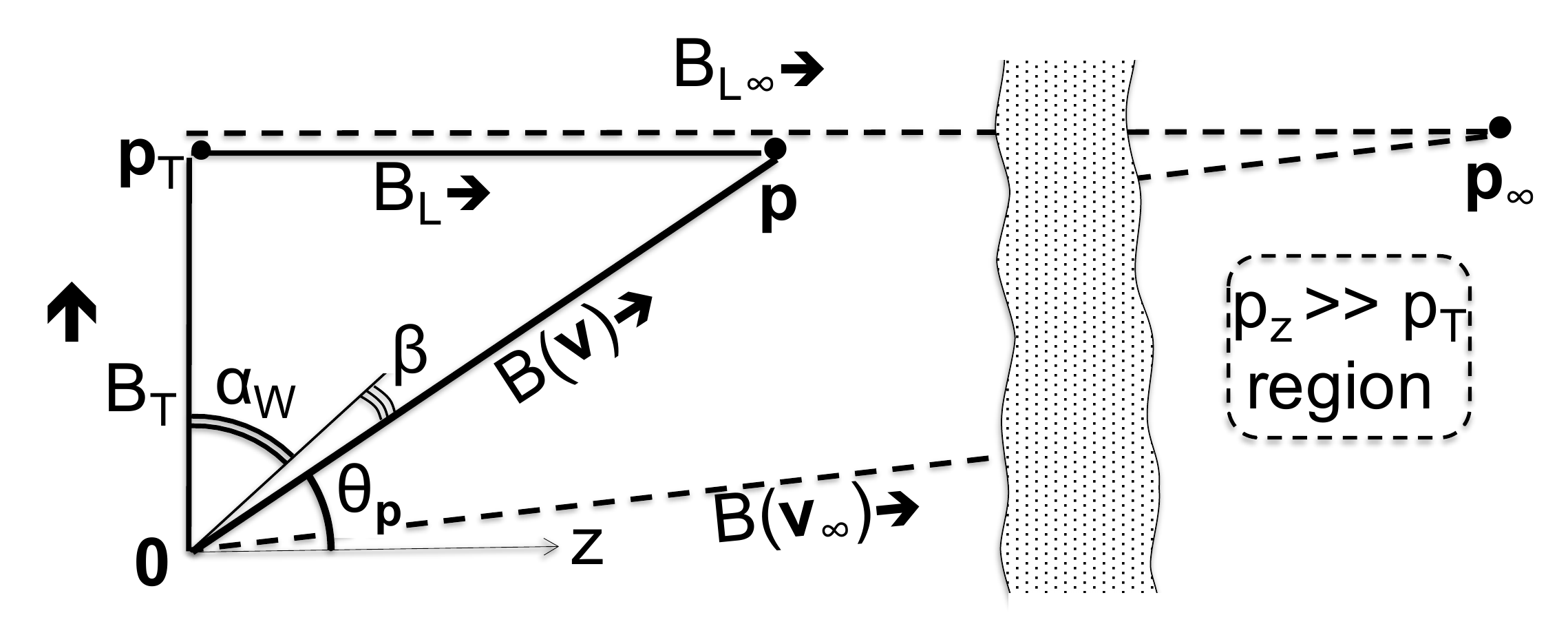}
\end{minipage}
\caption{ Boost compositions involved in the definition of $\Vv$ and the Wigner angle $\alpha_{\rm W}$, following
Eqs. (\ref{BL,BT}) - (\ref{Wigner}).  
 $\vv=\pv/E$. 
The dashed lines figure the $p_z\to\infty$ limit, Eq. (\ref{W-infini}).
}\label{COMPOSE_BOOSTS}
\end{figure}

$\Vv$ is obtained from the covariant 4-vector $A^\mu$ of the VM wave function by bringing the VM at rest via two successive Lorentz boosts: a longitudinal one $B\L\inv$ which suppresses $p_z$ and a transverse one $B\T\inv$ which suppresses $\pt$, where
\be  \label{BL,BT}
B\T=B(\pt/\ET) \,, \quad B\L=B(p_z\zu/E)
\ee
and the argument of $B$ is the velocity vector of the boost. The action of $B\T$ and $B\L$ is shown in Fig. \ref{COMPOSE_BOOSTS}.
%\be
% (0,\Vv) = \Lambda(\pt\to{\bf0}) \,   \Lambda(\pv\to\pt)\, (A^0,{\bf A})
%\ee
%where $\Lambda(\pv\to\pv')$ is the matrix of the Lorentz boost changing $p=(E(\pv),\pv)$ into $p'=(E(\pv'),\pv')$, with $E(\pv)=(m^2+\pv^2)^{1/2}$. %$ (0,\Vv) =\Lambda(\pv\to{\bf0})\,(A^0,{\bf A})$. 
Thus $(0,\Vv) = B\T\inv B\L \inv A^\mu$. This transformation preserves the longitudinal Lorentz invariance and 
the LR symmetry \cite{Lund1983} of the model. We call {\it LR symmetric (rest) frame} the resulting reference frame for $\Vv$ (also named $PL$ $frame$ in Ref. \cite{Minaenko:1993ne}).

The VM could have been put at rest with the direct boost $B\inv(\pv/E)$, leading to a different vector $\Vv_{\rm hl}$ (``hl'' refers to the so-called {\it helicity frame}). We have $(0,\Vv) =  {\cal R}_{\rm W}\,(0,\Vv_{\rm hl}) $ where 
${\cal R}_{\rm W} =B\T\inv B\L \inv B(\pv/E)$  
is a {\it Wigner rotation} about $\zu\times\pv$, of angle $\alpha_{\rm W}(\pv/E)$ given by
\begin{eqnarray}   \label{Wigner}
\alpha_{\rm W}(\pv/E) 
 &=& % \nonumber 
\arcsin\left(\frac{p_z p\T}{E\ET+\Mh\ET}\right)
\\ \nonumber &=&
\frac{\pi}{2} - \theta_\pv - \beta \,; 
\quad \beta =  % &=&
\arcsin\left(\frac{\Mh p_z}{\ET|\pv|}\right)   
\end{eqnarray}
%
%% =\arcsin[p_z p\T/(E\epsilon+\Mh\epsilon)] \,,    
and represented in Fig. \ref{COMPOSE_BOOSTS}%
\footnote{
Other expressions are: 
$\alpha_{\rm W} = \Nv\cdot \int_C (\pv\times d\pv) /(E^2+mE)  = \Mh\int\!\int_A d^2\pv/E^3 $, 
where $C$ is the closed path run by the vector $\pv$ in the successive boosts   $B(\pv/E)$,  $B\L \inv$,   $B\T\inv$ and $A$ is the area enclosed by $C$, and $\Nv$ is defined in Eq. (\ref{LMN}).
}.
$\Vv_{\rm hl}$ and $\alpha_{\rm W}(\pv/E)$ are not invariant under a longitudinal boost. In particular, in the SIDIS process they change from the target frame to the $\gamma^*$-nucleon frame. When $p_z\to+\infty$ (dashed lines of Fig. \ref{COMPOSE_BOOSTS}) the helicity frame becomes the {\it null plane (rest) frame} and
\begin{eqnarray}
\Vv_{\rm hl} &\to& \Vv\npl \,. \label{Vhl,infini}\\
\label{W-infini}
\alpha_{\rm W}(\pv/E) &\to& \alpha_{\rm W \infty}=\arctan(p\T/\Mh) \, .
\end{eqnarray}
%
%The Wigner rotation $\Lambda^{\pv\,``\infty''} =B\T\inv B\L \inv(\pv\,``\infty'')  B(\pv\,``\infty'')$
$\Vv\npl$ is longitudinally Lorentz invariant but not LR symmetric. 

%%%%%%%%%%%%%%%%%%%%%%%%%%%%%%%%%%%%%%%%%%%%%
 %%%%%%%%%%%%%%%%%%%%%%%%%%%%%%%%%%%%%
%%%%%%%%%%%%%%%%%%%%%%%%%%%%%%%%%%%%%%%%%%%

The above transformations also serve to adjust the individual momenta $\pv^*_1$,  $\pv^*_2$, ... or the relative momenta of the decay products in the rest frame of the VM. 
For a 2-body decay, the relative 4-momentum 
$p_{rel}^\mu=(E^*_2\, p_1^\mu - E^*_1 \, p_2^\mu)/M$
(with $E^*_i=P\cdot p_i/M$ and $M^2=P^2\equiv(p_1 \!+\! p_2)^2$) transforms like $A^\mu$~; 
\be \label{p-relatif}
\rv\equiv\pv_{rel}^* = \pv^*_1 = - \pv^*_2
\ee
is the analogue of $\Vv$ and
\be  \label{RT}
\Rv=  \left(
\begin{array}{c}
 \Rv\T  \\
 R_z   
\end{array}
\right)
= 
\frac{1}{z_1+z_2} \left(
\begin{array}{c}
 z_2\,\pv_{1\rm T}-z_1\,\pv_{2\rm T}  \\
E^*_2\,z_1-E^*_1\,z_2 
\end{array}
\right)
% (z_2\,\pv_{1\rm T}-z_1\,\pv_{2\rm T}  , \, E^*_2\,z_1-E^*_1\,z_2 )}
\ee  
is the analogue of $\Vv\npl$. 
$\Rv$ is obtained from $\rv$ in the LR symmetric frame by the rotation of angle $-\alpha_{\rm W\infty}$ about 
$\zu\times\pt$. 
$\Rv\T$ is the relative transverse momentum involved in the dihadron asymmetry, 
whether or not $h_1$ and $h_2$ come from a resonance.
It was introduced and named $\boldsymbol{\kappa}_\tau$ in Ref. \cite{LEUTWYLER197894}.  
%$\Rv\T=(z_2\pv_{1\rm T}-z_1\pv_{2\rm T})/z_1+z_2)$ is the $(x,y)$ part 
%of the relative momentum $\pv^*_1=-\pv^*_2$ obtained with $\Lambda^{(+\infty)}$.

\subsubsection{Coordinates in the rest frame}      % $\Vv^{(+\infty)}$
Independently of the choice of frame (LR symmetric or null-plane) we need three basic vectors to define the  coordinates of $\Sv_q$, $\Vv$, $\pv^*_1$ or $\Rv$. A natural basis, $\{\Lv,\,\Mv,\,\Nv\}$, is linked to the meson transverse momentum $\pt$~:
%%  , is 
%
\be \label{LMN}
\Lv=\zu, \ \Mv= \pt/|\pt|, \  \Nv = \zu \times \Mv \,. 
\ee 
%
%%with the general notations
%%%
%%\be \label{tilde}
%%\univec({\bf v}) \equiv \hat{\bf v} \equiv  {\bf v}/|{\bf v}|\,; \  \tilde{\bf v} \equiv \zu\times {\bf v}\,.
%%\ee
%%%
We will also use the $\{\lkp,\mkp,\nkp\}$ basis linked to the antiquark momentum $-\kpt$,
\be \label{lmn}
\lkp=\zu, \ \mkp= -\kpt/|\kpt|, \ \nkp = \zu\times\mkp \,, 
\ee 
and the $\{\XS,\YS,\ZS\}$ basis linked to the quark transverse polarization,
\be \label{XYZ}
 \YS= {\Sv_q}\T/|{\Sv_q}\T| \,, \  \ZS=\zu \,, \ \XS = \YS\!\times\!\ZS\,.
\ee 
These bases are simply related by rotations about $\zu$. 
From now on, $\Sv_q$, $\Sv_q\!\cdot\!\nkp$, $\Sv_q\!\cdot\!\XS$ etc. will shortly be written $\Sv$, $S_n$, $S_X$ etc. 
To a vector $\Vv$ is associated the pure spin state of the vector meson 
 \begin{eqnarray}
\big|\Vv\big\rangle &=& V_L \big|\Lv\big\rangle + V_M \big|\Mv\big\rangle + V_N \big|\Nv\big\rangle 
\nonumber \\
&=&
V_l \big|\lkp\big\rangle + V_m \big|\mkp\big\rangle + V_n \big|\nkp\big\rangle \,, \text{ etc.}
\end{eqnarray}
Note that $\big|\Vv\big\rangle$ 
and $\big|-\Vv\big\rangle$ are the same state.

\subsubsection{Splitting function for pseudoscalar mesons} 

Before studying the splitting function for vector mesons, let us first recall the one for pseudoscalar mesons (cf. Eq. (26) of \cite{kerbizi-2019}). Removing the argument $\Vv$ in Eq. (\ref{polar-split-mat}) and using Eqs. (\ref{eq: T final
vm}-\ref{gamma(q)}) and (\ref{eq: Gamma vm}-\ref{Proba(h)}), one gets 
\begin{eqnarray}\label{eq:F_explicit PS}
\nonumber  & & F_{q',h({\rm PS}),q}(Z,\pt;\kt,\textbf{S}_q)= \frac{\hat{u}_{q,h}}{\hat{u}_{q}}
\, \frac{f^2\T(\kptkpt)}{|\mu|^2+\langle \ktkt\rangle_{f\T}}  
%\nonumber \\ & & \times 
% \left[ (|\mu|^2+\kptkpt) - 2 \,\IM(\mu)\,k'\T \, \Sv_{q}\cdot\tilde{\textbf{n}}(\kpt) \right] 
\\  & & \times \ 
{N_a\inv(\ET^2)} \left(\frac{1-Z}{\ET^2}\right)^a \,{\exp{(-\bl \ET^2/Z)}} 
\\  \nonumber  & & \times \ 
(|\mu|^2+\kptkpt)   \left[ 1+\hat a \, \Sn \right]  \,,%\Sv_{q}\cdot\Nx \right]  \,,
\end{eqnarray}
with 
\be \label{hata}
\hat a \equiv % {2\,\IM(\mu)\,\rm{k'}_{\rm{T}}}/({|\mu|^2+\kptkpt}) .% 
\frac{2\IM(\mu)\,\rm{k'}_{\rm{T}}}{|\mu|^2+\kptkpt} > 0 \,.
\ee
The square bracket of Eq. (\ref{eq:F_explicit PS}) is responsible for the Collins effect, since $\nkp$ is correlated with  $\Nv$. In particular, 
$\nkp=\Nv$ for a rank one hadron.

\subsubsection{Splitting function for vector mesons} 

In the case of vector mesons, selecting one polarization of the $\{\Lx,\Mx,\Nx\}$ basis, we obtain from 
Eqs. (\ref{polar-split-mat}-\ref{gamma(q)}), Eqs. (\ref{eq: Gamma vm}-\ref{Proba(h)}) and Eq. (\ref{hata})
\begin{eqnarray} \label{Flmn}  
&&F_{q',h,q}(\Mh,\Vv,Z,\pt,\kt,\Sv_q) = 
\nonumber \\  && % \times \ 
\frac{\hat{u}_{q,h}}{\hat{u}_{q}} \,|D_h(\Mh)|^2
\, f^2\T(\kptkpt)\, \frac{(|\mu|^2+\kptkpt) }{|\mu|^2+\langle \ktkt\rangle_{f\T}}  
 \nonumber \\   && \times \ 
{N_a\inv(\ET^2)} \left(\frac{1-Z}{\ET^2}\right)^a \,{\exp{\left(-\bl \ET^2/Z\right)}} 
 %  \left[ 1 + \fL \, \hat a \, \Sv_{q}\cdot\tilde{\textbf{n}}(\kpt) \right].
 \nonumber \\   &&\times 
 \left\{ \begin{array}{ll}
    (1-\hat a \, \Sn) \,  \fL  & \text{ for }  \Vv=\Lx   
    \\
     (1-\hat a \, \Sn) \,  (1-\fL) /2 & \text{ for }  \Vv=\Mx  
    \\
     (1+\hat a \, \Sn) \,  (1-\fL) /2 & \text{ for }  \Vv= \Nx  
     \\
     1 - \fL \, \hat a \, \Sn & \text{ for the sum over }  \Vv  
    \end{array} \right.
\end{eqnarray} 

The last line after the brace is for the case where the VM polarization is not analyzed. 
Equation (\ref{Flmn}) with this choice and Eq. (\ref{eq:F_explicit PS})
are used in simulations to generate first the vector or pseudoscalar meson species of the emitted particle, then its transverse momentum $\pt=\kt-\kpt$, then its $Z$. 
% $\Sv_{q}\cdot\Nx$ <-> \Sn
\paragraph{Global Collins effect.}
It is the Collins effect of the vector meson and comes from the $\Sn$ term of Eq. (\ref{Flmn}).
It is to be distinguished from the dihadron asymmetry (or relative Collins effect) of the decay products. These have individual Collins effects resulting from both the global and the relative one.

\begin{figure} % [b!]  % =Fig.1 %%%%%%%%%%   FIGURE   %%%%%%%%%%%%%
\begin{minipage}{.5\textwidth}
 \includegraphics[width=0.9\textwidth]{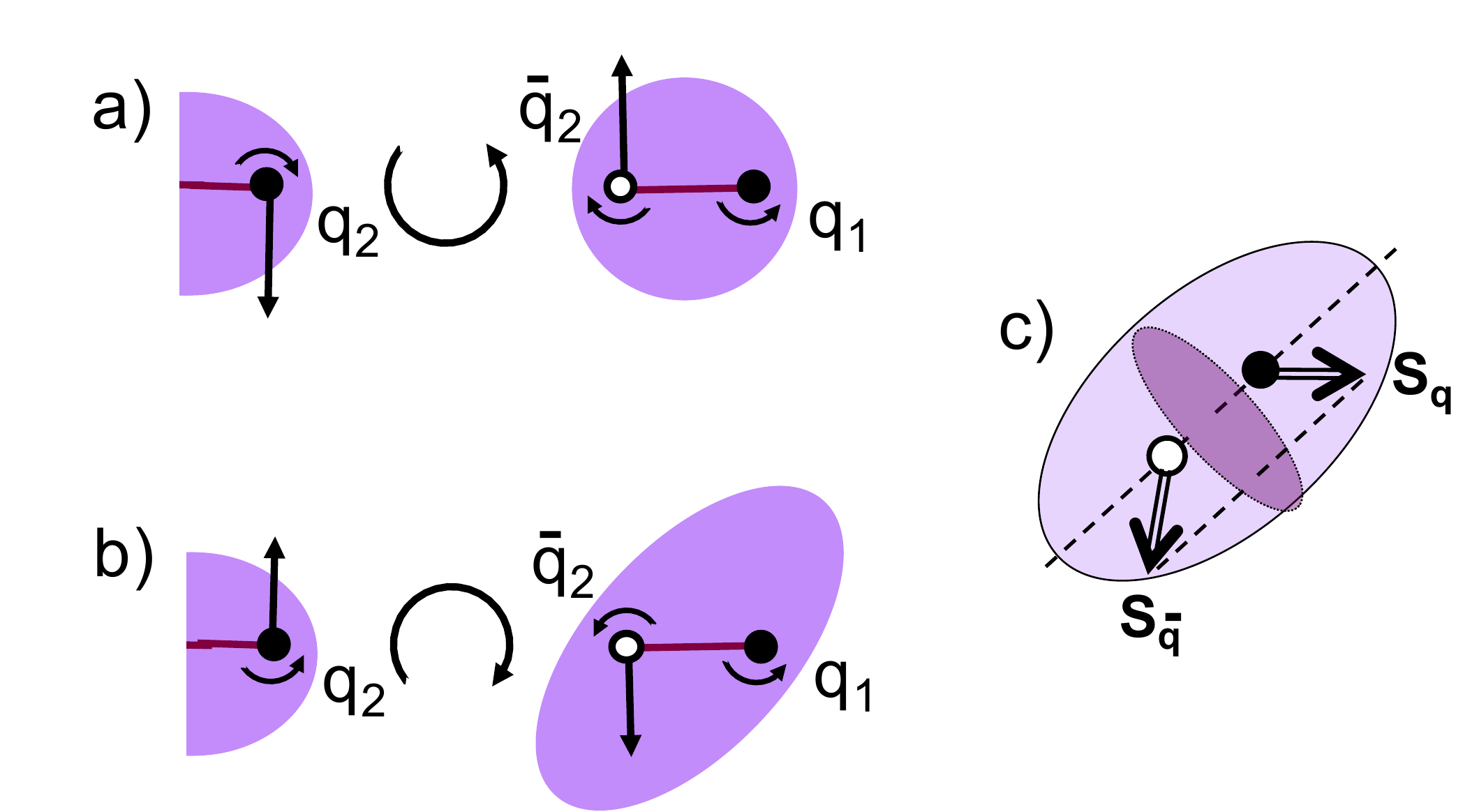}
\end{minipage}
\caption{ Production mechanism of a first-rank meson in the string+$^3P_0$ model. 
a) PS meson or VM of linear polarization perpendicular to the figure plane (the $\zu,\pt$ plane). b) VM of linear polarization in this plane. 
 Straight simple arrows represent quark or antiquark momenta. Circular arrows represent quark spins or $q\bar q$ relative orbital momenta. 
  Linearly polarized VM's are represented by ellipsoids.  
 c) Correlations between the ellipsoid major axis, the $q$ spin and the $\bar q$ spin. 
%The general correlation between $\Sv_q$ and $\Sv_{\bar q}$ inside a linearly polarized vector meson (represented by an elongated ball).
}\label{leadingVM}
\end{figure}

For a rank one meson, $-\kpt=\pt$ and $\Nx=\Nv$. Then the analysing power $a^{q_A\uparrow \rightarrow h+X}(z,\ptv)$
is equal to the coefficient of $ \Sn$ in Eq. (\ref{eq:F_explicit PS}) for PS mesons or (\ref{Flmn})
for VM. It is maximum for $|\pt|=|\mu|$. For the VM, it depends on the linear polarization, as pictured in Fig. \ref{leadingVM}, which gives a semi-classical description of PS and VM production in the model. 
 If the VM polarization is normal to the production plane as in Fig. \ref{leadingVM}a, the Collins asymmetry equals that of a pion of the same $|\pt|$. If the polarization is in the production plane as in Fig. \ref{leadingVM}b, the asymmetry is opposite to that of a pion.

\begin{figure} % [b!]  % =Fig.1 %%%%%%%%%%   FIGURE   %%%%%%%%%%%%%
\begin{minipage}{.5\textwidth}
 \includegraphics[width=0.9\textwidth]{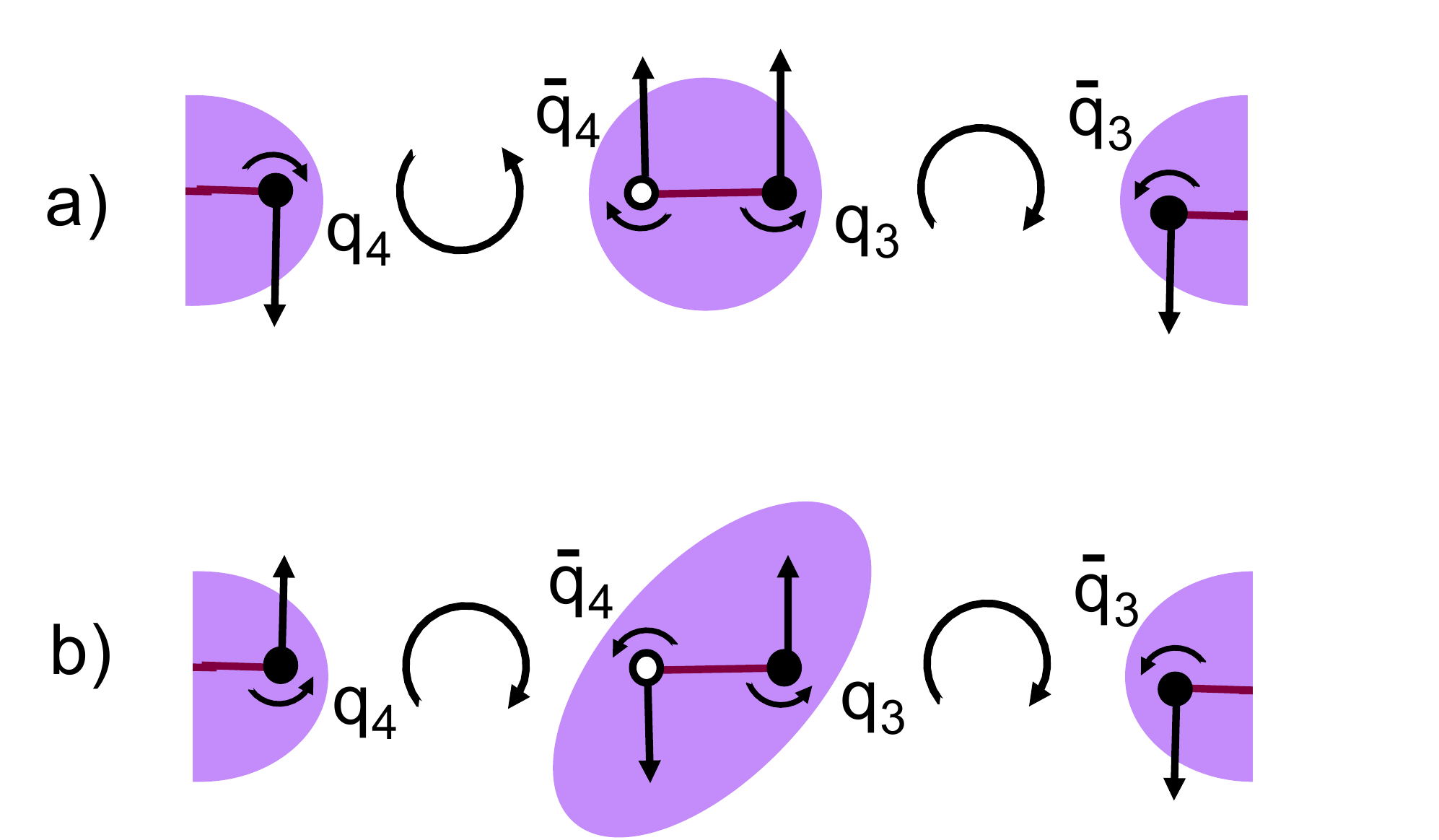}
\end{minipage}
\caption{ Production mechanism of a rank $\ge$ 2 meson, here 
the $q_3\bar q_4$ meson.
Same notations as in Fig. \ref{leadingVM}.
}\label{prodVM}
\end{figure}

\paragraph{A ``hidden spin'' effect.}
Figure \ref{prodVM} is the analogue of Fig. \ref{leadingVM} for a meson of rank $\ge2$. It shows that the quark and the antiquark transverse momenta 
are on the same side for a PS meson (Fig. \ref{prodVM}a).  The same occurs for a VM with probability $(1-\fL)/2$. In the case of Fig. \ref{prodVM}b, which occurs with probability $(1+\fL)/2$, the $q$ and $\bar q'$ momenta are on the opposite sides. So, $\langle \pt^2 \rangle$ is expected to be larger  for PS mesons than for VMs. This prediction is independent on the polarization of the initial quark and specific of the $^3P_0$ mechanism. It could be tested in unpolarized 
experiments, looking at ``unfavored'' quark fragmentation or at the central rapidity region.

%%%%%%%%%%%%%%%%%%%%%%%%%%%%%%%%%%%%%%%%%%%%%
 %%%%%%%%%%%%%%%%%%%%%%%%%%%%%%%%%%%%%
%%%%%%%%%%%%%%%%%%%%%%%%%%%%%%%%%%%%%%%%%%%

%%%%%%%%%%%%%%%%%%%%%%%%%%%%%%%%%%%%%%%%%%%%%
 %%%%%%%%%%%%%%%%%%%%%%%%%%%%%%%%%%%%%
%%%%%%%%%%%%%%%%%%%%%%%%%%%%%%%%%%%%%%%%%%%

\subsection{The density matrix of the vector meson} \label{theVMdensitymatrix}

Rewriting $T(\q',\h,\q)$, defined in Eq. (\ref{eq: T final vm}), as  $T_\alpha(\q',\h,\q)\, V_\alpha$, 
   %   (\ref{eq: T final vm})  (\ref{polar-split-mat})
the relative probability to find the VM in a state $|\Vv\rangle$ is of the form 
\be \label{VrhoV}
\big\langle\Vv\big| \hat\rho(h) \big|\Vv\big\rangle
= V^*_{\alpha} \, \hat\rho_{\alpha\alpha'}(h) \, V_{\alpha'}
\ee
where $\alpha$ and $\alpha' \in \{L,M,N\}$ or $\{l,m,n\}$ or $\{X,Y,Z\}$, depending on the basis, and  
\begin{eqnarray}\label{eq:rho vm matrix elements}
&& \hat\rho_{\alpha\alpha'}(h) =  
\frac{ \Tr \left\{T_\alpha \,\hat\rho(q)\,T_{\alpha'} ^\dag\ \right\} }
{\sum_{\beta}\Tr  \left\{T_\beta \, \hat\rho(q)\, T_{\beta} ^\dag \right\}} 
 \\ \nonumber && = \frac{
\Tr  \left\{(\mu + \sigma_z\, \vecsigma\!\cdot\!\kpt) \,\Gamma_{h,\alpha} \, \hat\rho(q) \, 
\Gamma^\dag_{h,\alpha'} \, (\mu^* +  \vecsigma\!\cdot\!\kpt \, \sigma_z) \right\} }
{(|\mu|^2+\kptkpt) \,   \NG(\Sv) } %  \overline{G^2} }
\end{eqnarray}
with 
\be \label{NG}
%N(\Sv_q) =  2|G\T|^2  + |G\L|^2 + |G\L|^2\,\hat a\, \Sv_q\!\cdot\!\tilde\nv(\kpt) \,.
\NG({\Sv}) =  2|G\T|^2  + |G\L|^2 - |G\L|^2\,\hat a\, \Sn \,.
\ee
$\hat\rho_{\alpha\alpha'}(h)$ is the polarization matrix or {\it (spin) density matrix} of the VM, normalized to $\Tr \hat\rho(h)=1$.

\begin{figure} % [b!]  % =Fig.1 %%%%%%%%%%   FIGURE   %%%%%%%%%%%%%
\begin{minipage}{.45\textwidth}
 \includegraphics[width=0.8\textwidth]{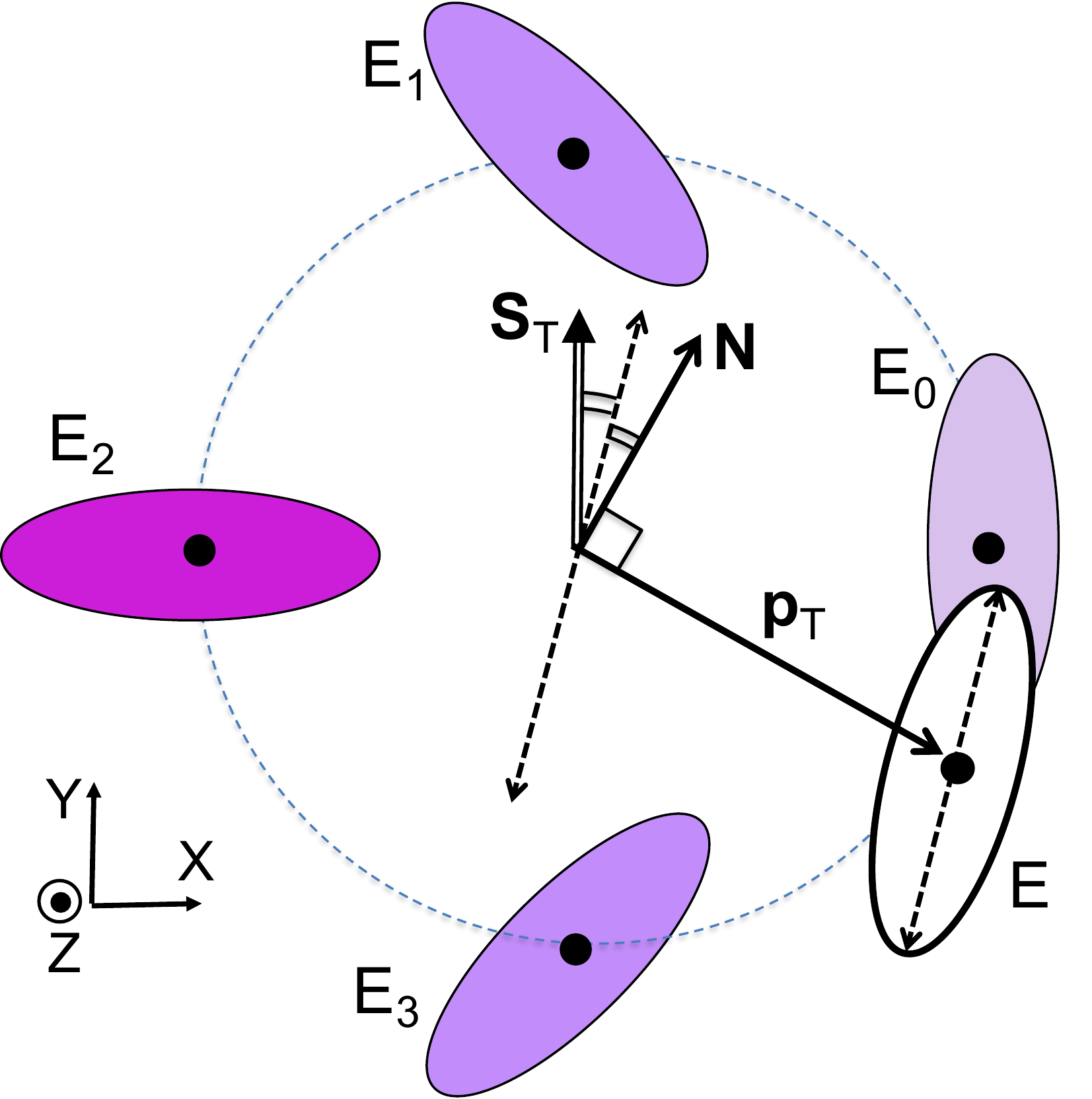}
\end{minipage}
\caption{ 
Transverse polarization of a first-rank vector meson, for azimuths $\phi(\pt)=j\pi/2$ ($j=0,...3$) of $\pt$  
in the (X,Y) frame. %relative to ${\Sv_q}\T$.  
The ellipses are projections of the polarization ellipsoids. %  . 
%E$_i$ are the ellipses for %i=1,...4 are for the $\pt$ 
%azimuths $(i-1)\pi/2$. 
The darknesses of E$_j$ figure the depths $\propto(\hat{\rho}_{ZZ})^{1/2}$ of the ellipsoids in the $Z$ direction. Ellipse E is for an ordinary azimuth. 
Its major axis (dashed) is parallel to the bisector of ${\Sv}\T$ and $\Nv$.       
}\label{fig.poltransVM}
\end{figure}

\begin{figure} % [b!]  % =Fig.1 %%%%%%%%%%   FIGURE   %%%%%%%%%%%%%
\begin{minipage}{.45\textwidth}
 \includegraphics[width=0.95\textwidth]{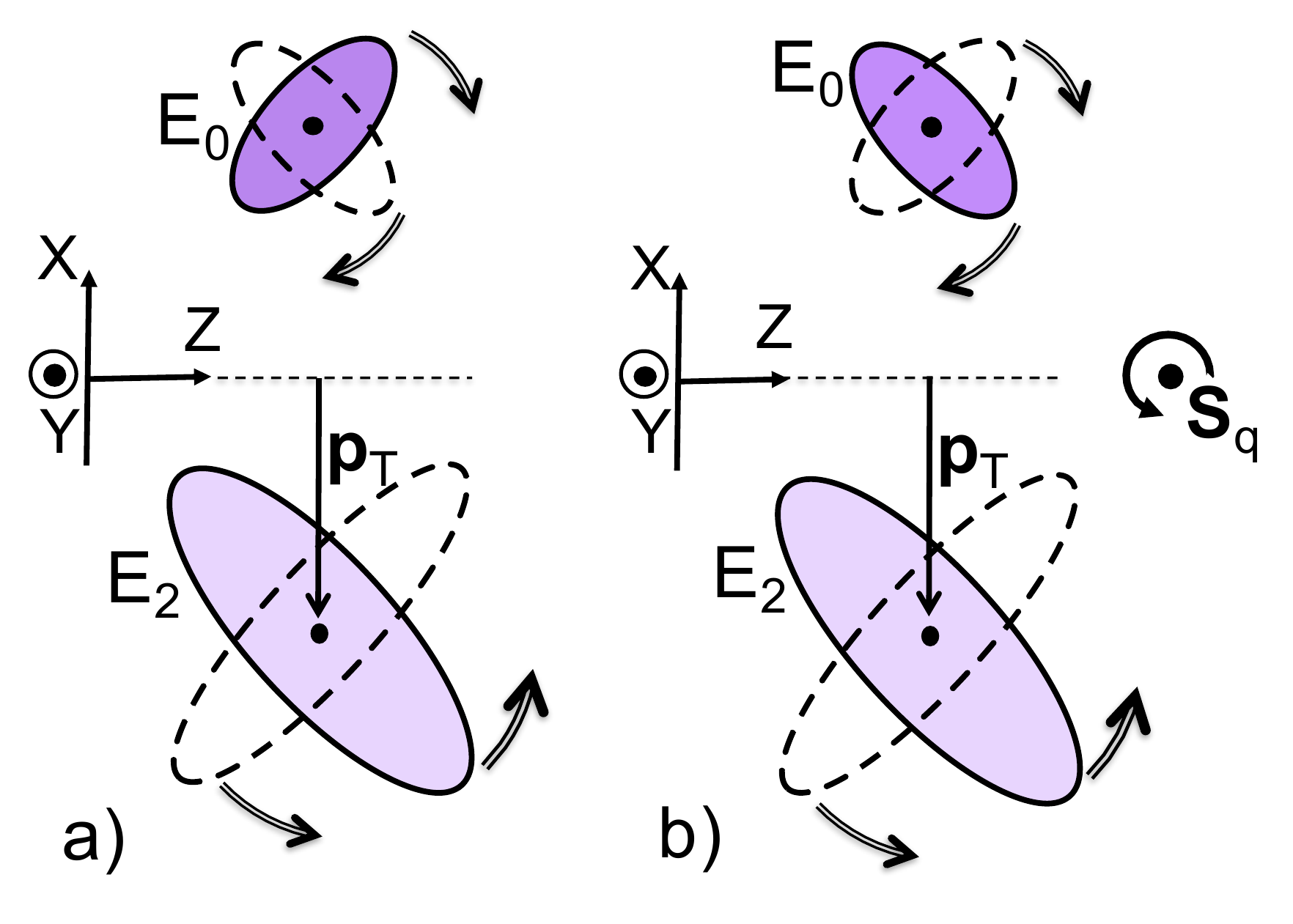}
\end{minipage}
\caption{Oblique polarization in the $(\ZS,\XS)$ plane of a first-rank vector meson,
corresponding to cases E$_0$ and E$_2$ of Fig. \ref{fig.poltransVM}.
a) with only the $\hat a \sin\theta\LT$ term in $\RE\,\hat\rho_{\rm{ml}}$ in Eq. (\ref{eq:rho vm MNL decomposition}).
b)  with only the $\sin\theta\LT \, S_Y$ term in $\RE\,\hat\rho_{\rm{XZ}}$ in Eq. (\ref{eq:rho vm XYZ decomposition}).
The continuous (dashed) contours are for positive (negative) $S_Y\sin\theta\LT$. Ellipsoid E$_0$ is drawn darker to figure its larger depth $\propto(\rho_{YY})^{1/2}$ in the $Y$ direction.  
The bent arrows indicate the Wigner rotation when passing from the LR symmetric to the null-plane frame, according to Eqs. 
(\ref{Vhl,infini})-(\ref{W-infini}). % (\ref{W-infini},\ref{Vhl,infini}).
}\label{fig.oblique}
\end{figure}

The real part of $\hat\rho$ is the {\it tensor}, or {\it linear} polarization \footnote{A general description of the density matrix for spin $1$ particles can be found in Ref. \cite{X.A_et_al_spin_observables}.}. It is convenient to represent it by a {\it polarization ellipsoid} as in Figs. \ref{leadingVM}-\ref{fig.oblique}.
The axes of this ellipsoid are parallel to the eigenvectors of $\RE\,{\hat\rho}$ and their half-lengths are equal to the square roots of the eigenvalues (see Appendix \ref{app:polarization ellipsoid}).
$\RE\,{\hat\rho}$ governs the angular distribution of the decay product. Thus, in the VM$\to h_1h_2$ decay, the relative $h_1h_2$ momentum tends to be aligned with the major axis (but without preferred sense).

In the $(\Lx,\Mx,\Nx)$ basis $\RE\, \hat\rho_{\alpha\alpha'}(h)$ writes:
\begin{eqnarray}\label{eq:rho vm MNL decomposition}
\hat\rho_{\rm ll} &=&   \left(1- \hat a \, \Sn\right) \, {|G\L|^2} / \NG(\Sv) 
\\ \nonumber 
\hat\rho_{\rm mm} &=&  \left(1- \hat a \,\Sn\right)  \, {|G\T|^2} / \NG(\Sv) 
\\ \nonumber 
\hat\rho_{\rm nn} &=&  \left(1+ \hat a \,\Sn\right) \, {|G\T|^2} / \NG(\Sv) 
\\ \nonumber 
\RE\,{\hat\rho}_{\rm mn} &=& \hat a \,\Sm \ {|G\T|^2} / \NG(\Sv) 
\\ \nonumber 
\RE\,{\hat\rho}_{\rm ml} &=& \left(\hat a -  \Sn)\, \sin\theta\LT  \, |G\L G\T| / \NG(\Sv\right)
\\  \nonumber 
\RE\,{\hat\rho}_{\rm nl} &=&  
\left( \hat a \,\cos\theta\LT \,\Sl - \sin\theta\LT\,\Sm\right)  \, |G\L G\T| / \NG(\Sv) ,
\end{eqnarray}
together with  $\RE\,{\hat\rho}_{\alpha\alpha'} = \RE\, \hat\rho_{\alpha'\alpha}$ \footnote{Equation (27) of \cite{DS09} is in agreement with Eq. (\ref{eq:rho vm MNL decomposition}) except for its wrong sign in front of 
$2 \IM(\mu) (\Vv_T.\tilde{\bf t} \, \Vv_T.\Sv
+\Vv_T.{\bf t} \, \Vv_T.\tilde\Sv)$.}. %The diagonal ones are in accordance with (\ref{Flmn}). The complete matrix elements are given in appendix A.   
%Their imaginary parts play no role in the decay processes considered here.
These are in accordance with Eq. (\ref{Flmn}) %The complete matrix elements are given in appendix A.
and satisfy automatically the positivity conditions.

The imaginary, antisymmetric part of $\hat\rho(h)$ is the {\it vector} or 
{\it circular} polarization. It plays no role in the decay processes considered here. The complete matrix elements are given in appendix A. 

\paragraph{Aligned and transverse linear polarizations.}
The element $\hat\rho_{\rm ll}\equiv\hat\rho_{\rm LL}$ is related to the {\it alignment} parameter $(3\hat\rho_{\rm LL}-1)/2$. The elements $\hat\rho_{\rm mm}$, $\hat\rho_{\rm nn}$ and $\hat\rho_{\rm mn}$ define the transverse linear polarization whereas $\hat\rho_{\rm ml}$ and $\hat\rho_{\rm nl}$ depend on $\theta\LT$ and describe {\it oblique} polarizations.  
Note, however, that our separation in {\it aligned}, {\it transverse} and 
{\it oblique} is linked to our choice of  the Lorentz transformations bringing the meson at rest.  

Figure \ref{fig.poltransVM} represents, for a first-rank VM and various orientations of $\pt$ with respect to ${\Sv}\T$, the transverse linear polarization defined by the 2$\times$2 restricted matrix 
$\RE\, {\hat{\rho}}^{(\rm T)}_{\alpha\alpha'} = \RE\, {\hat{\rho}}_{\alpha\alpha'}$  for $\alpha$ and $\alpha'\ne z$. The ellipses are the projections of the polarization ellipsoids. Their axes are parallel to the eigenvectors of ${\hat{\rho}^{(\rm T)}}$ and have lengths equal to the square roots of the eigenvalues 
$(1\pm\hat a \, |{\Sv}\T|) \, {|G\T|^2} / \NG({\Sv})$ of ${\hat{\rho}^{(\rm T)}}$.
Note that for a 2-body decay and $\phi(\pt)=\pi$ (case of ellipse E$_2$),  %  in the left side of fig. \ref{fig.poltransVM}, 
the left-moving decay hadron gets a large transverse momentum and the same Collins effect as the VM itself.

\paragraph{Oblique polarizations.} 
They are interferences between transverse and longitudinal amplitudes, therefore depend on $|G\L G\T|$ and $\theta\LT$ and correspond to the elements $\hat\rho_{\rm ml}$ and $\hat\rho_{\rm nl}$.  We analyze it in the basis 
 $\{\XS,\YS,\ZS\}$ introduced in Eq. (\ref{XYZ}). %linked to the quark transversity,
%
%\be \label{XYZ}
%\XS = \YS\!\times\!\ZS\,, \, \YS= {\Sv_q}\T/|{\Sv_q}\T| \,, \  \ZS=\zu %\,.
%\ee 
%
Let us consider separately, in Eq. (\ref{eq:rho vm MNL decomposition}) or (\ref{eq:rho vm XYZ decomposition}), 
the effects of the terms $\hat a\sin\theta\LT$,  % of $\hat\rho_{\rm ml}$, 
 % of $\hat\rho_{\rm nl}$ 
 $\hat a\cos\theta\LT S_z$ % of  $\hat\rho_{\rm ml}$ and $\hat\rho_{\rm nl}$. 
and $\sin\theta\LT \, \Sv\T$ in $\hat\rho_{\rm ml}$ and $\hat\rho_{\rm nl}$. 

The term $\hat a \sin\theta\LT$ in $\hat\rho_{\rm ml}$ is independent of the quark polarization and gives an oblique polarization in the $\lkp,\mkp$ plane, projected on the $(\XS,\ZS)$ plane in Fig. \ref{fig.oblique}a. 
 For the 2-body decay VM$\to h_1h_2$ it acts upon the dependence of $\langle\pt_i^2\rangle$ on $z_i$~: 
  at large $z_1$, $p_{1z}^*$ is likely positive and, for rank one, Fig. \ref{fig.oblique}a indicates a larger $\langle \pv_{1\rm T}^2\rangle$ for positive than for negative $\sin\theta\LT$. It comes from the $\pt$ composition law
 \be \label{compose-pT}
 \pv_{i\rm T} =  \pv_{i\rm T}^* + \left[E_i^* + \pt\cdot \pv_{i\rm T}^* (\ET+M)\right] \pt/M
 \ee
 and the fact that the sign of $\pt\cdot \pv_{i\rm T}^*$ is most likely that of $p_{iz}^* \times \sin\theta\LT$. 
 % Thus $\pv_{i\rm T}^2$ 
 \\
 The term $\hat a \cos\theta\LT S_l$ in $\hat\rho_{\rm nl}$ % is proportional to the quark helicity and 
gives an oblique polarization in the $(\lkp,\nkp)$ plane. This is a {\it jet handedness} effect like the one with only direct pions treated in section VI of \cite{kerbizi-2018}.  
For the decay of a first-rank VM in two mesons $h_1,h_2$ we have at fixed $\pv^*_{1z}$ 
\be
\pv^*_{1z} \, \langle (\pv_{1} \!\times\! \pv_{2} )_z\rangle = -
(2/5) \, \RE\,\hat\rho_{\rm nl} \, |\pv^*_1|^2 |\pt| \,. 
\ee
 The terms in $\sin\theta\LT \,{\Sv}\T$, gathered in one term of (\ref{eq:rho vm XYZ decomposition}), are independent of $\IM \mu$ and produce an oblique polarization in the $(\XS,\ZS)$ plane (Fig. \ref{fig.oblique}b).
In a 2-body decay it contributes to the individual Collins asymmetry of the decay products. 
Considering Eq. (\ref{compose-pT}), we see that at fixed $z_i$ this obliquity adds to or subtracts from the part inherited from the global Collins effect. This effect will be studied in more detail in Sec. \ref{sec: sensisivity spin asymmetries}.

The oblique polarization is also a source of dihadron asymmetry, 
which bears on the variable $\Rv\T$ defined in Eq. (\ref{RT}). The asymmetry sign is not simply deduced from the orientation of the ellipses in Fig. \ref{fig.oblique}, due to the Wigner rotation.
In Eq. (\ref{RT}) the distinction between $h_1$ and $h_2$ must not be done according to their charges (like $h_1\,$=$\,\pi^+$, $h_2\,$=$\,\pi^-$) but between ``fast'' and ``slow'',  for instance, by $z_1>z_2$. A distinction by the charges gives no dihadron asymmetry because of the $\pv_1^*\leftrightarrow\pv^*_2$ invariance of the decay angular distribution. 
This dihadron asymmetry is not the result of an interference with a non-resonant amplitude. It is related to the fragmentation function $H_{1\,LT}$ of Ref. \cite{Bacchetta:Spin1}. The asymmetry will be discussed in more detail in Sec. \ref{sec:new 2h asymmetries}.

%%%%%%%%%%%%%%%%%%%%%%%%%%%%%%%%%%%%%%%%%%%%%
 %%%%%%%%%%%%%%%%%%%%%%%%%%%%%%%%%%%%%
%%%%%%%%%%%%%%%%%%%%%%%%%%%%%%%%%%%%%%%%%%%

\subsection{The decay of a polarized VM}\label{sec:decay of a vector meson}

\paragraph{Decay in two pseudoscalar mesons.}

We suppose that, by the Monte Carlo method, we have generated the species 
$h$ of the VM, its running mass $\Mh$, its momentum $\pv$ and calculated its density matrix $\hat\rho_{\alpha\alpha'}$ in the 
$\{\Lx,\Mx,\Nx\}$ basis, for instance, with Eq. (\ref{eq:rho vm MNL decomposition}). 

First, one choses the $h_1$ and $h_2$  species, \eg, $K^+\, \pi^0$ for a $K^{*+}$, following the known branching ratio. This fixes the modulus of the relative momentum $\rv=\pv^*_1=-\pv^*_2$ in the VM rest frame, 
\be
%|\rv| = (2\Mh)\inv \left[\Mh^2-(m_1+m_2)^2 \right]^{1/2} \times \left[\Mh^2-(m_1-m_2)^2 \right]^{1/2} \,,
|\rv| = (2\Mh)\inv \left[\Mh^2-m_+^2 \right]^{1/2} \times \left[\Mh^2-m_-^2 \right]^{1/2} \,,
\ee
where $m_{\pm}=m_1 \pm m_2$. It remains to generate its direction $\hat\rv$. 
The decay amplitude is 
\be \label{decay-amp2}
{\cal M}(\VMM \to h_1\, h_2) = g_{h\,h_1\,h_2} \, A^\mu(p_1-p_2)_\mu = 
- 2g_{h\,h_1\,h_2} \,  \Vv\cdot\rv \,.
\ee
Then, the  resonant $q\to h_1+h_2 + q'$ amplitude is proportional to 
\be
 \sum_{\Vv = \Lx,\Mx,\Nx}  T_\alpha \, V_\alpha \, V_\beta \, r_\beta = T_\alpha \, r_\alpha \,.
\end{equation}
and the angular distribution takes the form
\be \label{angular}
 {d{\cal N}(\hat\rv)}/{d\Omega}  
=  3 \, (4\pi)\inv \, \hat r_\alpha \, \hat\rho_{\alpha\alpha'}(h) \, \hat r_{\alpha'} \,,
\ee
reminiscent of Eq. (\ref{VrhoV}).
A corresponding formula is found in Eq. (B10) of Ref. \cite{Bacchetta:Spin1}.

\paragraph{Decay $\omega\to\pi^0\gamma$.} The decay amplitude is
\be \label{decay-amp2'} % \|abel{decay-amp2'}
{\cal M}(\omega \to \pi^0\gamma) \propto  \Vv_\omega\cdot (\Vv_\gamma\times\rv) \,,
\ee
where $\Vv_\gamma\perp\rv$ is the vector amplitude of the photon and $\rv=\pv^*_\gamma$. Averaging over $\Vv_\gamma$, we have to replace in Eq. (\ref{angular})  $\hat r_{\alpha}\, \hat r_{\alpha'}$ 
by the tensor $(1/2) (\delta_{\alpha\alpha'}-\hat\rv_{\alpha} \, \hat\rv_{\alpha'})$.
% on the plane transverse to $\rv$ and the coefficient $3/(4\pi)$ by $3/(8\pi)$. The same method applies to $\phi\to\pi^0\gamma$. 

\paragraph{Decay $\omega$ or $\phi\to\pi^+\pi^-\pi^0$.} 
%\paragraph{Decay VM $\to$ 3 pseudoscalar mesons.} 
Due to parity conservation the invariant decay amplitude in three pseudoscalars is of the form
%  au lieu de  (E^*_1,E^*_2,E^*_3) 
\be \label{decay-amp3}
{\cal M}(\VMM \to h_1\,h_2\,h_3) \propto  {\cal F}(s_1, s_2 , s_3)\, \Vv\cdot {\boldsymbol{\tau}}\,,
\ee
%
% with. 
where  ${\boldsymbol{\tau}} = \pv^*_1\times\pv^*_2$ is normal to the {\it decay plane},
$s_i= (p_j+p_k)^2 $ and $\{i,j,k\}$ is a cyclic permutations of  $\{1,2,3\}$. 
%The $s_i$ are linked to the variable mass $\Mh$ of the resonance by
%%
%\be \label{stu}
%s_1+s_2+s_3 = \Mh^2 + m_1^2 + m_2^2 + m_3^2 \,.
%\ee
%%
% In the VM rest frame, the $\pv^*_i$ span the {\it decay plane}.
From energy-momentum conservation, $\pv^*_1+\pv^*_2+\pv^*_3={\bf0}$
and  $E^*_1+E^*_2+E^*_3=\Mh$ which is the variable mass of the resonance.  
The $E^*_i$ are linearly related to the $s_i$ by  % ir moduli are given by % the $\pv^*_1$,   $\pv^*_2$, $\pv^*_3$ triangle   
\be
E^*_i = [\pv^*_i+ m_i^2]^{1/2}   = (\Mh^2+m_i^2-s_i)/(2\Mh) \,. %\quad |\pv^*_i| = [(E^*_i)^2-m_i^2]^{1/2} \,.
\ee
Taking into account energy-momentum conservation, the 3-body phase space element reduces to 
\be\label{eq:angular omega}
d\Phi(\pv^*_1,\pv^*_2) \propto  d\Omega(\boldsymbol{\tau}) \, d\phi_{1|\boldsymbol{\tau}} \, dE^*_1 \, dE^*_2 \,,  
\ee
where $\phi_{1|\boldsymbol{\tau}} $ is the azimuth of $\pv^*_1$ about ${\boldsymbol{\tau}}$. 
In the $(E^*_1,E^*_2)$ plane (Dalitz  plot) the physical phase space is limited to the domain 
\be
{\boldsymbol{\tau}}^2 \equiv \pv^{*2}_1\,\pv^{*2}_2-(1/4) \left(\pv^{*2}_1+\pv^{*2}_2-\pv^{*2}_3\right)^2 \ge0\,.
\ee
The {\it form factor} ${\cal F}(s_1, s_2 , s_3)$ depends on the dynamics, in particular on final state 2-body interactions. Following the  isobar model, we assume that the VM decay occurs in two steps, $h\to \pi^i + \bar\rho^{i}$, then $\bar\rho^{i}\to \pi^j+\pi^k$, where now $\{+,0,-\}$ replace  $\{1,2,3\}$. So, we take
\begin{equation}\label{eq:form factor F}
    \mathcal{F}(s_+, s_- , s_0) = \sum_{i=+,0,-} \frac{g_{h\bar\rho^i\pi^i}\,g_{\bar\rho^i\pi^j\pi^k}}
    {s_i-\overline{m}^2_{\rho^i}+i\overline{m}_{\rho^i}\hat{\gamma}_{\rho^{i}}} \,.
\end{equation}
 By isospin symmetry the coupling constants $g_{h\rho\pi}$ and $g_{\rho\pi\pi}$ do not depend on the charge of the intermediate $\rho$ meson: $g_{h\rho^+\pi^-}=g_{h\rho^-\pi^+}=g_{h\rho^0\pi^0}$ and $g_{\rho^+\pi^0\pi^+}=g_{\rho^-\pi^-\pi^0}=g_{\rho^0\pi^+\pi^-}$.

To generate the pion momenta $\pv_i^*$, we proceed in three steps.
First we draw $E^*_1$ and $E^*_2$ according to the (not normalized) distribution
\begin{equation}\label{eq:w(m+-,m+0)}      
    w(E^*_1,E^*_2) = {\boldsymbol{\tau}}^2 \, 
    {|\mathcal{F}(s_1,s_2,s_3)|^2} \,.    
\end{equation}
and calculate $|\pv_1|$, $\pv_1\!\cdot\! \pv_2$ and $|{\boldsymbol{\tau}}|$.

Then we generate $\hat{\boldsymbol{\tau}} = {\boldsymbol{\tau}} /|{\boldsymbol{\tau}}|$ according to Eq. (\ref{angular}) with $\hat\rv \to \hat{\boldsymbol{\tau}}$. Indeed, $\hat{\boldsymbol{\tau}}$ plays the same role as $\hat\rv$ in the two-body decay. 

Then we draw at random $\phi_{1|\boldsymbol{\tau}}$ in $[0,2\pi]$
and build
\be
\pv^*_1 = |\pv^*_1|\, \mathcal{R}_{\zu\times \boldsymbol{\tau}}(\theta^*_{\boldsymbol{\tau}}) % R_{\zu}(\varphi)
(\cos\phi_{1|\boldsymbol{\tau}},\sin\phi_{1|\boldsymbol{\tau}} ,0)^{\rm{T}}
\ee
where $\theta^*_{\boldsymbol{\tau}}$ is the polar angle of $\boldsymbol{\tau}$~;

Finally we build $\pv^*_2$ and  $\pv^*_3 = - \pv^*_1 - \pv^*_2$, using
\be
\pv^*_2 = |\pv^*_1|^{-2}  \left[ (\pv_1\!\cdot\! \pv_2) \, \pv^*_1 + \boldsymbol{\tau} \times \pv^*_1 \right] \,.
\ee

\subsubsection{Boosting the decay mesons }

\begin{figure} % [b!]  % =Fig.1 %%%%%%%%%%   FIGURE   %%%%%%%%%%%%%%
\begin{minipage}{.5\textwidth}
%\centering
 \includegraphics[width=0.9\textwidth]{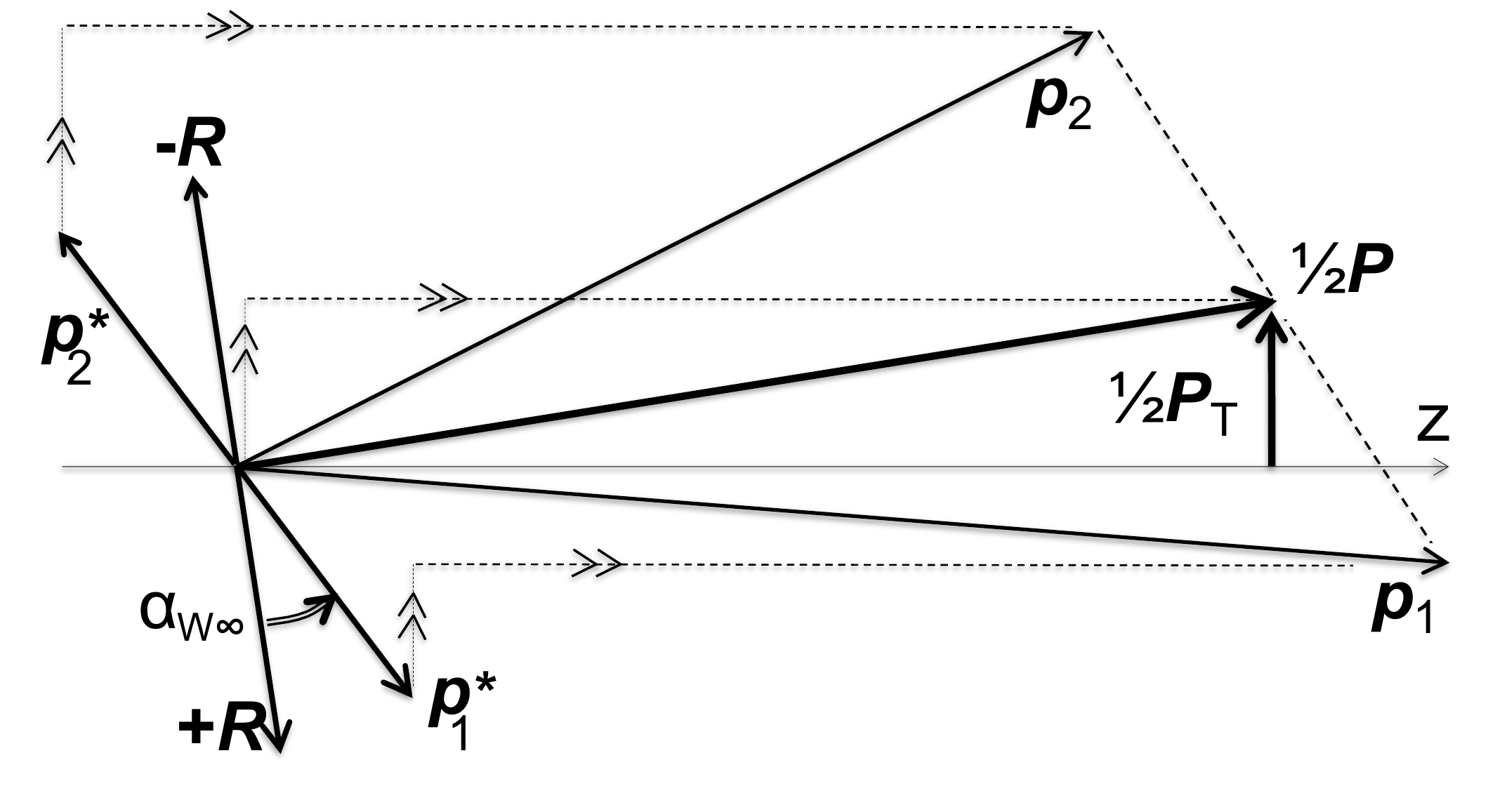}
 \end{minipage}
\caption{ Boosts transforming the pion momenta of $\rho$ decay, from the LR symmetric frame (${\bf p}_1^*=\textbf{r}$ and ${\bf p}_2^*=-\textbf{r}$) to the string frame ($\pv_1$ and $\pv_2$). Also shown is the relative momentum $\Rv$ in the null-plane frame, related to $\textbf{r}$ by a Wigner rotation of angle $\alpha_{{\rm W}\infty}$. The line ``$...._{>>}....$'' represents the move of the extremity of a vector $\pv$ during the boosts $B\T$ and $B\L$.
The figure is calculated for $|\pt|/M_\rho$ = 8/15, $P\L/\ET(\rho)$ = 35/12 and ${\bf p}_1$ and ${\bf p}_2$ in the $(\zu,\pt)$ plane.
} \label{R-and-p1cm} 
 \end{figure}

Once $(E^*_i,\pv^*_i)$ have been generated, the momenta $(E_i,\pv_i)$ in the string frame are obtained by the inverse of the boosts which serve to 
define $\Vv$ in section \ref{defineV}:
\be\label{eq:boost}
(E_i,\pv_i) =  B\L \, B\T \, (E^*_i,\pv^*_i),
\ee
where $B\L$ and $B\T$ are the boosts defined in Eq. (\ref{BL,BT}). The effect of these boosts on the momenta of the decay pions from a $\rho$ is illustrated in Fig. \ref{R-and-p1cm}.

%%%%%%%%%%%%%%%%%%%%%%%%%%%%%%%%%%%%%%%%%%%%%%

%\begin{center}
 %%%%%%%%%%%%
% \end{center}
%%%%%%%%%%%%%%%%%%%%%%%%%%%%%%%%%%%%%%%%%%%%%%

\subsection{Spin density matrix of $q'$}
When a VM has been generated by the splitting $q\to\VMM+q'$, the information about the spin state of $q'$, encoded in its density matrix $\hat\rho(q')=(\un+\boldsymbol{\sigma}\cdot\Sv_{q'})/2$, depends on the information about the decay products of the VM.  

\subsubsection{Case without information about the decay products}

Suppose that the VM is not analyzed (only $p$ is recorded, not the momenta of the decay products). Then
\begin{eqnarray} \label{eq:rho(q') after vm decay averaged}
&& \hat\rho(q') =   \left[\sum_{\Vv = \Lx,\Mx,\Nx}   T(\q',\h,\q) \ \hat\rho(q)  \ T^\dagger(\q',\h,\q)\right] 
\big/ \Tr \bigg[\cdots\bigg]
\nonumber \\ && = \frac{
(\mu + \sigma_z\, \vecsigma\cdot\kpt) \, \Gamma_{h,\alpha} \, \hat\rho(q) \, 
\Gamma^\dag_{h,\alpha} \, (\mu^* +  \vecsigma\cdot\kpt \, \sigma_z)
}{ 
(|\mu|^2+\kptkpt) \, N(\Sv_q) % [ 2|G\T|^2  + |G\L|^2 + |G\L|^2\,\hat a\, 
%\Sv_q\cdot\tilde\nv(\kpt) ] 
} \,,
\end{eqnarray}
where $[\cdots]$ repeats the numerator. The second expression looks like in Eq. (\ref{eq:rho vm matrix elements}), but summing over $\alpha=\alpha'$ 
and removing the symbol $\Tr$  in the numerator.

For the emission of a pseudoscalar meson, the spin density matrix of $q'$ can be calculated using Eq. (\ref{eq:rho(q') after vm decay averaged}) with $\Gamma_h=\sigma_z$ as in M19 \cite{kerbizi-2019}.

\paragraph{Depolarization of the recurring quark.}
As in the model with only pseudoscalar mesons (cf. Eqs. (31)-(32) of \cite{kerbizi-2019}), the recurring quark becomes less and less polarized as its rank increases, if the transverse momenta of the emitted hadrons are integrated over. The depolarization coefficients $D_{\rm TT}^{\rm VM} \equiv \Sv_{q'\rm T}/ \Sv_{q\rm T}$
and $D_{\rm LL}^{\rm VM} \equiv S_{q'\rm L}/ S_{q\rm L}$
following the emission of a VM are obtained by replacing the numerator and the denominator of Eq. (\ref{eq:rho(q') after vm decay averaged}) by their averages on $\kpt$ weighted by $f\T^2(\kpt)$:
 %
 % \def\rLT{r_{\rm L/T}}
% \def\rkmu{r_{kT/\mu}}
%%
% \begin{eqnarray}\label{eq:DTT and DLL for vm emission}
%    D_{\rm TT}^{\rm vm} &=& \frac{r\LT}{r\LT+2} \,\frac{1}{1+\rkmu}  = 
%\frac{-\,r\LT}{r\LT+2} \, D_{\rm{TT}}^{\rm ps} 
% \\
%D_{\rm LL}^{\rm vm} &=& \frac{r\LT-2}{r\LT+2}\,\frac{1-\rkmu }{1+\rkmu}  = \frac{r\LT-2}{r\LT+2}\, D_{\rm{LL}}^{\rm ps} 
%\end{eqnarray}
%%
%
 \begin{eqnarray}\label{eq:DTT and DLL for vm emission}
    D_{\rm TT}^{\rm VM} &=& \fL \, ({1+ \langle\ktkt\rangle_{f\T} / |\mu|^2})\inv  = -\fL  \, D_{\rm{TT}}^{\rm PS} \,,
 \\ \nonumber
D_{\rm LL}^{\rm VM} &=&(2\fL-1)\,\frac{|\mu|^2- \langle\ktkt\rangle_{f\T}  }{|\mu|^2+ \langle\ktkt\rangle_{f\T} }  =(2\fL-1)\, D_{\rm{LL}}^{\rm PS} \,,
\end{eqnarray}
%
% where   $\rkmu\equiv \langle\ktkt\rangle_{f\T} / |\mu|^2$ % $\rLT \equiv {|G\L/G\T|^2}$,
% %
% \begin{eqnarray}\label{eq:DTT and DLL for vm emission}
%    D_{\rm{TT}}^{\rm{vm}} &=& \frac{|G\L|^2}{2|G\T|^2+|G\L|^2} \,\frac{|\mu|^2}{|\mu|^2+\langle\ktkt\rangle_{f\T} }   %|D_{\rm{TT}}^{ps}| 
% \\
%D_{\rm{LL}}^{vm} &=& \frac{|G\L|^2-2|G\T|^2}{2|G\T|^2+|G\L|^2}\,\frac{|\mu|^2-\langle\ktkt\rangle_{f\T} }{|\mu|^2+\langle\ktkt\rangle_{f\T} }  
%\end{eqnarray}
%%
where  $D_{\rm{TT}}^{\rm PS}$ and $D_{\rm{LL}}^{\rm PS}$ are given in Eqs. (31)-(32) of \cite{kerbizi-2019}. 
Note that these coefficients are smaller for VM than for PS. This is due to the loss of information when the momenta of the decay products are not 
measured. 
Note also the opposite signs of $D_{\rm{TT}}^{\rm VM}$ and $D_{\rm{TT}}^{\rm PS}$. %For a given rank $r>1$ the number of antecedent PS mesons is fluctuating, therefore  $|\Sv_{q_r}|$ is smaller than in the model with only PS.

\subsubsection{Case where the momenta of decay products are known}

\paragraph{The VM decay matrix.}

As already said, $\hat\rho(q')$ depends on the information about the decay products of the VM. This information is encoded in a matrix $\check\rho(h)$ called {\it decay matrix} (also indicated with $D$ in literature) \cite{collins-corr,knowles-corr} or {\it acceptance (density) matrix} \cite{X.A_et_al_spin_observables}. 
$\check\rho(h)$ can be seen as the density matrix of the VM running backward in time, contrarily to the {\it emittance} density matrix $\hat\rho(h)$ studied in section \ref{theVMdensitymatrix}. 
For a definite state $|\pv^*_1,\pv^*_2\cdots\rangle$ of the decay products, 
\be
\check\rho_{\alpha'\alpha}(h) = {\cal M}^\dag_{\alpha'}(p_1,p_2\cdots) \, {\cal M}_\alpha(p_1,p_2\cdots) \,,
\ee
writing the decay amplitude as ${\cal M}_\alpha(p_1,p_2\cdots)V_\alpha$.
For the decays that we consider, 
\be\label{eq:acceptance matrix}
\check\rho_{\alpha'\alpha}(h) \propto 
\left\{
\begin{array}{ll}
\hat\rv_{\alpha'} \, \hat\rv_\alpha & (\VMM\to2 {\rm PS})     \\
\hat{\boldsymbol{\tau}}_{\alpha'} \, \hat{\boldsymbol{\tau}}_\alpha & (\VMM\to3 {\rm PS})  \\
\delta_{\alpha\alpha'}-\hat\rv_{\alpha} \, \hat\rv_{\alpha'} & (\omega\to\pi^0\gamma) \\
\delta_{\alpha\alpha'} & \text{(VM not analyzed}) \,.
\end{array}
\right.
\ee
%
% Eqs. (\ref{decay-amp2}) (\ref{decay-amp2'})  (\ref{decay-amp3}) give 
We write '$\propto$' instead of '=' because we do not fix the trace of 
$\check\rho(h)$. The angular distribution in Eq. (\ref{angular}) or in Eq. (\ref{eq:angular omega}) is proportional to $\Tr\lbrace \hat\rho(h)\check\rho(h)\rbrace$. In the third line of Eq. (\ref{eq:acceptance matrix}), the decay state is an incoherent superposition of the states with photon helicities $\pm1$. The last line is for the case where $\hat\rv$ or $\hat\nv$ is not recorded. In lines 1 and 2, $\check\rho(h)$ is the density matrix of a pure state, instead in lines 3 and 4 it is that of a statistical mixture. 

\paragraph{Combining $\check\rho(h)$ and $\hat\rho(q)$.}
Taking into account the information encoded in $\check\rho(h)$, we replace Eq. (\ref{eq:rho(q') after vm decay averaged}) by 
\begin{eqnarray} \label{eq:rho(q')_recordedVMdecay}
\hat\rho(q') &\propto&   T_\alpha \,\rho(q)\,T_{\alpha'} ^\dag  \, \check\rho_{\alpha'\alpha}(h)   
% \big/ \Tr_q \bigg[\cdots\bigg]
\nonumber \\ &\propto& % \frac{
(\mu + \sigma_z\, \vecsigma\cdot\kpt) \, \Gamma_{h,\alpha} \, \hat\rho(q) 
\\ \nonumber  &\times&
\Gamma^\dag_{h,\alpha'} \, \check\rho_{\alpha'\alpha}(h) \, (\mu^* +  \vecsigma\cdot\kpt \, \sigma_z) \,.
%}{ (|\mu|^2+\kptkpt) \, [ 2|G\T|^2  + |G\L|^2 + |G\L|^2\,\hat a\, \Sv_q\cdot\tilde\nv(\kpt) ] } \,,
\end{eqnarray}
Compared to Eq. (\ref{eq:rho(q') after vm decay averaged}), there are two indices $\alpha$ and $\alpha'$, which we contract with those of $ \check\rho_{\alpha'\alpha}(h)$. Again the use of '$\propto$' instead of '=' means that we have not yet fixed the trace of $\hat\rho(q')$.

\textit{Note:} Carrying information ``backward in time'' with $\check\rho(h)$ is necessary to generate the correct correlations between the spin of $q'$ and the momenta of the decay products  when the joint density matrix 
\be   % _{i,\alpha,i',\alpha'}(h,q')         | i \rangle  \,  \langle i | 
 \big\langle \alpha\big|   \otimes \langle s_{q'} | \hat\rho(h,q')  | s'_{q'} \rangle  \otimes \big| \alpha' \big\rangle
= \langle s_{q'} | T_\alpha \, \hat\rho(q) \, T_{\alpha'} ^\dag | s'_{q'} \rangle
\ee
is entangled. This is the general case: for instance, if $q$ is in the pure spin state $|\Sv_{q}=+\yu\rangle$, the $\VMM+q'$ system is in the entangled (non-separable) state 
\begin{eqnarray}
&& \sum_\alpha \big|\alpha\big\rangle \otimes T_\alpha \, |\Sv_{q}\rangle 
\propto
G\T \left[ \, \big|\xv\big\rangle \otimes  \sigma_x  
| \yu \rangle
+
 \big|\yv\big\rangle \otimes  \sigma_y  | \yu \rangle \right]
 \nonumber \\ &&+ \ 
G\L \, \big|\zv\big\rangle \otimes    | \yu \rangle
\nonumber \\ &&= \
G\T \big|\xv\big\rangle \otimes   | -\!\yu \rangle
 +  \left[G\T \big|\yv\big\rangle % \otimes  | \yu \rangle 
+G\L \, \big|\zv\big\rangle \right] \otimes    | \yu 
\rangle .
\end{eqnarray}

\section{Monte Carlo implementation}\label{sec:MC implementation}
The structure of the stand alone MC implementation of M20 is the same as that of M19 \cite{kerbizi-2019}. First the flavor $u$, $d$ or $s$, the four-momentum and the spin density matrix of the fragmenting quark $q_A$ are defined. In the simulations of the fragmentation process in a SIDIS event the initial quark energy has been taken from a sample of SIDIS events collected by the COMPASS experiment with a $160\,\rm{GeV}/c$ muon beam, and having $Q^2>1\,(\rm{GeV}/c)^2$ and the invariant mass of the final hadronic system larger than $5\,\rm{GeV}/c^2$, as in Ref. \cite{kerbizi-2019}. For the comparison with $e^+e^-$ data a fixed center of mass energy $\sqrt{s}=10\,\rm{GeV}$ has been used to compare with the BELLE experiment. Once the initial quark state has been set up, the fragmentation chain is simulated by repeating recursively the elementary splitting $q\rightarrow h + q'$ until the condition for the termination of the fragmentation chain is reached. The hadron $h$ is assigned to the vector or pseudoscalar multiplet according to the relative probability $\fvmps$. This parameter is fixed and taken as in PYTHIA 8, namely for light mesons containing only $u$ and $d$ quarks it is $\fvmps=0.62$ whereas for mesons containing at least one strange quark it is $\fvmps=0.725$.

The simulation of the elementary splittings in M20 proceeds in the following steps:
\begin{itemize}\itemsep0em
    \item [(1)] Generate a new $q'\bar{q}'$ pair with $q'=u,d,s$ taking into account the suppression of $s$ quarks according to the relative probabilities $P(u\bar{u}):P(d\bar{d}):P(s\bar{s})=3/7:3/7:1/7$ as in \cite{kerbizi-2019} with M19.
    \item [(2)] Form $h=q\bar{q}'$ and choose the VM multiplet with probability $\fvmps/(1+\fvmps)$.
    If $(q\bar{q}')$ is flavor neutral, choose the meson species $h$ with probability proportional to $|C_{q',h,q}|^2$, according to Eqs. (\ref{uchapeau}) and (\ref{Proba(h)}). If $h$ is PS assign the corresponding mass. If $h$ is a VM generate its mass $\Mh$ with the $|D_h(\Mh)|^2$ distribution, according to Eq. (\ref{Flmn}) summed over $\textbf{V}$ and integrated over $Z$ and $\kpt$.   
    \item [(3)] According to Eqs. (\ref{eq:F_explicit PS}) and (\ref{Flmn}) generate $\kptkpt$ with the distribution $f\T^2(\kptkpt) \, (|\mu|^2+\kptkpt)/\langle|\mu|^2+\kptkpt\rangle_{f\T}$, and the azimuthal angle $\phi(\kpt)$ following the factor $(1+\hat a \, \Sn)$ for a PS, $(1-f\L\hat a \, \Sn)$ for a VM. Construct $\pt=\kt-\kpt$ (with $\kt=0$ for $q=q_A$).
    
    \item [(4)] Generate $Z$ with the distribution given in the second line of Eq. (\ref{eq:F_explicit PS}) or third line of Eq. (\ref{Flmn}).

    \item [(5)] Calculate $p^+=Zk^+$, $p^-$ imposing the mass shell condition $p^+p^-=\ET^2$ and $k'^{\pm}=k^{\pm}-p^{\pm}$.
    \item [(6)] Test the exit condition (see below) on the remaining mass squared $M_{\rm X}^2=(k'+k_{B})^2$. If it is not satisfied continue with the next step, otherwise the current hadron is removed and the fragmentation chain ends. We do not treat the decay of the remaining string piece.
    \item [(7)] Construct the hadron four-momentum $p=(E,\pt,p_z)$ by calculating $E=(p^++p^-)/2$ and $p_z=(p^+-p^-)/2$. Store the hadron in the event record.

    \item [(8)] If $h$ is a PS calculate the spin density matrix of $q'$ using Eq. (\ref{eq:rho(q') after vm decay averaged}) with $\Gamma_{h}=\sigma_z$ and return to step (1). If $h$ is a VM:
        \begin{itemize}\setlength\itemsep{0.1em}
        \item [(8.1)] Calculate the spin density matrix $\hat{\rho}(h)$ of $h$ using Eq. (\ref{eq:rho vm matrix elements}).
        \item [(8.2)] Chose the decay channel (if more than one) as specified below. Construct the momenta of the decay hadrons using $\hat{\rho}(h)$ to generate the angles as explained in Sec. \ref{sec:decay of a vector meson}.
        \item [(8.3)] Boost the decay products according to Eq. (\ref{eq:boost}). Store the decay hadrons in the event record.
        \item[(8.4)] Build the acceptance matrix $\check{\rho}(h)$ of Eq. (\ref{eq:acceptance matrix}).
        \item[(8.5)] Calculate the spin density matrix of $q'$ using Eq. (\ref{eq:rho(q')_recordedVMdecay}). Go to step (1).
    \end{itemize}
\end{itemize}
The probabilities used to determine the PS meson species at step (2) are the same as in M19. The probabilities of the VM species are obtained from the corresponding wave functions in flavor space. Unlike the PS case, for VM production there is no suppression factor among flavor neutral states, e.g. a spin-1 $u\bar{u}$ or $d\bar{d}$ pair is assigned to a $\rho^0$ or to an $\omega$ with the same probability (see also Ref. \cite{Kerbizi:PhD}).

The exit condition in step (6) is satisfied when not enough remaining mass squared is left in the string to produce at least one resonance (baryonic in SIDIS, mesonic in $e^+e^-$) as in M19.

The decay channels considered in (8.2) are $\rho\rightarrow \pi\pi$, $K^*\rightarrow K\pi$, $\omega\rightarrow\pi^+\pi^-\pi^0$, $\omega\rightarrow \pi^0\gamma$, $\omega\rightarrow \pi^+\pi^-$, $\phi\rightarrow K^+K^-$, $\phi\rightarrow K_S^0K_L^0$, $\phi\rightarrow \pi^+\pi^-\pi^0$, $\phi\rightarrow \eta\gamma$ and $\phi\rightarrow \pi^0\gamma$. The corresponding branching ratios are taken from the PDG \cite{PDG2019}. In the case of the $K^*\rightarrow K\,\pi$ decay, we take the branching ratios given by isospin symmetry, e.g. $K^{*0}\rightarrow K^+\pi^-$ with branching ratio $2/3$ and $K^{*0}\rightarrow K^0\pi^0$ with branching ratio $1/3$. Concerning $K^0$ and $\bar{K}^0$ we keep the quantum state as it is immediately after emission without evolving with mixing and oscillations.

\section{Results of SIDIS simulations}\label{sec:results}
This section is dedicated to the results obtained from the Monte Carlo simulations of the fragmentation of $u$ quarks with full transverse polarization along the $\YS$ axis (hence $|\textbf{S}_{{q_A} \rm T}|=1$). Results for $d$ quarks can be deduced from isospin and charge conjugation arguments. The primordial transverse momentum of the fragmenting quark has been switched off. Its effect on transverse spin asymmetries was studied for M18 in Ref. \cite{kerbizi-2018}.

Concerning the free parameters of the model $a$, $\bl$, $\bt$ and $\mu$, the same values as in M19 are used, namely $a=0.9$, $\bl = 0.5\,(\rm{GeV}/c^2)^{-2})$, $\bt=5.17\,(\rm{GeV}/c)^{-2}$ and $\mu=(0.42+\,i\,0.76)\rm{GeV}/c^{2}$.
For the two new free parameters, we take first  $\glgt=1$ (\ie{} $\fL=1/3$) and $\theta_{LT}=0$, in agreement with Ref. \cite{Czyzewski-vm} (see the text concerning Fig. \ref{fig:collins rank pi rho}). The sensitivity of the observables on the values of the new parameters is then discussed in Sec. \ref{sec:sensitivity}.

\subsection{Kinematic distributions}

In the study of the distributions of the hadrons fractional energy $\zh$ and transverse momentum $\ptv$ we apply the cuts $p_{\rm{T}}>0.1\,\rm{GeV}/c$ when looking at the $\zh$ distribution and $\zh>0.2$ when looking at the $\ptv$ distribution, in analogy with real data analyses.

\begin{figure}[tb]
\centering
\begin{minipage}{0.46\textwidth}
\centering
  \includegraphics[width=1.0\linewidth]{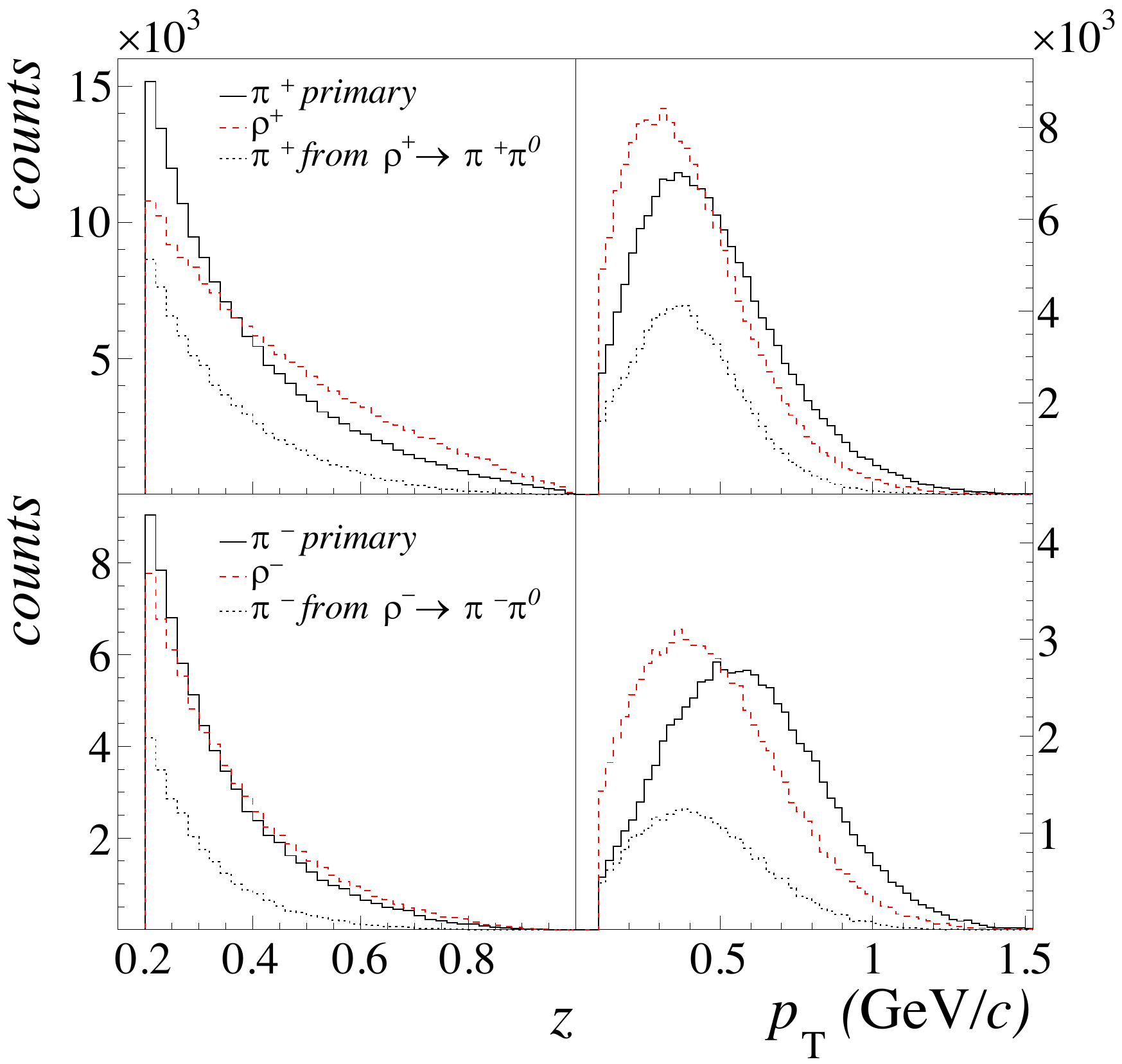}
  \end{minipage}
\caption{Upper row: comparison between the $\zh$ (left) and $\ptv$ (right) distributions for primary $\pi^+$ (continous histogram), $\rho^+$ (dashed histogram) and $\pi^+$ from the $\rho^+$ decay (dotted histogram). Lower row: same distributions for $\pi^-$ and $\rho^-$. \case}
\label{fig:kinematic prim rho pi from rho}
\end{figure}

In the top row of Fig. \ref{fig:kinematic prim rho pi from rho} we compare the $\zh$ (left plots) and $\ptv$ (right plots) distributions for the primary $\pi^+$, the $\rho^+$ and the $\pi^+$ produced in $\rho^+$ decays. The analogue distributions for $\pi^-$ and $\rho^-$ are given in the bottom row.

As can be seen, vector mesons carry typically larger fractions of the initial quark energy than primary pseudoscalar mesons. It is due to the exponential factor in Eq. (\ref{eq:F_explicit PS}) which favors large $Z$ for large $M$. 

Concerning the $\ptv$ distributions, VMs have typically smaller transverse momenta than primary PS mesons. This is due to the hidden spin effect described in Fig. \ref{prodVM}: for rank $r\geq 2$ in the String+${}^3P_0$ model the transverse momenta of the quarks that constitute the vector meson have on the average opposite directions while in the pseudoscalar case where they lay along the same direction. We have then $\langle \ptpt\rangle_{\rm VM}<2\langle \ktkt\rangle<\langle \ptpt\rangle_{\rm PS}$. This is at variance with \texttt{PYTHIA}, where the $Z$-integrated splitting function is the same for vector mesons and for primary pseudoscalar mesons.

Coming to PS mesons from a VM decay, they carry smaller fractional energies and comparable transverse momenta with respect to their parent. They inherit only part of the parent transverse momentum, but to this is added a contribution from the PS momentum $\textbf{p}_{i\rm T}^*$ in the VM rest frame, following Eq. (\ref{compose-pT}).

\begin{figure}[tb]
\centering
\begin{minipage}{0.46\textwidth}
  \hspace{-2em}
  \includegraphics[width=1.0\linewidth]{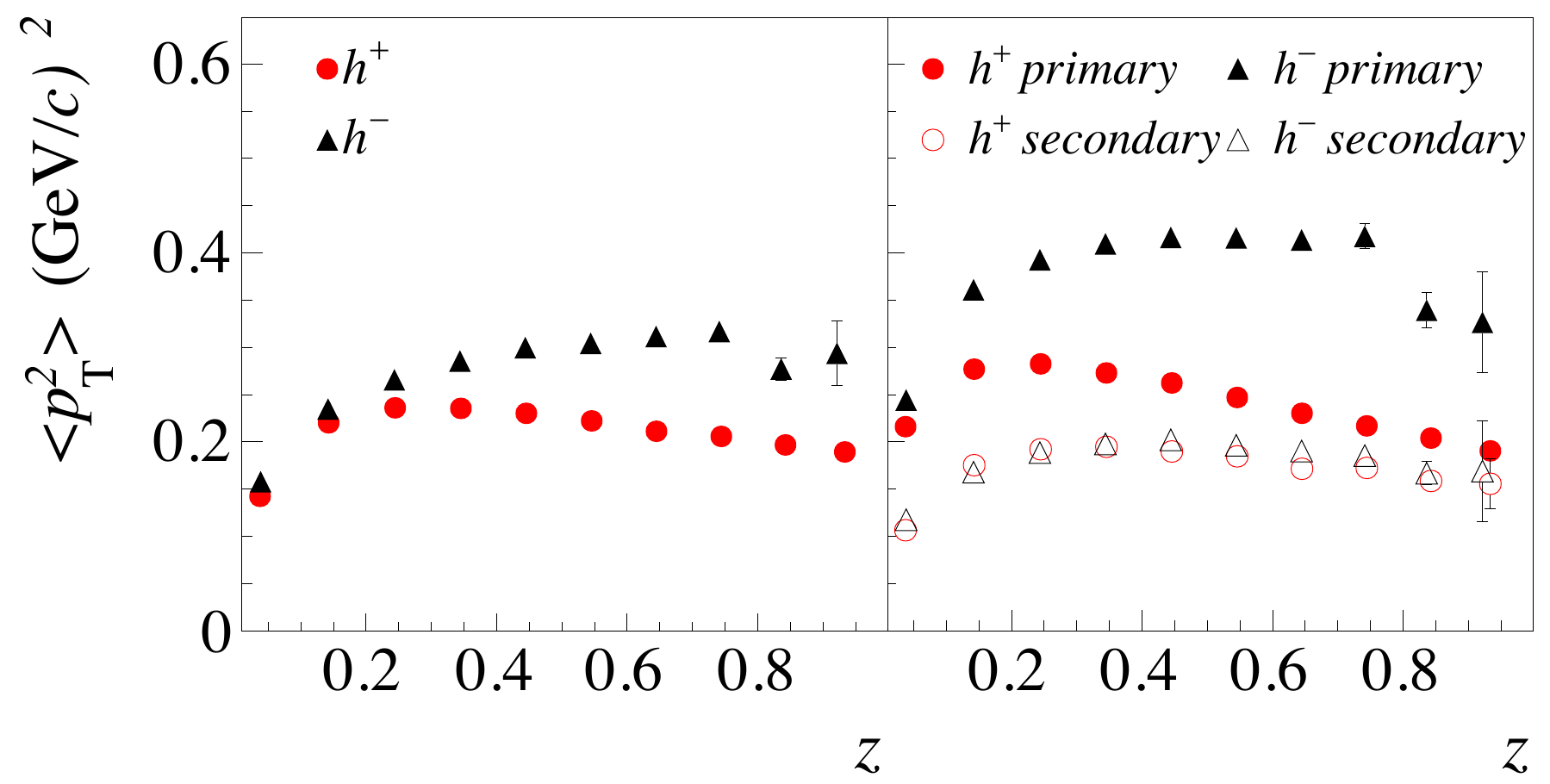}
  \end{minipage}
\caption{Left panel: $\langle \ptpt \rangle$ as function of $\zh$ for positive (circles) and negative (triangles) hadrons. Right panel: same for primary (closed markers) and secondary (open markers) positive (circles) and negative (triangles) hadrons. \case}
\label{fig:kinematic pt2 vs z}
\end{figure}

The hierarchy among the transverse momenta of the different final hadrons is more clearly seen in Fig. \ref{fig:kinematic pt2 vs z}. The left panel shows $\langle \ptpt\rangle$, namely the mean of the $\ptpt$ distribution, as function of $\zh$ for positive and negative hadrons. The same quantity for primary and secondary (from VM decay) mesons is shown in the right panel. Among the primary mesons, the negative ones have larger transverse momenta than the positive ones, as expected from recursive fragmentation models and discussed in Ref. \cite{kerbizi-2019}. Positive and negative secondary mesons, instead, have nearly the same $\langle \ptpt\rangle$, thus the large difference between the positive and negative hadrons at large $\zh$ is reduced when looking at all hadrons but it is still there, at variance with the experimental data \cite{compass-pt2}.

\begin{figure}[b]
\centering
\begin{minipage}[tb]{0.46\textwidth}
\hspace{-3em}
 \includegraphics[width=0.9\linewidth]{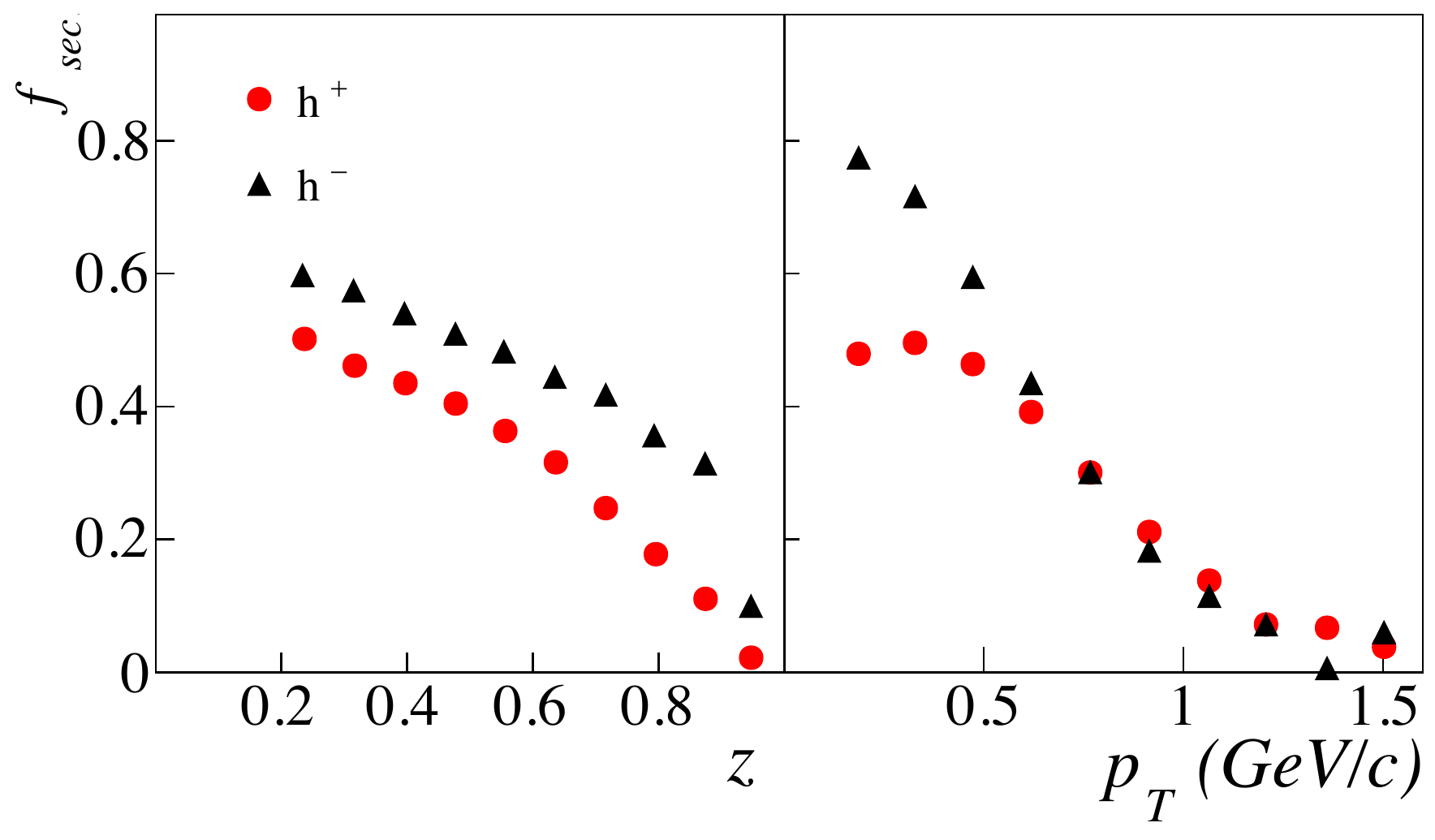}
  \end{minipage}
 \caption{Fraction of secondary positive (circles) and negative (triangles) hadrons produced in decays of vector mesons as function of $\zh$ (left panel) and $p_{\rm{T}}$ (right panel). \case}\label{fig:pi+ from rho / prim. pi+}
\end{figure}

Figure \ref{fig:pi+ from rho / prim. pi+} shows the fraction of secondary charged hadrons as function of $\zh$ and of $p_{\rm{T}}$ in the final sample.
Again, the contribution of VM decay decreases with $\zh$. Also, the fraction of the secondary mesons is large at  $\ptv<0.5\,(\gevc)$, rising up to $0.8$ for negative hadrons at small transverse momenta.

\subsection{Transverse spin asymmetries}
\subsubsection{Collins asymmetry}
In the fragmentation process of transversely polarized quarks, the final state hadrons are produced with an azimuthal distribution given in Eq. (\ref{eq:distribution collins}).
When looking at the simulated events, the Collins analysing power $a^{q\uparrow \rightarrow h+X}$ is extracted as
\be \label{an.pow}
a^{q_A\uparrow \rightarrow h+X}(z,\ptv) = 2 \frac{\langle\sin(\phi_h-\phi_{S_{q_A}})\rangle }{ |{\Sv_{q_A\rm T}}|}.
\ee
It has been studied as function of the hadron rank $r$, of the fractional energy $\zh$ and of the transverse momentum $\ptv$ for primary and secondary PS and for VM.
Also, we apply the kinematic cuts $\zh>0.2$ (when looking at $\ptv$) and $\ptv>0.1\,\rm{GeV}/c$. We remind that for these simulations the values $\glgt=1$ and $\thetalt=0$ have been used and that other choices give different Collins analysing powers, as will be shown in Sec. \ref{sec:sensitivity}. 

\begin{figure}[tbh]
\centering
\begin{minipage}[t]{.48\textwidth}
\centering
  \hspace{-4em}
  \includegraphics[width=0.75\linewidth]{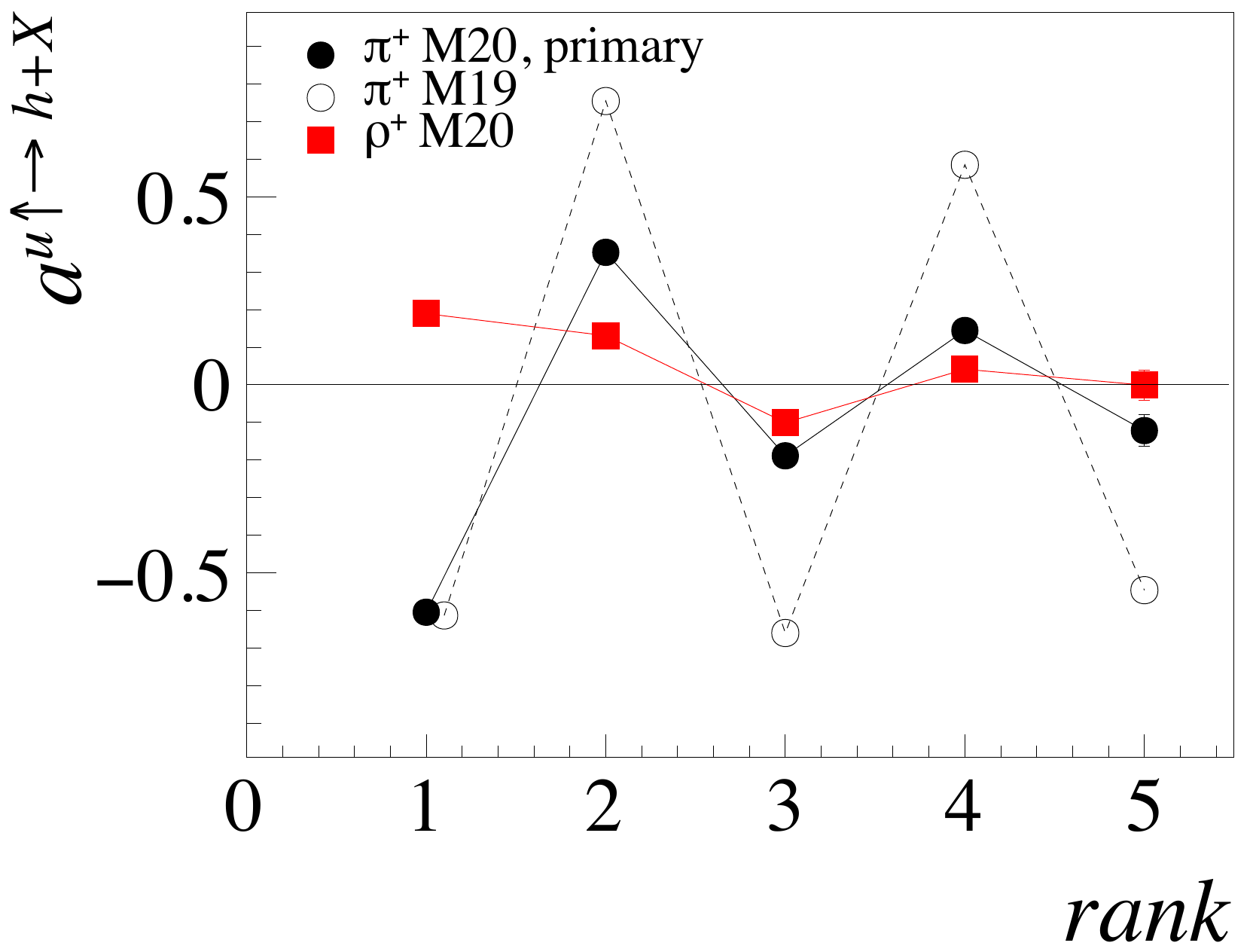}
  \end{minipage}
  \caption{Comparison between the Collins analysing power as function of rank for $\pi^+$ from M19 (open circles), primary $\pi^+$ from M20 (closed circles) and $\rho^+$ (squares). \case}\label{fig:collins rank pi rho}
\end{figure}

Figure \ref{fig:collins rank pi rho} shows the rank dependence of the Collins analysing power for primary $\pi^+$ and for $\rho^+$. It is compared with the analysing power for $\pi^+$ from M19. 
Rank one $\rho^+$ have a Collins analysing power of opposite sign with respect to rank one $\pi^+$ and a factor of $3$ smaller. This is expected combining Eq. (\ref{eq:F_explicit PS}) with Eq. (\ref{Flmn}) which give for the rank $1$ the relation $a^{u\uparrow\rightarrow VM+X}/a^{u\uparrow\rightarrow PS+X}=-\fL$. For $\glgt=1$ this ratio is $-1/3$ \cite{Czyzewski-vm}. For $r\geq 2$ the $\rho^+$ analyzing power has the same sign as the $\pi^+$ analysing power but is smaller. Indeed, both for VM and PS with $r\geq 2$, $\pt$ is more likely on the same side as $\kt$, but it is reduced by $\kpt$ for a VM polarized in the $(\zu,\pt)$ plane.
Also, the analysing power of $\pi^+$ mesons decays faster with the rank as compared to M19. This is expected from the opposite signs of the $D_{\rm TT}$ depolarization factors and from the fact that, for a given rank, the number of antecedent PS mesons is not fixed.

\begin{figure}[b]
\centering
\begin{minipage}[c]{.46\textwidth}
  \hspace{-1em}
  \includegraphics[width=1.0\linewidth]{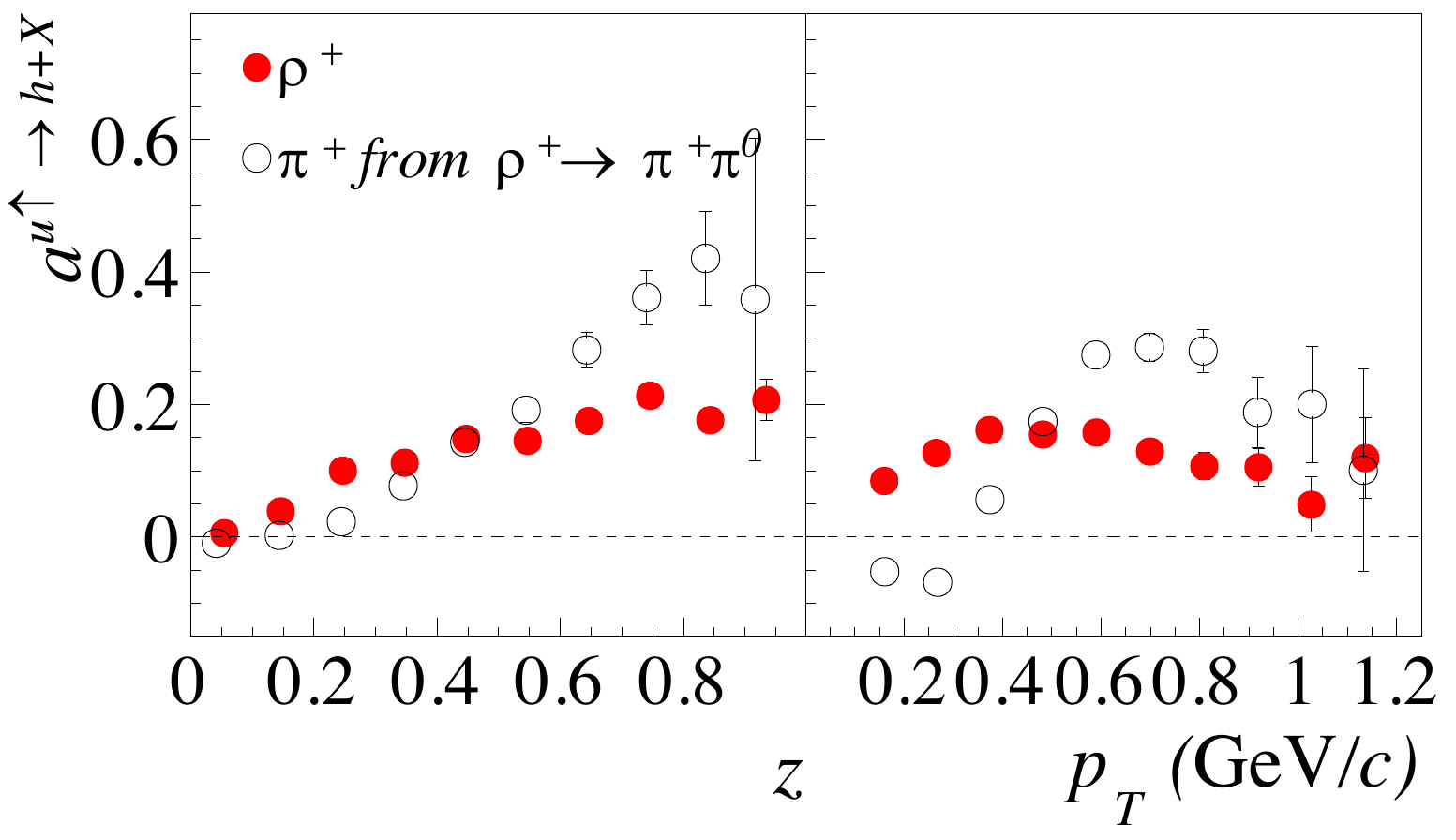}
  \end{minipage}
  \caption{Comparison between the Collins analysing power as function of $\zh$ (left panel) and as function of $p\T$ (right panel) for $\rho^+$ (closed points) and the analysing power of $\pi^+$ produced in the decay $\rho^+\rightarrow \pi^+\pi^0$ (open points). \case}\label{fig:collins rho+ vs decay pi+ pi0}
\end{figure}

Coming back to the observable quantities, Fig. \ref{fig:collins rho+ vs decay pi+ pi0} shows the rank-averaged Collins analysing power as function of $\zh$ (left panel) and as function of $p\T$ (right panel) for $\rho^+$ and $\pi^+$ produced in the $\rho^+$ decay. The $\rho^+$ analysing power is positive as expected from Fig. \ref{fig:collins rank pi rho}.
The analyzing power of the decay $\pi^+$, inherited from the $\rho^+$, exceeds the $\rho^+$ one at large $\zh$. This is due to the fact that large $\zh$ decay pions come mostly from longitudinally polarized vector mesons, which have an analysing power three times larger than the not analyzed ones, according to Eq. (\ref{Flmn}) with $f\L=1/3$.
Looking at the $\ptv$ dependence, decay $\pi^+$ have negative analyzing power at low $\ptv$ that becomes positive at large $\ptv$. This is due to the fact that decay $\pi^+$ with large $\ptv$ can be produced from a rank one $\rho^+$ polarized along $\Mv$ or from rank $\geq 2$ $\rho^+$ polarized along $\Nv$. In the former case, the transverse momentum that the pion acquires in the decay adds constructively (see Eq. (\ref{compose-pT})) to the transverse momentum inherited from the $\rho^+$, which has positive analysing power. In the latter case, the $\rho^+$ has a large transverse momentum and the same Collins analysing power as a PS meson (see Eq. (\ref{Flmn})). The negative analyzing power of the decay $\pi^+$ at low $\ptv$ is interesting. It is probably due to pions which have a transverse velocity in the $\rho^+$ rest frame larger than, but opposite to, the transverse velocity of the $\rho^+$ in the string rest frame.

The $\pi^0$ produced in the same decay has the same analysing power as the positive pion because of parity invariance. With the present choice of parameters, the $\rho^0$ and $\rho^-$ mesons have a similar analysing power as $\rho^+$ and the same features are seen also in the decay of $\rho^-$ and $\rho^0$ mesons.
\begin{table}[h]
\centering
\begin{tabular}{ p{2.7cm}|p{2.5cm}|p{2.5cm} }
 \hline
 M20       &   $\langle a^{u\uparrow \rightarrow \pi^+ +X}\rangle$     & $\langle a^{u\uparrow \rightarrow \pi^- +X}\rangle$             \\
 \hline \hline
 no VM decay                                     &   $-0.308 \pm 0.003$       & $0.218 \pm 0.005$             \\
 with $\rho^+$ decays                              &   $-0.178 \pm 0.003$       & $0.216 \pm 0.005$             \\
 with $\rho^-$ decays                              &   $-0.307 \pm 0.003$       & $0.172 \pm 0.004$             \\
 with $\rho^0$ decays                               &   $-0.210 \pm 0.003$       & $0.151 \pm 0.004$             \\
 with $\rho^{\pm, 0}$ decays                        &   $-0.136 \pm 0.003$       & $0.140 \pm 0.004$             \\
 with all VM decays                                &   $-0.124 \pm 0.003$       & $0.124 \pm 0.003$             \\
\hline
 \hline
 M19                                            &   $-0.251 \pm 0.004$       & $0.257 \pm 0.006$            \\
 \hline
 
\end{tabular}
\caption{Average values of Collins analysing power for charged pions obtained with M20 and M19. For each hadron the cuts $\zh>0.2$ and $p\T>0.1\,\rm{GeV}/c$ have been applied. \case}\label{tab:collins ap averages}
\end{table}

The effects of the decays of different VMs have been investigated separately. The results are summarized in Tab. \ref{tab:collins ap averages} where the integrated analysing power for positive and negative pions is given for all decays switched off, after switching on the $\rho$ meson decays separately or at the same time, and after switching on VM decays. The corresponding values of the analysing power as obtained with the model M19 are also given.
From Tab. \ref{tab:collins ap averages} one can see that primary $\pi^+$ in M20 have larger analysing power than $\pi^+$ in M19 while the reverse is true for $\pi^-$. This is due to the fact that rank two primary pions in M20 have smaller analysing power than in M19, as shown in Fig. \ref{fig:collins rank pi rho}.
The largest reduction of the analyzing power comes from decays of $\rho$ mesons while switching on $\omega$, $K^*$ and $\phi$ decays does not have a large impact. All things considered, after switching on decays of all VMs the analysing power of charged pions is reduced by a factor of two compared to M19. It is also important to note that in this model the absolute values of the analyzing power of $\pi^+$ and $\pi^-$ are different if restricted to primary mesons, but after switching on vector meson decays they become the same, as it is the case also in M19 and as seen in the experimental data \cite{interplay}.

\begin{figure}[b]
\begin{minipage}[t]{0.46\textwidth}
\centering
  \hspace{-1em}
  \includegraphics[width=1.0\linewidth]{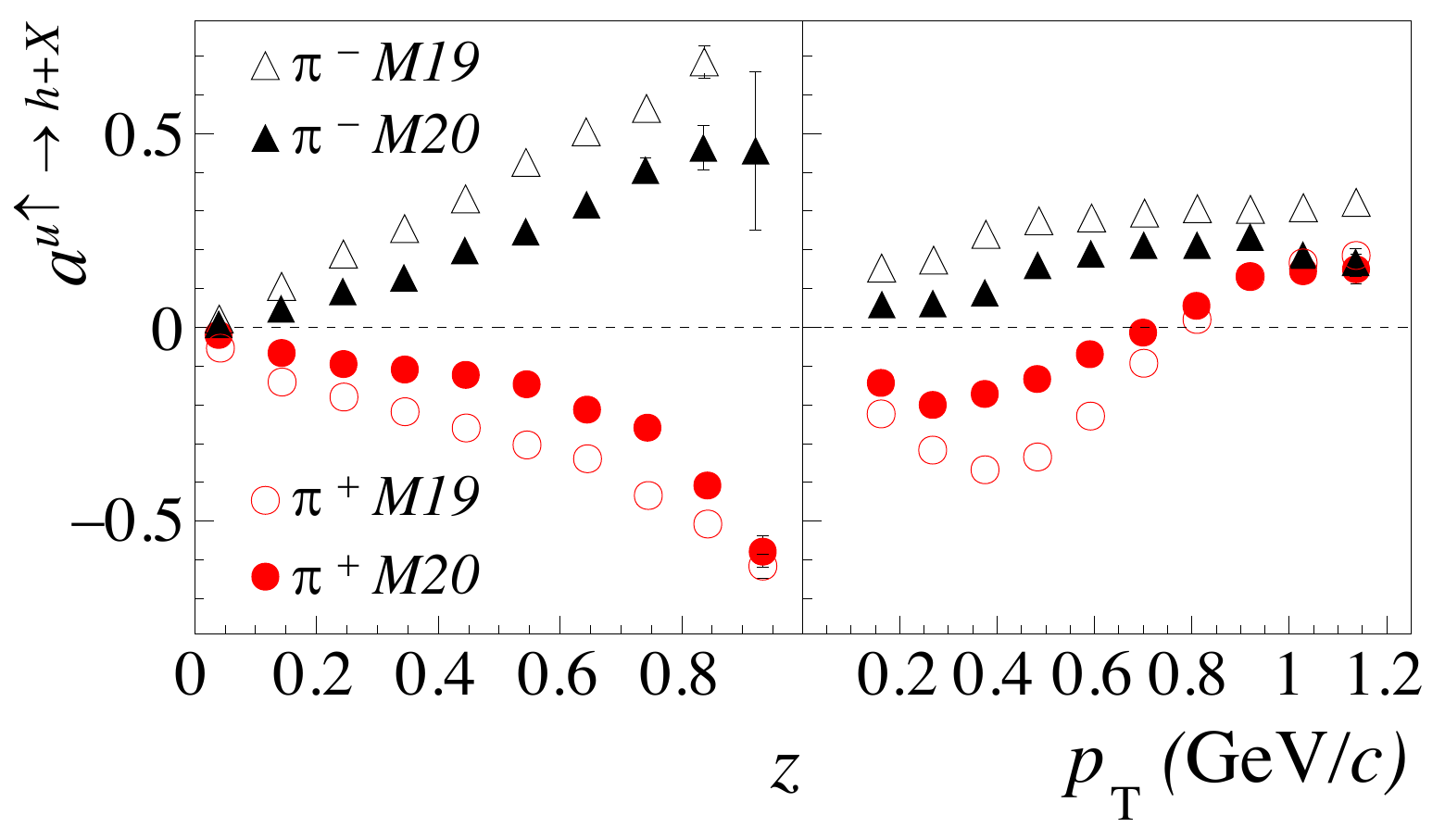}
  \end{minipage}
  \caption{The Collins analysing power of positive (circles) and negative (triangle) pions as function of $\zh$ (left panel) and of $p\T$ (right panel). The closed (open) markers are obtained with M20 (M19). \case}\label{fig:effetto rho pioni}
\end{figure}

The effect of vector mesons on the $\pi^{+}$ and $\pi^-$ Collins analysing power is shown in Fig. \ref{fig:effetto rho pioni} where the analysing powers for charged pions obtained with M20, when the decays of all VMs are simulated, and with M19 are compared. The effect is large for both charges and as function of $\zh$ and of $\ptv$. The $\zh$ dependence of the $\pi^+$ analysing power is not linear any longer, at variance with M19.

The same considerations hold for the analysing power of charged kaons. In this case, the effect of vector mesons is smaller than for pions.

\subsubsection{Dihadron asymmetry}\label{sec:dihadron asymmetry}

Dihadron transverse spin asymmetries are studied looking at hadron pairs $h_1h_2$, where $1$ ($2$) refer to the positive (negative) charged hadron. The azimuthal angle $\phi_{R}$ of $\textbf{R}_T$ is distributed according to
\begin{eqnarray}\label{eq:2h asymmetry}
\nonumber \frac{d^3N_{hh}}{d\z\,d\Mhh\,d\phi_R} \propto  1 + a^{q\uparrow \rightarrow h_1h_2+X}\,|\textbf{S}_{q_A\rm{T}}|\,\sin(\phi_R-\phi_{S_{q_A}}).  \\
\end{eqnarray}
The dihadron analysing power $a^{q\uparrow \rightarrow h_1h_2+X}$ is calculated as $2\langle \sin(\phi_R-\phi_{S_{q_A}})\rangle/|\textbf{S}_{q_A\rm{T}}|$ averaged on $|\textbf{R}\T|$. It has been evaluated as functions of the fractional energy $\z=\zone+\ztwo$ and the invariant mass $\Mhh$ of the pair.
In addition for each hadron of the pair we apply the kinematic cuts $\zh>0.1$, $x_{\rm F}>0.1$ and we ask for $R\T>0.07\,\rm{GeV/c}$, in analogy with the COMPASS analysis \cite{compass-dihadron}. The Feynman $x_{\rm F}$ variable is defined as $x_{\rm F}=2p^z_{cm}/\sqrt{s}$, $p^z_{cm}$ being the hadron longitudinal momentum in the string rest frame.
\begin{figure}[b]
\centering
\begin{minipage}[t]{.46\textwidth}
  \hspace{-1em}
  \includegraphics[width=1.0\linewidth]{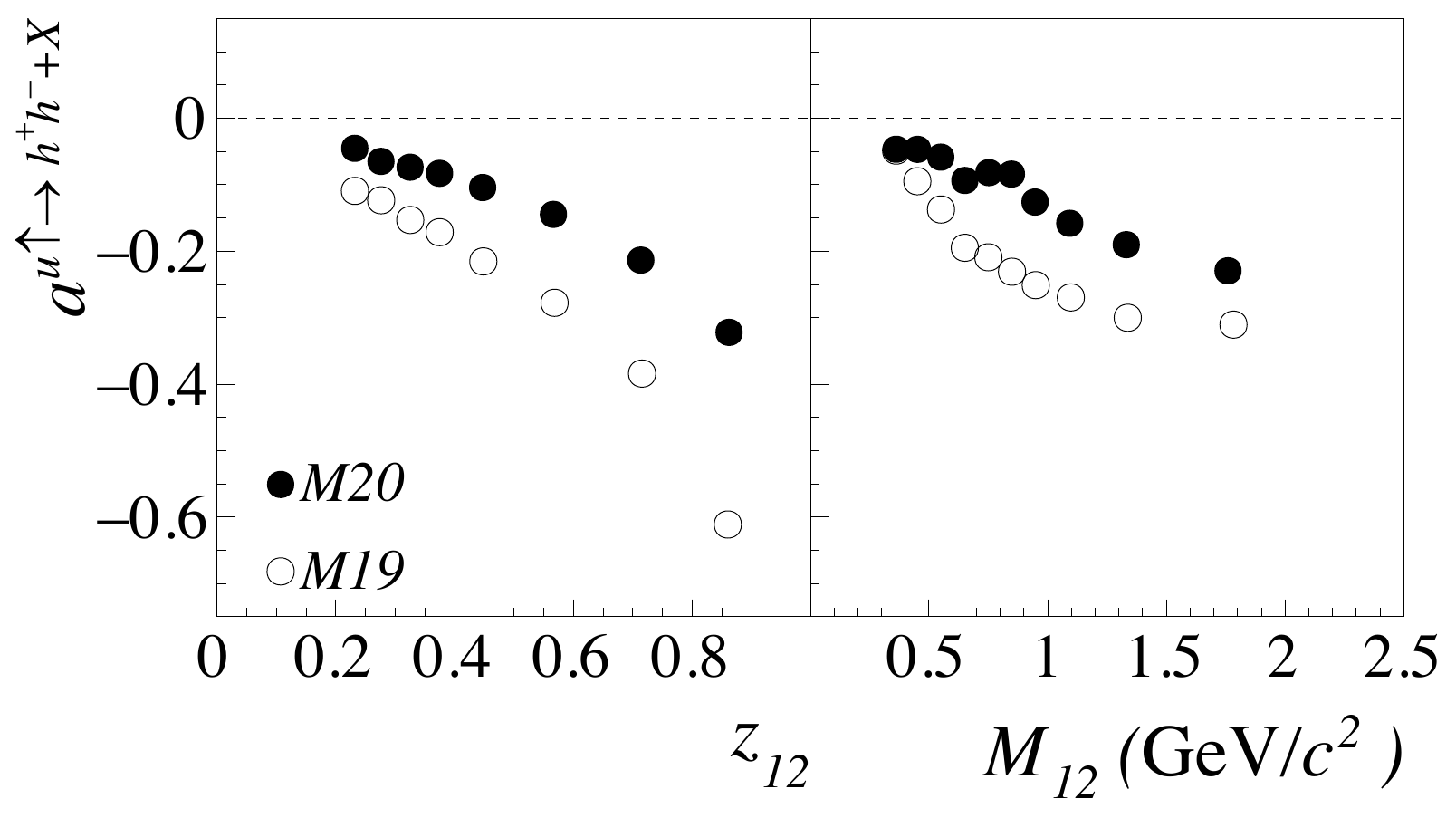}
\end{minipage}
  \caption{Dihadron analysing power for charged pions as function of $\z$ (left panel) and $\Mhh$ (right panel), as obtained with M19 (open markers) and with M20 (closed markers). \case}\label{fig:2h M19 vs M20}
\end{figure}

The result for is shown in Fig. \ref{fig:2h M19 vs M20} as function of $\z$ and $\Mhh$ when switching on the decays of all vector mesons.
The comparison with the dihadron analyzing power obtained by M19 is shown there and summarized in Tab. \ref{tab:2h ap averages}. We see that the introduction of VMs reduces the analyzing power obtained with only PS mesons in M19 by more than a factor of two.
One reason is that the dihadron asymmetry is strongly linked to the Collins effect \cite{interplay} and the latter is smaller for M20 than for M19 (see Fig. \ref{fig:effetto rho pioni}). The other reason is that the VM decay process is invariant by $\textbf{R}\rightarrow -\textbf{R}$, thus secondary mesons do not contribute to this dihadron analyzing power. Instead, they dilute it.
We note also that both in M19 and in M20 the average values of the dihadron and the Collins analysing powers are comparable, like in the experimental result \cite{interplay}. Still it must be reminded that these results are obtained without primordial transverse momentum, which reduces the Collins analysing power but does not affect the dihadron analysing power \cite{kerbizi-2018}.

\begin{table}[h]
\centering
\begin{tabular}{ p{3.5cm}|p{3.cm} }
 \hline
   model     &   $\langle a^{q\uparrow \rightarrow \pi^+\pi^-+X} \rangle$     \\
 \hline
 \hline
 M19                                            &   $-0.246 \pm 0.005$       \\
 %M20 no decays                                     &   $-0.306 \pm 0.010$       \\
 M20                                &   $-0.111 \pm 0.005$       \\
 \hline
\end{tabular}
\caption{Average values of dihadron analysing power for charged pions obtained with M19 and with M20. \case}\label{tab:2h ap averages}
\end{table}

\subsection{Case of $|G_{\rm L}|\neq |G_{\rm T}|$ and $\thetalt \neq 0$}\label{sec:sensitivity}
In this subsection and in Sec. \ref{sec:new 2h asymmetries} we show the effect of changing the values of the parameters $\glgt$ and $\thetalt$ on the relevant observables, namely kinematic distributions and spin asymmetries. We have selected three values for $\glgt$: $5$, $1$ and $1/5$, corresponding to $\fL=0.93$, $\fL=1/3$ and $\fL=0.02$. For each value of $\glgt$ we set $\thetalt=-\pi/2$, $\thetalt=0$ and $\thetalt=+\pi/2$. The values $\thetalt=\pm\pi/2$ maximise the oblique polarization whereas $\thetalt=0$ gives no oblique polarization (in the LR symmetric frame) as can be seen from Eq. (\ref{eq:rho vm MNL decomposition}). The values of the other parameters are the same as given in the previous section.

\subsubsection{Effect on the kinematic distributions}
The effect of changing the values of $\glgt$ and $\thetalt$ on the $\zh$ and $\ptv$ distributions of the produced hadrons is small and is not shown here.
More sizeable effects can be seen in the kinematic distributions of hadron pairs.

\begin{figure}[tb]
\centering
\begin{minipage}{0.48\textwidth}
  \hspace{-2em}
  \includegraphics[width=1.0\linewidth]{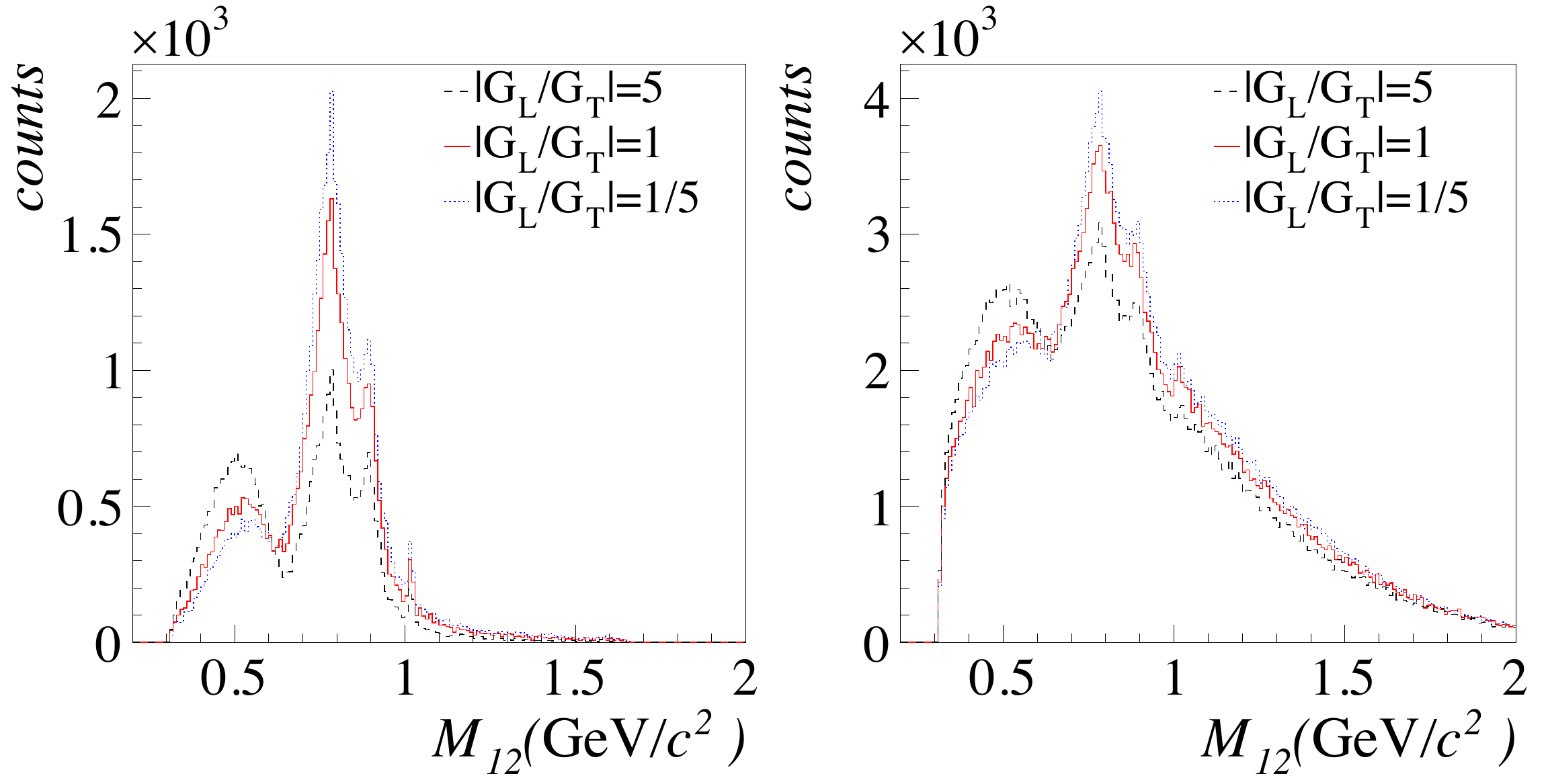}
\end{minipage}
  \caption{Distributions of $\Mhh$ for $h^+h^-$ produced in VM decays (left panel) and for all pairs (right panel), for $\glgt=5$ (dashed line), $\glgt=1$ (continous line) and $\glgt=1/5$ (dotted line) and $\thetalt=0$.}\label{fig:z Mhh glgt effect}
\end{figure}

Figure \ref{fig:z Mhh glgt effect} shows the $\Mhh$ distribution for hadrons coming from decays of vector mesons (left panel) and for all hadrons (right panel), for the parameter values $\glgt=5, 1, 1/5$. The parameter $\thetalt$ has a weaker influence and it is set to zero. The peaks corresponding to the decays $\rho^0\rightarrow \pi\pi$, $\phi\rightarrow KK$ and $K^*\rightarrow K\pi$ can be seen. The shoulder visible on the left of the $\rho^0$ peak is due to the decay $\omega\rightarrow \pi\pi\pi$. 
In the left panel, it is clearly seen that the peaks corresponding to $\rho^0$, $K^*$ and $\phi$ decrease by increasing $\glgt$. This is due to the $\ptv$ and $\zh$ cuts applied to the decay products which make the "acceptance" for VMs depend on its polarization, therefore on these parameters. In fact, in the 2-body decay of a longitudinally polarized vector meson one of the decay products has a low $\zh$ and can easily be rejected when applying the cut $\zh>0.1$. One the contrary, the $\omega$ shoulder increases with $\glgt$, due to the fact that the decay pions of a $\omega$ are emitted preferentially perpendicular, instead of parallel, to the linear polarization of the $\omega$ (see Eq. (\ref{decay-amp3})).

These effects can be seen also in the invariant mass distribution of all hadron pairs, shown in the right panel of Fig. \ref{fig:z Mhh glgt effect}. In this case, an other contribution to the increase of the shoulder on the left of the $\rho^0$ region for $\glgt=5$ %is given by the hadrons coming from decays of different vector mesons which constitute part of the combinatorial background. In fact, these hadrons can be produced close in rapidity, hence at small invariant mass, when the respective parent vector mesons are longitudinally polarized.
is given by pairs of hadrons from the combinatorial background. Indeed,
the decay pions of a longitudinal $\rho$ are separated in rapidity from their parent
by typically more than one unit and can be easily associated with other pions
to form low mass pairs.
For $\glgt=1$ the invariant mass distribution is similar to that obtained with PYTHIA. Instead, comparing with the distribution measured in SIDIS (see eg. Ref. \cite{compass-dihadron}), the combinatorial background is lower than in the data.

From these examples it is clear that the VM polarization has a non negligible role in the "spin-independent" kinematic distributions of the observed hadron pairs, when the experimental cuts are applied, and should be taken into account in the description of all fragmentation processes, the quark being polarized or not.

\subsubsection{Effect on the transverse spin asymmetries}\label{sec: sensisivity spin asymmetries}
In this paragraph we consider the Collins effects for the VMs ("global Collins effect") and for their decay products. The effects on the dihadron asymmetries are illustrated in Sec. \ref{sec:new 2h asymmetries}.

\begin{figure}[tb]
\centering
\begin{minipage}{0.46\textwidth}
\centering
\hspace{-2em}
  \includegraphics[width=1.0\linewidth]{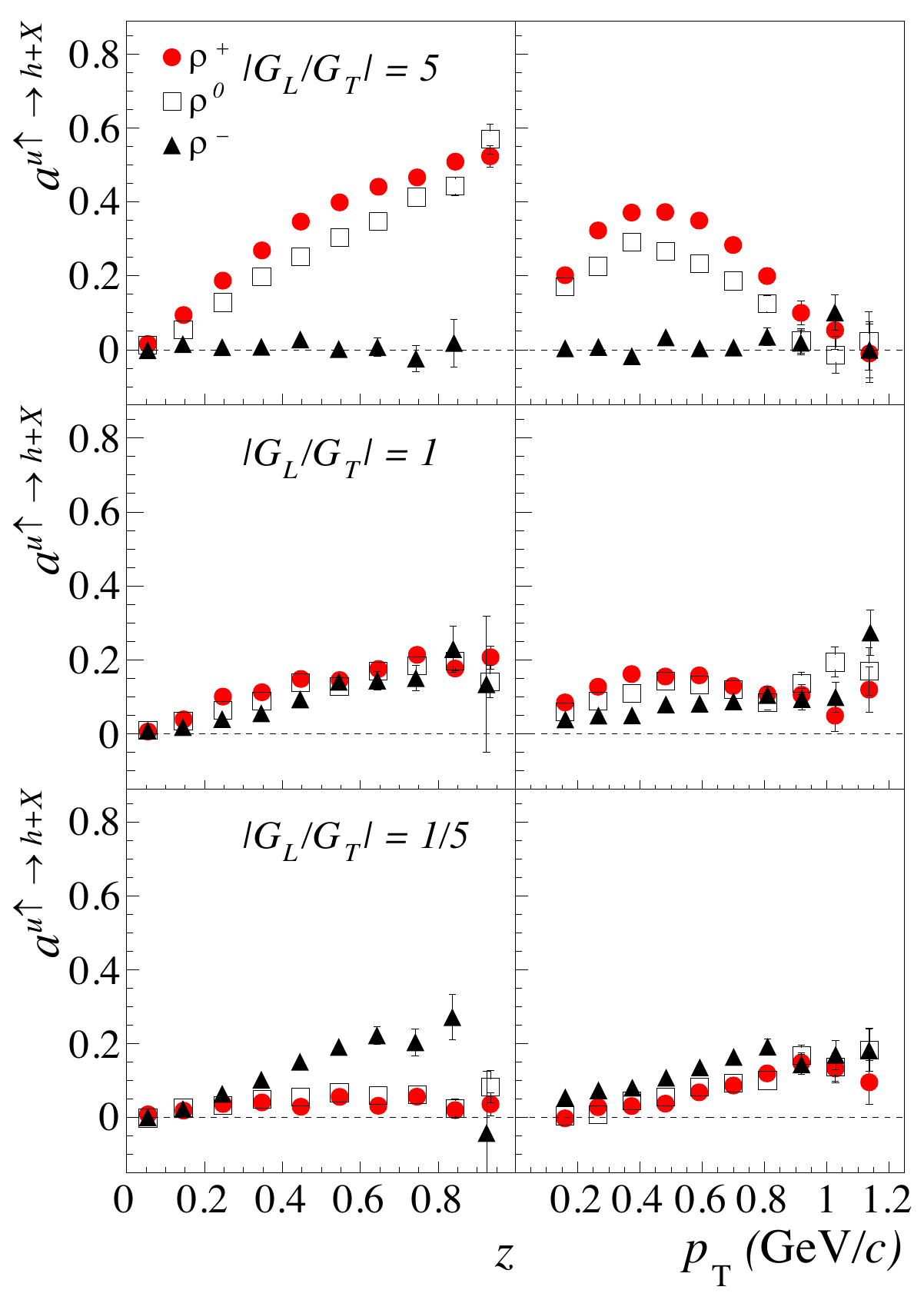}
  \end{minipage}
  \caption{Collins analysing power for $\rho^+$ (circles), $\rho^0$ (squares) and $\rho^-$ (triangles) as function of $\zh$ (left plots) and of $\ptv$ (right plot). The upper row is obtained with $\glgt=5$, the middle row with $\glgt=1$ and the lower row with $\glgt=1/5$. The parameter $\thetalt$ is taken zero. }\label{fig:collins ap rho glgt effect}
\end{figure}

\begin{figure}[tb]
\centering
\begin{minipage}{0.48\textwidth}
\centering
  \includegraphics[width=1.0\linewidth]{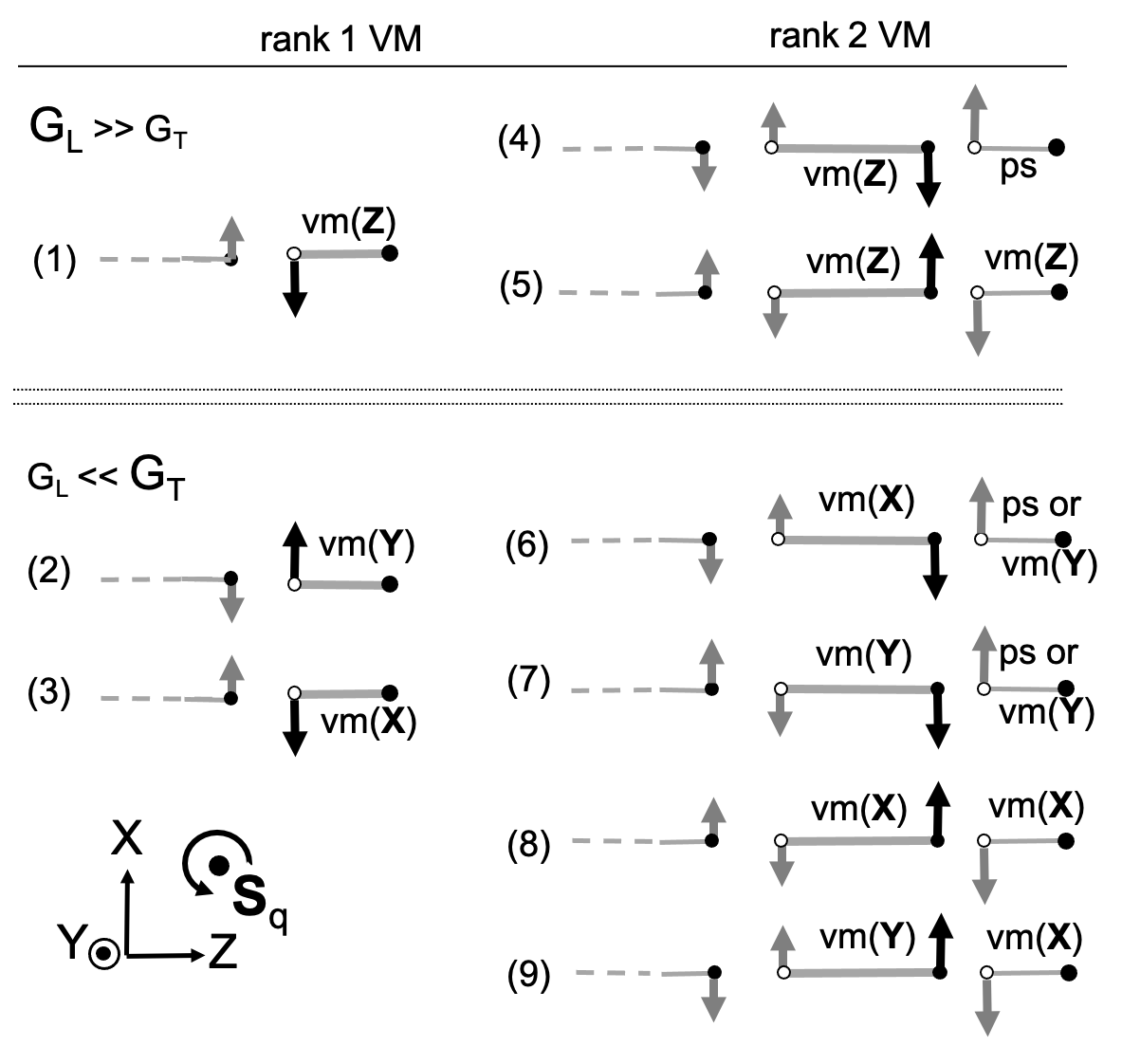} %rank1_2_gray
  \end{minipage}
  \caption{Classical String+${}^3P_0$ picture applied to the production of rank $1$ (left diagrams) and $2$ (right diagrams) vector mesons for $\glgt\gg 1$ (upper part) and $\glgt\ll 1$ (lower part). The quark $q_A$ is polarized along $\YS$; vm($\XS$), for instance, indicates a VM polarized along $\XS$.}\label{fig:rank 1 and 2}
\end{figure}

The effect of varying the value of $\glgt$ on the Collins analysing power of $\rho^+$, $\rho^0$ and $\rho^-$ mesons is shown in Fig. \ref{fig:collins ap rho glgt effect}. The parameter $\thetalt$ does not affect the global Collins analysing power of vector mesons and is set to zero. In each row the analysing power is given as function of $\zh$ (left plot) and of $\ptv$ (right plot).
To interpret these results it is useful to look at the production of rank $1$ and $2$ VMs in the classical String+${}^3P_0$ model illustrated in Fig. \ref{fig:rank 1 and 2} for $\glgt\gg 1$ (upper part) and $\glgt\ll 1$ (lower part). Each diagram shows the application of the ${}^3P_0$ mechanism to the production of VMs polarized along $\ZS$ (upper part), and $\XS$ or $\YS$ (lower part) for an initial quark polarized along $\YS$.

As it can be seen in Fig. \ref{fig:collins ap rho glgt effect}, varying $\glgt$ produces large effects for all $\rho$ mesons. In particular for $\glgt=5$ the Collins analysing power of $\rho^+$ mesons as function of $\zh$ is large rising up to $0.5$. It is then dominated by the rank one diagram (1) in Fig. \ref{fig:rank 1 and 2}. Rank one longitudinally polarized $\rho^+$ have opposite but equal in magnitude analysing power compared to rank $1$ PS meson (compare Eq. (\ref{eq:F_explicit PS}) and Eq. (\ref{Flmn})). For low values of $\glgt$ the analyzing power of $\rho^+$ mesons is reduced due to the presence of two transverse polarization states with opposite analyzing powers (diagrams (2) and (3) in Fig. \ref{fig:rank 1 and 2}).

Concerning $\rho^{-}$, its analyzing power is small at large $\glgt$ and it increases at small $\glgt$, becoming larger than $\rho^+$ and $\rho^0$ for $\glgt=1/5$. For $\glgt=5$ the analysing power is in fact dominated by diagrams (4) and (5) in Fig. \ref{fig:rank 1 and 2} which have opposite signs. For $\glgt=1/5$ the asymmetry is essentially given by diagrams (7) and (6) associated to rank one PS. When associated to rank one VM, (6) and (7) are cancelled by (8) and (9). For $\glgt=1$ the analysing power of all $\rho$ mesons are very similar, as already mentioned. Also the analysing power for $\rho^0$ mesons is the weighted average of the analysing powers for $\rho^+$ and $\rho^-$ due to isospin invariance of the production amplitude.

The effect as function of $\ptv$ is also strong for $\rho^+$ and $\rho^0$, which for $\glgt=5$ behave similarly to PS mesons but with opposite analysing power. %Note that for a $\rho^+$ the analyzing power does not change sign with $\ptv$ as in the case for $\pi^+$ (see Fig. \ref{fig:effetto rho pioni}). This is because the analysing power for rank $1$ and $2$ $\rho$ mesons does not change sign with $\ptv$ as seen in Fig. \ref{fig:collins rank pi rho}, as expected from Fig. \ref{fig:rank 1 and 2} where diagram (4) has the same sign as diagram (1) and dominates over diagram (5).
Decreasing $\glgt$, transverse polarization states become dominant and the shape of the analysing power is changed. At large $\ptv$ $\rho$ mesons are mostly rank $2$ polarized along $\Nv$ and the analysing power is essentially given, in Fig. \ref{fig:rank 1 and 2}, by the diagram (7) associated to a rank one PS. When associated to a rank one VM, diagrams (7) and (9) cancel each other, only contributing to dilute the effect.

\begin{figure}[tb]
\centering
\begin{minipage}{0.46\textwidth}
\centering
  \hspace{-2em}
  \includegraphics[width=1.0\linewidth]{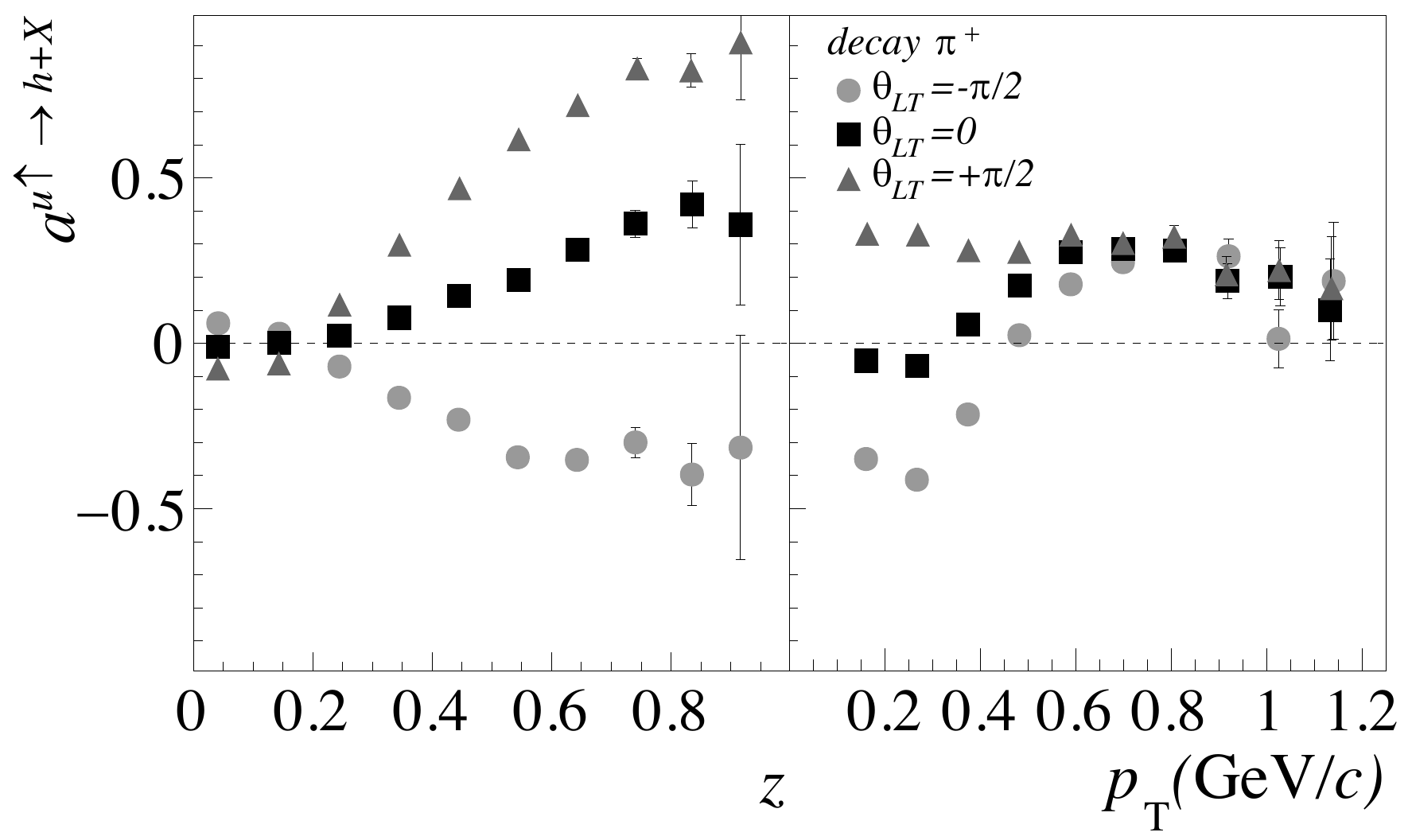}
  \end{minipage}
  \caption{Collins analysing power for $\pi^+$ mesons produced in $\rho^+$ decays with $\thetalt=-\pi/2$ (full circles), $\thetalt=0$ (squares) and $\thetalt=+\pi/2$ (triangles). We have taken $\glgt=1$.}\label{fig:collins decay pi}
\end{figure}

The parameter $\thetalt$ has little influence on the global Collins effect of the resonance, as said before, but a strong influence on the Collins effects of the decay products. This is seen in Fig. \ref{fig:collins decay pi} which shows the analyzing power of $\pi^+$ produced in $\rho^+\rightarrow \pi^0\pi^+$ decays, for $\thetalt=-\pi/2,0,+\pi/2$ and $\glgt=1$. For $\sin\thetalt<0$ the decay process acts as a source of a negative (positive) Collins effect for the fastest (slowest) decay product, as illustrated by the dashed contours in Fig. \ref{fig.oblique}b. For the fastest decay pion this contribution adds destructively with the Collins effect inherited from the $\rho$ and gains over it, giving an overall negative analysing power. The inverse is true for $\sin\thetalt>0$ (continuous contour in Fig. \ref{fig.oblique}).

Concerning the $\ptv$ dependence (right panel of Fig. \ref{fig:collins decay pi}) the largest effects can be seen for $\ptv<0.5\,(\gevc{})$. Decay pions of this domain are mostly emitted with relative transverse momenta opposite to that of the $\rho$ mesons. Besides the cut $\zh>0.2$ selects mainly pions of positive $p^*_z$. Then, for $\sin\theta\LT<0$, looking at the orientations of the dotted ellipses in Fig. \ref{fig.oblique}b, 
one guesses that the pion momentum $\textbf{p}^*_{\rm T}$ in the $\rho$ rest frame is most often on the side opposite to the $\rho$ one.  
Assuming the dominance of the first term in Eq. (\ref{compose-pT}), $\pt(\pi)$ also is on the side opposite to  $\pt(\rho)$. %  are one the same side, 
This explains the negative analyzing power of $\pi^+$ at not too large $\pt$. 
%$|\pt^*| > \pt(\rho)$, the $\pt$(pion) is given by 
%This is likely to be true also for the net pion $\pt$ (given by Eq. ($\simeq$30)),  
% explaining the negative analyzing power of $\pi^+$ at not too large $\pt$. 
As for large $\pt$, they are mainly obtained when $\textbf{p}\T^*$ and $\pt(\rho)$ are on the same side, thus producing a positive analyzing power. The Wigner rotation increases this effect by making the major axis of the dashed ellipse nearly perpendicular to the $\zu$ axis.
For $\sin\theta\LT>0$, according to Fig. \ref{fig.oblique}, the cut $\zh>0.2$ mainly rejects the $\textbf{p}^*_{\rm T}$ which are opposite to $\pt(\rho)$, explaining the positive analyzing power at all $\ptv(\pi)$.

\begin{figure}[tb]
\centering
\begin{minipage}{0.46\textwidth}
\centering
\hspace{-1em}
  \includegraphics[width=1.0\linewidth]{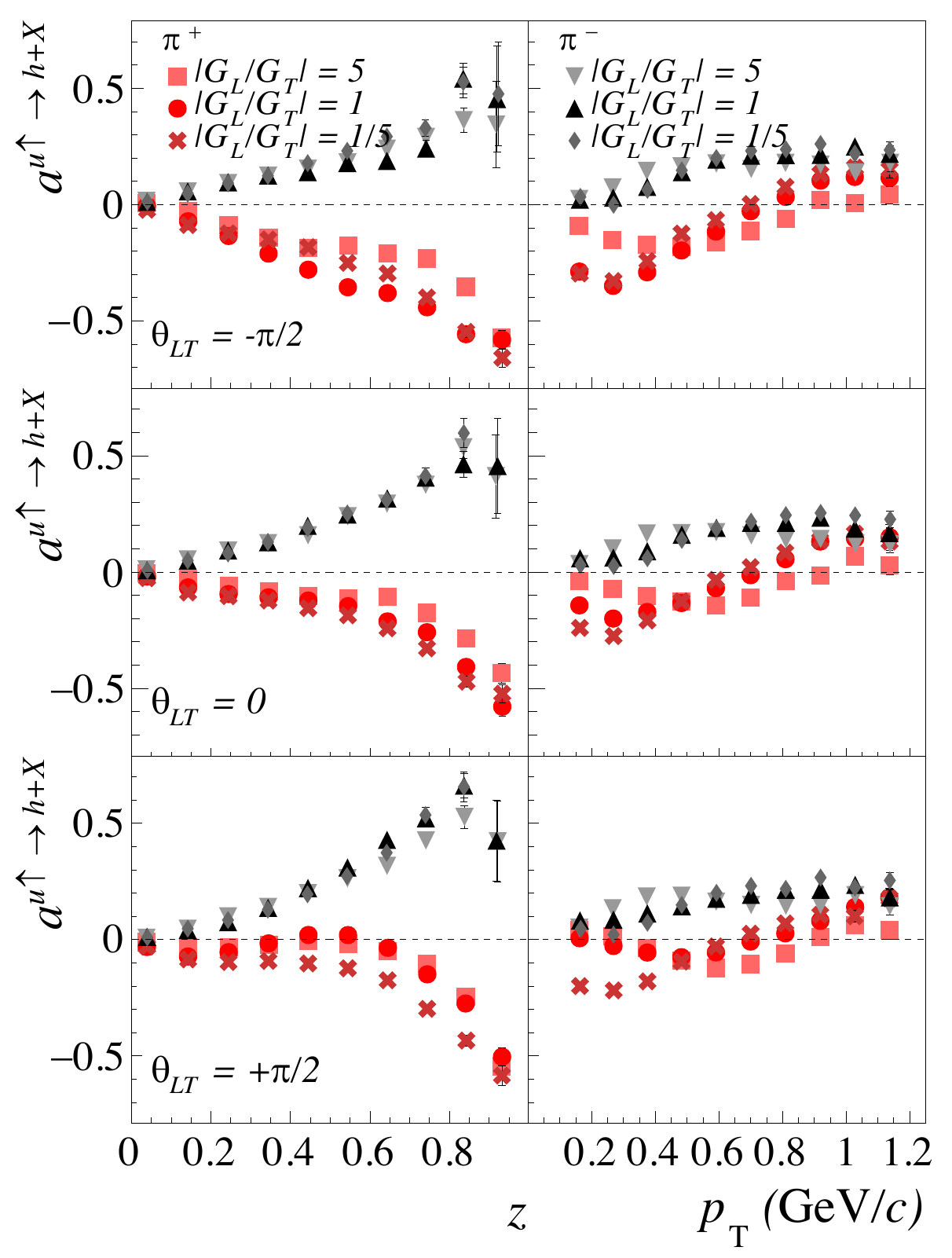}
  \end{minipage}
  \caption{Collins analysing power for $\pi^+$ and $\pi^-$ mesons as function of $\zh$ (left panel) and of $\ptv$ (right panel) for $\thetalt=-\pi/2$ (upper row), $\thetalt=0$ (middle row) and $\thetalt=+\pi/2$ (lower row). For each value of $\thetalt$ the analysing power is calculated for $\glgt=5$, $\glgt=1$ and $\glgt=1/5$.}\label{fig:effetto G collins ap}
\end{figure}

The sensitivity to $\glgt$ and $\thetalt$ of the Collins analysing power for all the final pions, with all VMs decays, is shown in Fig. \ref{fig:effetto G collins ap}. As can be seen the overall effect of vector meson is stronger for favoured fragmentation and weaker for unfavoured fragmentation. In particular the $\zh$ dependence of the $\pi^+$ analysing power is no more linear for both $\pi^+$ and $\pi^-$ as it was in M19. The positive value of $\sin\thetalt$ strongly decreases the size of the $\pi^+$ analysing power and increases the size of the $\pi^-$ analysing power.
As function of $\ptv$ the effect of changing the parameters is large for $\pi^+$ in the small $\ptv$ region, as expected from Fig. \ref{fig:collins decay pi}, whereas for $\pi^-$ only small differences can be seen.

Summarizing, variations in the free parameters $\glgt$ and $\thetalt$ produce large effects on the Collins analysing power of the observed pions, and changes in $\glgt$ can be competed by different choices of $\thetalt$. Precise measurements would allow to fix their values.

\begin{figure}[b]
\centering
\begin{minipage}{0.48\textwidth}
\centering
  \includegraphics[width=1.0\linewidth]{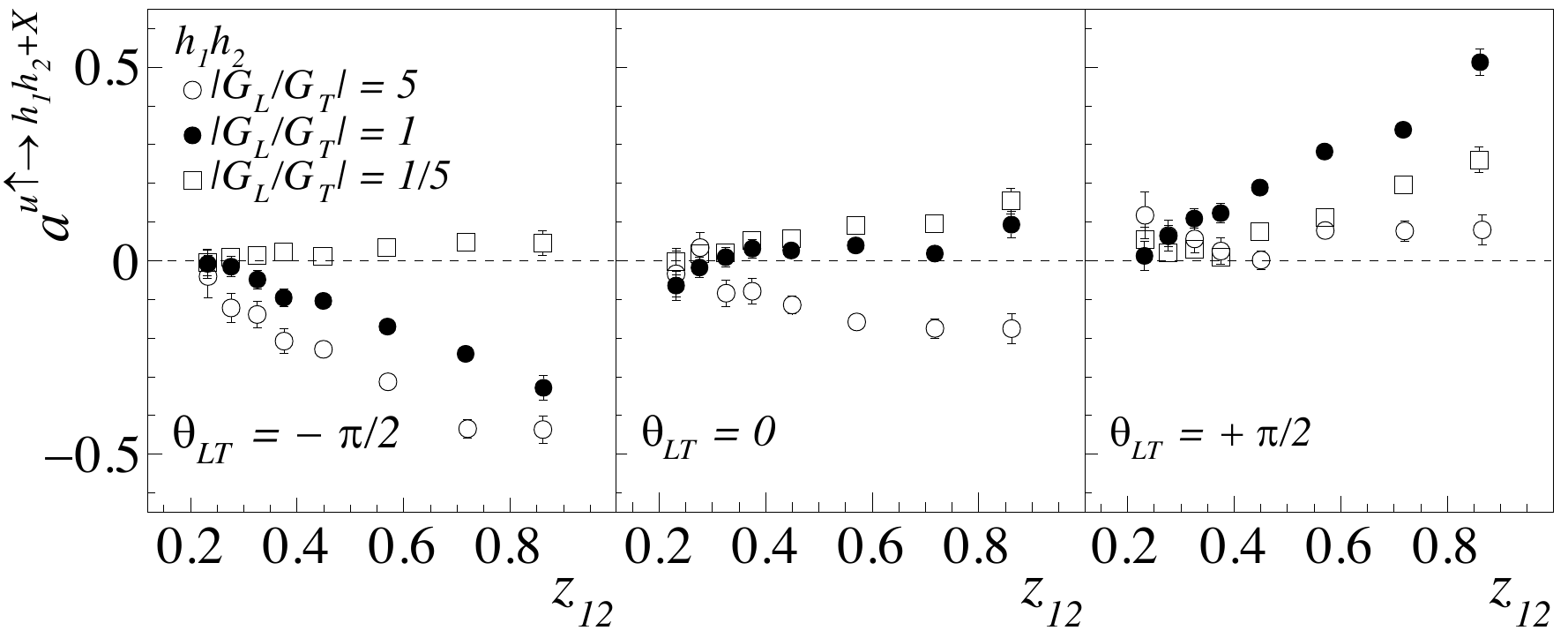}
  \end{minipage}
  \caption{Results for the analyzing power of the $\zh$-ordered dihadron asymmetry as function of the fractional energy $\z$ of the $\pi^+\pi^-$ pair produced in $\rho^0$ decay for the different values of $\thetalt$ and $\glgt$.}\label{fig:2h asymm z-ordered}
\end{figure}

\section{A new dihadron transverse spin asymmetry}\label{sec:new 2h asymmetries}
As mentioned in Sec. \ref{theVMdensitymatrix}, vector meson decays do not contribute to the dihadron asymmetry if in Eq. (\ref{eq:2h asymmetry}) $h_1$ is taken as the $h^+$ of a $h^+h^-$ pair (or the $h^{\pm}$ of a $h^{\pm}h^0$ pair) due to parity invariance.
This is not true when $h_1$ is taken to be the fastest hadron of the pair, namely the hadron such that $\zone>\ztwo$. In this case a dihadron asymmetry may appear, related to the oblique polarization of the vector meson, more precisely to the element $\hat{\rho}^{npl}_{\rm XZ}$ of the density matrix in the $\lbrace \XS,\YS,\ZS\rbrace$ basis of the null plane frame. We refer to this asymmetry as to the \textit{$\zh$-ordered dihadron asymmetry}. $\hat{\rho}^{npl}$ deduces from the density matrix in the LR symmetric frame by the Wigner rotation $\hat{\rho}^{npl}={\cal{R}}_{\Nv}(-\alpha_{\rm W\infty})\,\hat{\rho}\,\cal{R}_{\Nv}(\alpha_{\rm W\infty})$. The angular distribution of $\textbf{R}$ is given by Eq. (\ref{angular}), replacing $\hat{\textbf{r}}$ by $\hat{\textbf{R}}=\textbf{R}/|\textbf{R}|$ and $\hat\rho$ by $\hat\rho^{npl}$. The $z$-ordered dihadron asymmetry is measured by $2\langle\sin(\phi_R-\phi_{\textbf{S}_{q_A}})\rangle/ |\textbf{S}_{q_A\rm T}|$ with the restriction $R_z>0$. It occurs between the primary mesons as well.

The simulated asymmetry is shown in Fig. \ref{fig:2h asymm z-ordered} for pions produced in $\rho^0$ decay as function of the fractional energy of the pair $\z$. The same cuts as in the standard dihadron asymmetry have been applied.
For $\sin\thetalt\neq 0$, the large negative and positive asymmetries shown in the left and right panels are mainly due to the oblique polarization term $\sin\thetalt\,S_{Y}$ in $\RE{\,\hat\rho_{\rm XZ}}$ (see Eq. (\ref{eq:rho vm XYZ decomposition})). When going to $\RE{\,\hat\rho^{npl}}$ the Wigner rotation is, in average, not strong enough to change the sign of the $\rm XZ$ component.
As can be seen the largest positive asymmetry is obtained for $\glgt=1$ and $\thetalt =+\pi/2$. The combination $\glgt=5$ and $\thetalt=-\pi/2$ gives also an asymmetry of the same size but with opposite sign.

The small dihadron asymmetry shown in the middle panel of Fig. \ref{fig:2h asymm z-ordered} for $\thetalt=0$ and $\glgt=1$ requires another explanation. Indeed, with this choice of parameters it can be seen from Eq. (\ref{eq:rho vm MNL decomposition}) that there is no oblique polarization in the LR symmetric rest frame. There is however a non-vanishing $\hat{\rho}_{\rm mn}$ element which, after the Wigner rotation, produces $\hat{\rho}^{npl}_{\rm XZ}<0$ in the null-plane frame. Combined with $R_z>0$ this produces the small positive asymmetry at large $z$ shown in Fig. \ref{fig:2h asymm z-ordered} for $\thetalt=0$ and $\glgt=1$. The change of sign of the asymmetry at small $z$ is instead due to the cuts $z_1>0.1$ and $z_2>0.1$.

For $\glgt\neq 1$ but $\sin\thetalt=0$ the matrix element $\hat{\rho}^{npl}_{\rm XZ}$ receives, by the Wigner rotation, a contribution from $\rho_{\rm mm}-\rho_{\rm ll}\propto |G\L|^2-|G\T|^2$ responsible for the negative (positive) asymmetry for $\glgt=5$ ($1/5$).

It has been checked that the sensitivity to the free parameters as well as the size of the asymmetry remains still large when the $z$-ordered dihadron asymmetry is evaluated by using all final state hadron pairs in the $\rho^0$ mass region. Thus the $z$-ordered dihadron asymmetry depends strongly on the free parameters. The measurement of this asymmetry in SIDIS or $e^+e^-$ annihilation would help to understand whether vector mesons produced in polarized fragmentation processes possess oblique polarization and to determine the values of the free parameters $\glgt$ and $\thetalt$.

\section{Comparison with existing data}\label{sec:comparison}
In order to get hints on the values of the free parameters $\glgt$ and $\thetalt$ we have compared the model results for fully polarized $u$ quarks with the transverse spin asymmetries measured in SIDIS and in $e^+e^-$ annihilation. In particular we compare the simulated asymmetries with the COMPASS results of Ref. \cite{COMPASS-collins-sivers} and Ref. \cite{compass-dihadron}, which are also in good agreement with the HERMES results \cite{hermes-ssa,hermes-dihadron}. Concerning the $e^+e^-$ measurements we compare with the Collins asymmetries measured for oppositely charged back-to-back pions in $e^+e^-$ annihilation to hadrons at BELLE \cite{belle-2019} which are similar to the measurements performed by BABAR \cite{BABAR-Collins} and BESIII \cite{BESIII-Collins}.

\subsection{SIDIS}
\begin{figure}[b]
\centering
\begin{minipage}{0.46\textwidth}
\centering
\hspace{-1em}
  \includegraphics[width=1.0\linewidth]{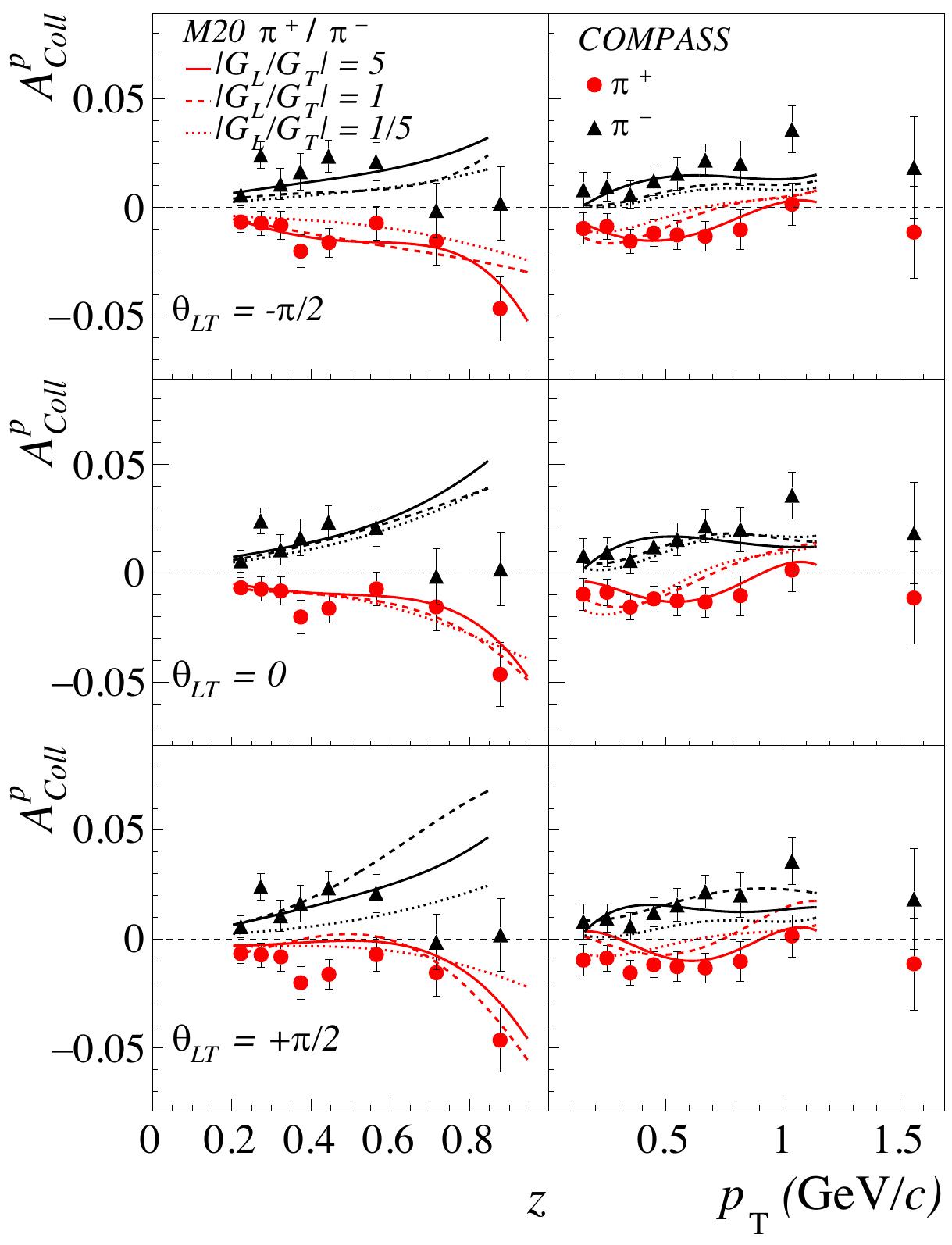}
  \end{minipage}
  \caption{Comparison between the scaled simulated Collins analysing power for $\pi^+$ and $\pi^-$ mesons (curves) and the Collins asymmetry as measured by COMPASS \cite{COMPASS-collins-sivers} (points) as function of $\zh$ (left panel) and of $\ptv$ (right panel) for different combinations of $\glgt$ and $\thetalt$.}\label{fig:collins comparison compass}
\end{figure}

Figure \ref{fig:collins comparison compass} shows the comparison between the Collins analysing power for charged pions as obtained from simulations with the Collins asymmetries measured by COMPASS \cite{COMPASS-collins-sivers}. In experiments, quarks are only partially polarized, following the transversity distribution (see Eq. (\ref{eq:Acoll})). To take into account this fact, for each combination of the free parameters the MC results have been scaled by a constant factor $\lambda$ estimated by a $\chi^2$ minimization procedure using the simulated and measured asymmetries for $\pi^-$ as function of $\ptv$. The factor $\lambda$ is similar for the different combinations of the free parameters and generally larger (up to a factor of two) than the value used for M18 in Ref. \cite{kerbizi-2018}, due to the fact that in M20 the average Collins analyzing power is decreased as a consequence of the introduction of vector mesons. This difference can be recovered by increasing $\IM{\mu}$ by a factor of two while keeping $|\mu|^2$ constant.

All in all, given the small differences of the analysing power for different parameter settings as compared to the experimental precision, no pair of values could be chosen. To exclude some combinations, a $\chi^2$ test at $5\%$ significance level considering the $\pi^+$ and $\pi^-$ asymmetries as function of $z$, and the $\pi^+$ asymmetry as function of $\ptv$ has been performed. For the test, the last two $z$ bins have been excluded since the trend at large $\zh$ is expected to change in simulations of SIDIS events where a realistic mixture of the fragmenting quark flavours is considered \cite{kerbizi-lonnblad}. We find that the test is passed by only three combinations of $\glgt$ and $\thetalt$: $\glgt=5$ with $\thetalt=-\pi/2$ or $0$ and $\glgt=1$ with $\thetalt=0$.

\begin{figure}[t]
\centering
\begin{minipage}{0.46\textwidth}
\centering
  \hspace{-1em}
  \includegraphics[width=1.0\linewidth]{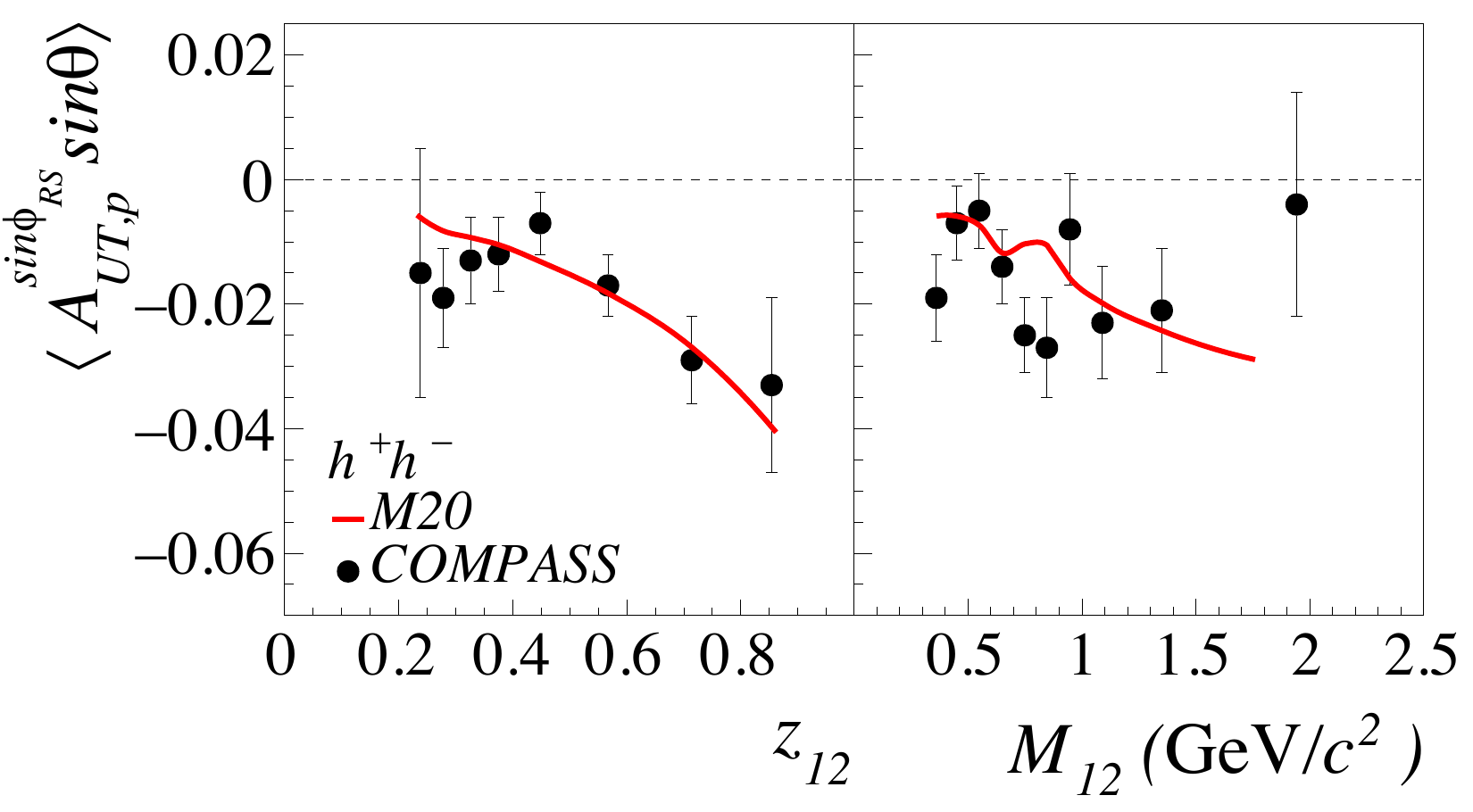}
  \end{minipage}
  \caption{Comparison between the scaled simulated dihadron analysing power for $h^+h^-$ pairs (curve) and the dihadron asymmetry measured by COMPASS \cite{compass-dihadron} (points), as function of $\z$ (left panel) and of the invariant mass (right panel). \case}\label{fig:dihadron comparison compass}
\end{figure}

Concerning dihadron asymmetries, the comparison between the simulated dihadron analysing power and the corresponding asymmetries measured by COMPASS \cite{compass-dihadron} is given in Fig. \ref{fig:dihadron comparison compass}. The asymmetries are shown as functions of $\z$ and of the invariant mass $\Mhh$. The sensitivity of the dihadron analysing power on the new parameters is small compared to the uncertainties of data, and in the figure only the results of the simulations obtained with $\glgt=1$ and $\thetalt=0$ have been used. Also the same scale factor as for the Collins asymmetry has been taken. The comparison is satisfactory apart from the invariant mass dependence in the $\rho^0$ region where the trend of the simulated analysing power seems to be opposite to the data. This could be due to the fact that in the current model we have neglected the interference between amplitudes for the resonant and direct productions of oppositely charged hadron pairs \cite{Collins:1993kq,Collins-Ladinsky,Bianconi-IFF,Bacchetta:IFF_rho0}.

Recent measurement of the Collins asymmetries for $\rho^0$ mesons produced inclusively in SIDIS on protons has been performed by COMPASS in Ref. \cite{Kerbizi:2021xjj}. The $\ptv$ dependence is similar to our simulated results for $\glgt=5$,
up to large statistical uncertainties coming from the combinatorial background under the $\rho^0$ invariant mass peak.

\subsection{$e^+e^-$ annihilation}
We consider now the $A^{UL}_{12}$ asymmetry as measured by BELLE for back-to-back charged pions in the annihilation process $e^+e^-\rightarrow q\bar{q}\rightarrow h_1h_2+X$ \cite{belle-2019}.
$A^{UL}_{12}$ asymmetry is related to $a_{12}$, introduced in Eq. (\ref{eq:AUL ee general}), by $A^{UL}_{12}=a_{12}^U-a_{12}^L$ where the superscript $U$ refers to pairs of pions with unlike charges ($h_1h_2=\pi^+\pi^-$ or $h_1h_2=\pi^-\pi^+$) and $L$ refers to pairs of pions with like charges ($h_1h_2=\pi^+\pi^+$ or $h_1h_2=\pi^-\pi^-$). Experimentally, $A_{12}^{UL}$ corresponds to the amplitude of the $\cos(\phi_1+\phi_2)$ modulation in the ratio $(1+a_{12}^U\,\cos(\phi_1+\phi_2))/(1+a_{12}^L\,\cos(\phi_1+\phi_2))$ between the normalized yields for unlike and like charge pion pairs, and is practically equivalent to $a_{12}^U-a_{12}^L$. $\phi_1$ and $\phi_2$ are the azimuthal angles of $h_1$ and $h_2$ about the thrust axis, measured from the plane defined by this axis and the $e^-$ beam (the thrust axis approximates the $q\bar{q}$ axis).

We restrict ourselves to the case $\zone=\ztwo=z$ and $\ptone=\pttwo=\ptv$. The asymmetry can be written as
\begin{eqnarray}\label{eq:A12}
A^{UL}_{12}(z,\ptv) &=& \langle \hat{a}_{\rm NN} \rangle\times |a^{fav}(z,\ptv)|^2\\
\nonumber && \times \left(\frac{5+5\,\alpha^2+2\,\alpha'^2}{5+5\,\beta^2+2\,\beta'^2}-\frac{5\,\alpha+\alpha'^2}{5\,\beta+\beta'^2}\right).
\end{eqnarray}
It includes the sum over the light quark flavors $q=u,d,s$. The quantities $\alpha=H_1^{unf}/H_1^{fav}$, $\alpha'=H_{1,s}^{unf}/H_1^{fav}$, $\beta=D_1^{unf}/D_1^{fav}$ and $\beta'=D_{1s}^{unf}/D_1^{fav}$ depend on $z$ and the transverse momentum $\ptv$ with respect to the thrust axis. By using isospin and charge conjugation invariance the favoured FF (“fav") is defined as $D_1^{fav}=D_{1u}^{\pi^+}=D_{1\bar{u}}^{\pi^-}=D_{1d}^{\pi^-}=D_{1\bar{d}}^{\pi^+}$ and similarly for the Collins function. Instead, the unfavoured FF (“unf") is defined as $D_1^{unf}=D_{1u}^{\pi^-}=D_{1\bar{u}}^{\pi^+}=D_{1d}^{\pi^+}=D_{1\bar{d}}^{\pi^-}$ for $u$ or $d$ quarks, and $D_{1,s}^{unf}=D_{1s}^{\pi^+}=D_{1s}^{\pi^-}=D_{1\bar{s}}^{\pi^+}=D_{1\bar{s}}^{\pi^-}$ for $s$ quarks (and similarly for the unfavoured Collins function). $a^{fav}$ is the Collins analysing power for the favoured fragmentation. $\langle \hat{a}_{\rm NN}\rangle = \langle \sin^2\theta\rangle /\langle 1+\cos^2\theta\rangle$ \cite{Artru-Collins} where the $\theta$ is the angle between the $e^-$ beam and the thrust axis.

\begin{figure}[t]
\centering
\begin{minipage}{0.46\textwidth}
\centering
  \hspace{-1em}
  \includegraphics[width=1.0\linewidth]{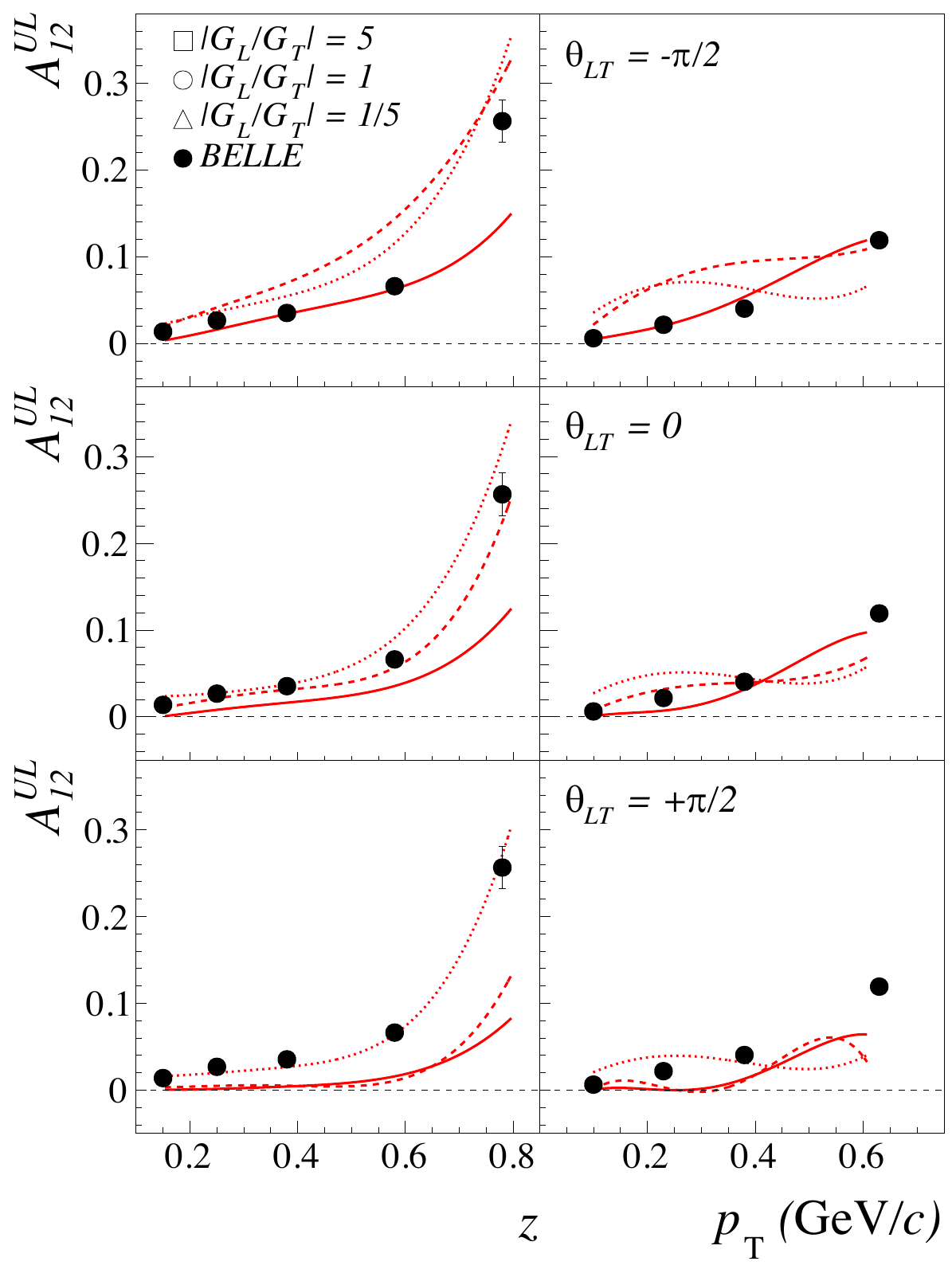}
  \end{minipage}
  \caption{Comparison between the $e^+e^-$ Collins asymmetry $A_{12}$ as measured by BELLE \cite{belle-2019} (full points) as function of $\zh$ (left panel) and of $\ptv$ (right panel), and the calculated Collins asymmetry from the simulation results for different values of the free parameters $\thetalt$ and $\glgt$ (curves).}\label{fig:A12}
\end{figure}

The $A^{UL}_{12}$ asymmetry measured by the BELLE collaboration is shown in Fig. \ref{fig:A12} as functions of $z$ and of $\ptv$. It has been corrected for the charm contribution by using the charm contamination factors provided by BELLE and assuming vanishing Collins asymmetries in events initiated by charm quarks \cite{belle-2019}.
In the figure, the curves are the result of Eq. (\ref{eq:A12}) evaluated using the fragmentation functions obtained from the simulated fragmentations of fully transversely polarized $u$ and $s$ quarks with $\glgt=5,1,1/5$, $\thetalt=-\pi/2,0,+\pi/2$ and the value $\langle \hat{a}_{\rm NN}\rangle = 0.91$ provided in Ref. \cite{belle-2019}. The simulation results have not been rescaled in this case. Each row refers to a different value of $\thetalt$. In each row the curves show the asymmetries from the simulations with different values of $\glgt$. The kinematic cut $z>0.1$ has been applied when looking at the asymmetry as function of $z$, and the cuts $z>0.2$ and $\ptv>0.1\,\gevc$ have been applied when looking at the asymmetry as function of $\ptv$, as in the BELLE analysis. Also, following the BELLE analysis, the cut $\alpha_O<0.3$ is applied on the opening angle $\alpha_O$ of the hadrons with respect to the string axis. %In this analysis we have reduced the cut to $\alpha_O<0.24$ to compensate a possible smearing of the transverse momenta distributions.
This cut is relevant for the $\ptv$ dependence of the asymmetry and has practically no effect when looking at the asymmetry as function of $z$.

As can be seen from Fig. \ref{fig:A12}, the simulated asymmetries are in satisfactory agreement with the BELLE measurements as function of $z$ and of $\ptv$ for the combinations $\glgt=5$ and $\thetalt=-\pi/2$, except for the last point in $z$, and for $\glgt=1$ and $\thetalt=0$.
This is consistent with the comparison with SIDIS measurements of the Collins asymmetries. For $\glgt=5$ and $\thetalt=-\pi/2$ quarks would couple preferentially to vector mesons with longitudinal polarization along the string axis but with some oblique polarization. Instead, for $\glgt=1$ and $\thetalt=0$ there is no preference for transversely or longitudinally polarized vector mesons and these would not have oblique polarization in the LR symmetric frame.

\section{Conclusions}\label{sec:conclusions}
Vector meson production in the polarized quark fragmentation process has been studied within the recursive String+${}^3P_0$ model and the new model M20 has been developed. It improves the previous version (M19) by treating both vector and pseudoscalar meson emissions. It preserves the LR symmetry and the quantum mechanical properties like positivity and entanglement. The production of longitudinally and transversely polarized vector mesons in the LR symmetric frame has been implemented by introducing two different couplings to quarks of complex coupling constants $G\L$ and $G\T$. To this aim, the new parameters $\glgt$ and $\thetalt$ have been added to the complex mass parameter $\mu$ already present in the model M19. Both $\glgt$ and $\thetalt$ enter the spin density matrix of the vector mesons producing angular modulations in the distribution of the decay products. 
The Wigner rotation relating the LR symmetric frame and the null-plane frame, where the decay products are recorded, has been studied.

M20 has been implemented in a stand alone Monte Carlo program which allowed to perform detailed simulations of the fragmentation process.
We have found that the quark spin degree of freedom enters both the kinematic distributions (\textit{hidden-spin} effects) and the spin dependent quantities like the Collins and dihadron asymmetries. The Collins asymmetries of vector mesons turns out to be opposite to their pseudoscalar analogues for the favoured fragmentation and strongly dependent on the $\glgt$ parameter.

The contribution of the decay hadrons to the Collins asymmetry has also been studied and found to depend on the oblique polarization of the vector meson, which is governed by the parameter $\thetalt$. The oblique polarization has also a relevant role in the $z$-ordered dihadron asymmetry, proposed here but not yet measured. Future precise measurements of these asymmetries in SIDIS will allow a better estimate of the free parameters of the model and more safe predictions.

Finally, the simulation results on the Collins and the dihadron asymmetries have been compared to the SIDIS and $e^+e^-$ annihilation data finding an encouraging similarity. The precision of the existing experimental data, however, does not allow to fix the values of the free parameters but give some indication that the values $\glgt\geq 1$ and $\thetalt \leq 0$ are the preferred ones, namely that quarks may couple preferentially to longitudinally polarized vector mesons with oblique polarization in the LR symmetric frame.

To summarize, this new version of the String+${}^3P_0$ model with vector meson production is rich in the predicted phenomena, like the oblique polarization and the hidden spin effects, and it is successful in the description of the experimental data.

\section*{Acknowledgement}
We would like to thank Franco Bradamante for his support and encouragement, and John Collins and Torbj\"orn Sj\"ostrand for the interesting and useful discussions. The work of A. Kerbizi has also been supported by the STRONG-2020 project.

\appendix
\section{The polarization ellipsoid}\label{app:polarization ellipsoid}
In the decay of a vector meson in two pseudoscalar mesons, the angular distribution of the decay products is given by
\begin{eqnarray}  \label{angular'}
 {d{\cal N}(\hat\rv)}/{d\Omega} &=& \frac{3}{4\pi}\,  A^2(\hat\rv) \,,
\nonumber \\
A^2(\hat \rv) &=& \hat r_\alpha \, \hat\rho_{\alpha\alpha'}(h) \,  \hat r_{\alpha'} \,,
\end{eqnarray}
where $\hat\rho_{\alpha\alpha'}$ is the density matrix of the VM, $\rv$ the relative momentum of the decay mesons and $\hat\rv=\rv/|\rv|$. One can replace $\hat\rho_{\alpha\alpha'}$ by the tensor polarization matrix $\RE\, \hat\rho_{\alpha\alpha'}$. 
From this matrix on can build the {\it polarization ellipsoid}, whose symmetry axes are along the eigenvectors, with half lengths equal to the square roots of the eigenvalues. It is the dual of the ellipsoid $r_{\alpha}\,\hat \rho_{\alpha\alpha'}\,r_{\alpha'}=1$ in the polar reciprocal transformation.
This is the 3-D generalization of the polarization ellipse of photons.

\begin{figure}[h]
%\vspace{-1em}
\begin{minipage}{.45\textwidth}
 \includegraphics[width=0.5\textwidth]{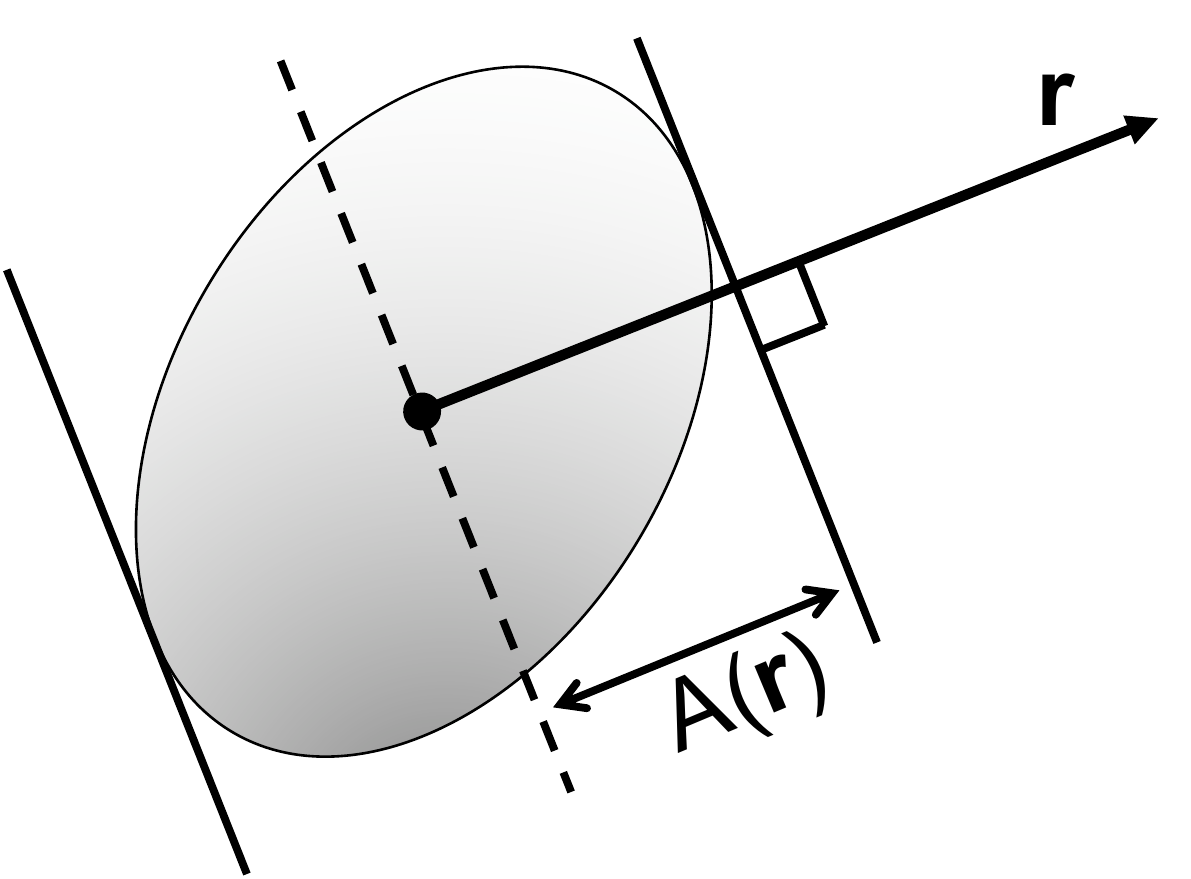} 
\end{minipage}
\caption{Polarization ellipsoid of a vector meson. 
}\label{ellipsoide}
\end{figure}

$A(\hat\rv)$ is obtained geometrically as shown in Fig. \ref{ellipsoide}: the distance between two planes orthogonal to $\hat\rv$ and tangential to the ellipsoid is $2A(\hat\rv)$. 
The projection of the ellipsoid on, for instance, the $(x,y)$ plane is the {\it polarization ellipse} associated to the 2$\times$2 reduced matrix of elements $\RE \, \hat\rho_{xx}$,  $\RE \, \hat\rho_{xy}$, $\RE \, \hat\rho_{yx}$ and $\RE \, \hat\rho_{yy}$.  

\section{The full VM density matrix}\label{appendix:full VM density matrix}

Including the imaginary, antisymmetric part of the VM density matrix, Eq. (\ref{eq:rho vm MNL decomposition}) generalizes as % (with $\Sv\equiv\Sv_q$)

\begin{eqnarray} \label{eq:rho vm MNL full} 
\hat\rho_{\rm ll} &=&   \left(1- \hat a \, \Sn\right) \, {|G\L|^2} / \NG(\Sv)  
\nonumber \\ \nonumber 
\hat\rho_{\rm mm} &=&  \left(1- \hat a \,\Sn\right)  \, {|G\T|^2} / \NG(\Sv) 
\\ \nonumber 
\hat\rho_{\rm nn} &=&  \left(1+ \hat a \,\Sn\right) \, {|G\T|^2} / \NG(\Sv) 
\\ \nonumber 
\hat\rho_{\rm ml} &=& 
i\,( -\Sn+ \hat a ) \, G\T G\L^* / \NG(\Sv) \,\,\,\,\,\,\,\,\,\,\,\,= (\hat\rho_{\rm lm})^*
\\ \nonumber 
\hat\rho_{\rm mn} &=& 
(i \, S_l - \hat a \, S_m) \ {|G\T|^2} / \NG(\Sv) \,\,\,\,\,\,\,\,\,\, = (\hat\rho_{\rm nm})^*
\\ 
\hat\rho_{\rm nl} &=& 
 (-i \, S_m + \hat a \, S_l) \, G\T G\L^* / \NG(\Sv)  ~~= (\hat\rho_{\rm ln})^* \,.
\end{eqnarray}
%
%and $\hat\rho_{\rm \alpha\alpha'} = (\hat\rho_{\rm \alpha'\alpha})^*$. 
\ni NOTE: If $|\Sv_q|=1$, $\hat\rho$ is a matrix of rank 2 (\ie{} $\det\hat\rho=0$). %Indeed, the emittance matrix of $q$ is of rank one in this case and the acceptance matrix of $q'$ is of rank 2.
Indeed, its rank is bounded by the rank of $\hat\rho(q)$, which is $1$, times the rank of the acceptance matrix $\check\rho(q')$, which is $2$ as long as the fragmentation of $q'$ has not yet been performed by the simulation.

The real part of $\hat\rho$ in the $\{\XS,\YS,\ZS\}$ basis linked to the quark transversity (\ie, $S_X$=0) is 
\begin{eqnarray}\label{eq:rho vm XYZ decomposition}
\hat\rho_{\rm ZZ} &=&   \left(1+ \hat a \, m_X\, S_Y\right) \, {|G\L|^2} / \NG(\Sv) 
\nonumber \\
\hat\rho_{\rm XX} &=&   \left(1+ \hat a \, m_X\, S_Y\right) \, {|G\T|^2} / \NG(\Sv) 
 \nonumber \\
\hat\rho_{\rm YY} &=&  \left(1- \hat a \,  \, m_X\, S_Y\right)  \, {|G\T|^2} / \NG(\Sv) 
\nonumber \\
\RE\,\hat\rho_{\rm XY} &=& \hat a \,  \, m_X\, S_Y \, {|G\T|^2} / \NG(\Sv) 
\nonumber \\
\RE\,{\hat\rho}_{\rm XZ} &=& - [ \sin\theta\LT \, (S_Y + \hat a \,  m_X) % \mkp\!\cdot\!\XS) 
\nonumber  \\  
&& + \cos\theta\LT \, \hat a \, m_Y\, S_Z ] \, |G\L G\T| / \NG(\Sv) 
\nonumber \\
\RE\,{\hat\rho}_{\rm YZ} &=& 
 (- \sin\theta\LT \, \hat a \, m_Y 
\nonumber \\ 
&& + \cos\theta\LT \, \hat a \, m_X \,  S_Z ) \, |G\L G\T| / \NG(\Sv) \,.
\end{eqnarray}

\bibliography{main.bbl}
\end{document}